\documentclass[aps,prl,preprint,groupedaddress]{revtex4-1}
\usepackage{bm,amsfonts,graphicx,cancel,psfrag}
\usepackage{latexsym,amssymb,amsmath,color}

\usepackage{theorem}
\usepackage{ulem}

\providecommand\bnabla{\mathbf{\nabla}}
\providecommand\bcdot{\mathbf{\cdot}}

\def\<{\langle}
\def\>{\rangle}
\newcommand\etal{\mbox{\textit{et al.~}}}
\newcommand\diff{\mbox{d}}            
            
\newcommand\p{\ensuremath{\partial}}

\newcommand{\vetU}[0]{\mathbf{U}}

\begin{document}

\title{Passive scalar transport in rotating turbulent channel flow}
\author{Geert Brethouwer$^1$}  %
\affiliation{
$^1$Linn\'e FLOW Centre, KTH Mechanics, SE-10044 Stockholm, Sweden
}

\date{\today}

\begin{abstract}
Passive scalar transport in turbulent channel flow subject to spanwise system
rotation is studied by direct numerical simulations. The Reynolds number
$Re = U_b h/\nu$ is fixed at $20\,000$ and the rotation number
$Ro = 2 \Omega h/U_b$ is varied from 0 to 1.2, where
$U_b$ is the bulk mean velocity, $h$ the half channel gap width and $\Omega$
the rotation rate.
The scalar value is constant but different at the two walls, leading to 
steady scalar transport across the channel. 
The rotation causes an unstable channel side with relatively strong turbulence
and turbulent scalar transport, and a stable channel side with
relatively weak turbulence or laminar-like flow, weak turbulent scalar
transport but large scalar fluctuations and steep mean scalar gradients.
The distinct turbulent-laminar patterns observed at certain $Ro$ on the stable channel
side induce similar patterns in the scalar gradient field.
The main conclusions of the study are that rotation reduces the similarity
between the scalar and velocity field and that the Reynolds analogy
for scalar-momentum transport does not hold for rotating turbulent channel flow.
This is shown by a reduced correlation between 
velocity and scalar fluctuations,
and a strongly reduced turbulent Prandtl number of less than 0.2
on the unstable channel side away from the wall at higher $Ro$.
On the unstable channel side,
scalar scales become larger than 
turbulence scales according to spectra and the turbulent scalar flux vector becomes
more aligned with the mean scalar gradient 
owing to rotation. Budgets in the governing equations of the
scalar energy and scalar fluxes are presented and discussed
as well as other statistics relevant for turbulence modelling.

\end{abstract}

\maketitle
\section{1. Introduction}

In this paper, I present a numerical study of passive scalar transport in
turbulent channel flow subject to spanwise system rotation. A passive scalar, 
a scalar that does not influence the flow, can represent e.g. a contaminant or small
temperature variation. The present investigation is therefore connected to
heat and mass transfer in rotating flows. Examples of these can be found
in many industrial apparatus such as turbo machinery, separators and chemical reactors.

Brethouwer (2005) and Kassinos \etal (2007) investigated passive scalar transport
in homogeneous turbulent shear flow subject to rotation about an axis
normal to the mean shear plane by direct numerical simulations (DNS).
They found that rotation has a large impact on the 
scalar fluctuation intensity and direction of turbulent scalar transport. 
With a transverse mean scalar gradient, 
scalar transport is predominantly in the streamwise direction if the
flow is non-rotating, but if the system rotation exactly counteracts 
the rotation by the mean shear the scalar transport is nearly
aligned with the mean scalar gradient. Brethouwer (2005) observed that
rotation can strongly reduce the turbulent Prandtl number. Many of these 
rotation effects are also shown by rapid distortion theory (Brethouwer 2005).

A generic case to investigate the influence of system rotation on mass or heat transfer
in wall flows is turbulent plane channel flow including a passive scalar
subject to spanwise system rotation.
The effect of spanwise rotation on turbulent channel flow has already been thoroughly investigated
numerically (see e.g. Kristoffersen \& Andersson 1993, Grundestam \etal 2008, Xia \etal 2016)
and experimentally (see e.g. Johnston \etal 1972). 
Most of the numerical studies are limited to 
$Re_\tau = u_\tau h / \nu \leq 194$, where
$u_\tau$ is the friction velocity, $h$ the half channel gap width, 
$\nu$ the viscosity, but recently I have extended 
the study of spanwise rotating channel flow to
$Re = U_b h/\nu = 31\,600$ and 
a wide range of rotation numbers $Ro = 2 \Omega h/U_b$ (Brethouwer 2017).
Here, $U_b$ is the bulk mean velocity and $\Omega$ the system rotation rate.
The main conclusions of these studies can be summarized as follows.
Turbulence and especially the wall-normal velocity fluctuations are augmented
on the channel side were the system rotation is anti-cyclonic, i.e. in the direction opposite 
to the rotation induced by the mean shear, while they are damped on the channel side
were the rotation is cyclonic at moderate $Ro$ (Xia \etal 2016). These channel sides are from now on
called the unstable and stable sides, respectively. 
Large streamwise roll cells, sometimes called Taylor-G{\"o}rtler
vortices, are observed on the unstable channel sides at low to moderate $Ro$
(Liu \& Lu 2007, Dai \etal 2016, Brethouwer 2017). The flow on the stable side relaminarizes 
partly or fully at moderate $Ro$ whereas in the limit of very high $Ro$ the 
whole channel relaminarizes (Grundestam \etal 2008, Wallin \etal 2013, Brethouwer 2017).
A key feature of spanwise rotating channel flow is the 
development of linear part in the mean streamwise velocity profile on the unstable side
where the absolute mean vorticity is nearly zero
(Grundestam \etal 2008, Xia \etal 2016).
Another remarkable feature 
is the occurrence of a linear instability in a range of $Re$ and $Ro$,
which leads to recurring strong bursts of turbulence on the stable
channel side (Brethouwer \etal 2014, Brethouwer 2016).

Passive scalar transport in non-rotating turbulent channel flow has been investigated
extensively as well, see e.g. the DNS studies by Kawamura \etal (1998, 1999)
and Johansson \& Wikstr{\"o}m (1999).
The velocity and scalar fields show a quite high degree of similarity in several aspects,
especially near the wall (Abe \& Antonia 2009, Antonia \etal 2009, Dharmarathne \etal 2016).
Pirozzoli \etal (2016) carried out DNS of passive scalar transport
in turbulent channel flow for $Re_\tau$ up to 4088 and
Prandtl number $Pr = \nu/ \alpha$ between 0.2 and 1, where 
$\alpha$ is the scalar diffusivity.
They observed that the mean scalar profiles follow a logarithmic law
and that large-scale structures are present in the scalar field 
in the outer layer. The turbulent Prandtl number was found to be close
to 0.85 in a large part of the channel, suggesting that the Reynolds analogy
for scalar-momentum transfer
is valid in a non-rotating turbulent channel flow.

Scalar transport in rotating channel flow has been much less examined.
Matsubara \& Alfredsson (1996) have experimentally investigated 
the momentum and heat transfer in
a laminar rotating channel flow. They found
that streamwise roll cells, present at higher rotation rates,
have a profound effect on
heat transfer, and concluded that the Reynolds analogy is not valid
since the Nusselt number changes but not the skin friction under the influence of rotation.
Nagano \& Hattori (2003) and Liu \& Lu (2007) carried out DNS of passive scalar transport in 
spanwise rotating turbulent channel flow at $Re_\tau = 150$ and 194, respectively, for 
$Ro_\tau = 2 \Omega h /u_\tau \leq 7.5$ corresponding to  
$Ro \lesssim 0.5$.
They observed a reduced turbulent scalar transport on the stable channel side,
resulting in a relatively steep mean scalar gradient and strong scalar fluctuations.
Liu \& Lu (2007) also observed that the streamwise turbulent scalar transport
is reduced by rotation, like in rotating homogeneous turbulent shear flow (Brethouwer 2005),
and that streamwise roll cells significantly contribute to
scalar transport, in particular at low rotation rates.
The Nusselt number showed a moderate decrease with rotation rate in their DNS.
Wu \& Kasagi (2004) studied scalar transport in turbulent channel flow at
$Re = 2280$ subject to system rotation with varying directions by DNS.
Also in their DNS, rotation
has a large impact on the direction and rate of turbulent scalar transport.
Yang \etal (2011) observed that the velocity and scalar field 
are quite strongly correlated on the unstable side but much less so on the stable
side in DNS of spanwise rotating channel flow at $Re=2666$.

The objective of this study is to investigate passive scalar transport
in plane turbulent channel flow subject to spanwise system rotation
by DNS at higher $Re$ and for a wider range of $Ro$ than in previous studies
in order to obtain a better understanding of turbulent heat and mass transfer in 
rapidly rotating wall-bounded flows. 
I will show that rotation weakens the similarity between the scalar and velocity field
and that the Reynolds analogy is not necessarily valid in rotating channel flow.
With my study, I aim
to assist the development of models for transport and mixing in rotating
turbulent flows.
Several models for turbulent scalar and heat transfer in rotating wall flows
have been proposed (Nagano \& Hattori 2003, Hattori \etal 2009,
M{\"u}ller \etal 2015, Hsieh \etal 2016) and data on scalar transport
in rotating channel flow may be used to validate them.

\section{2. Numerical procedure}

\begin{figure}[t]
\centering
\includegraphics[height=4.5cm]{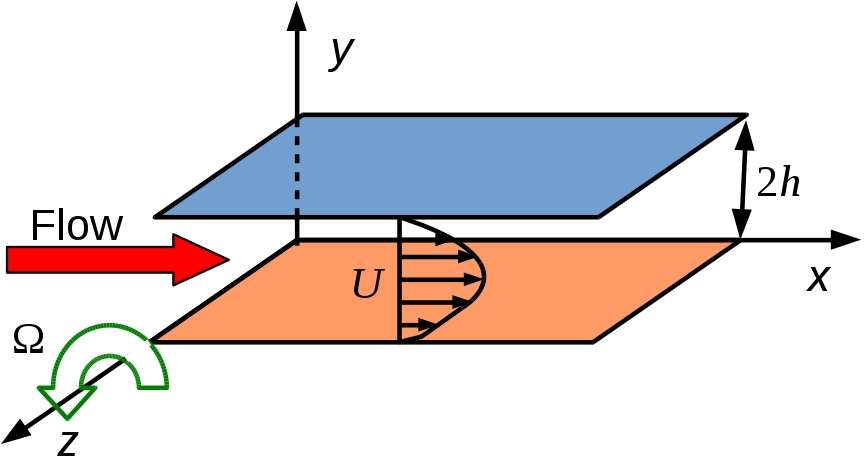}
\caption{Flow configuration and coordinate system.}
\label{geo}
\end{figure}

I have carried out DNS of plane turbulent channel flow with a passive scalar subject
to rotation about the spanwise direction. The flow geometry and coordinate system
are shown in figure \ref{geo}.
The incompressible Navier-Stokes equations (Brethouwer 2016) are solved with
a pseudo-spectral code, as in Brethouwer (2016, 2017), with Fourier expansions
in the streamwise $x$- and spanwise $z$-direction and Chebyshev polynomials
in the wall-normal $y$-direction (Chevalier \etal 2007). 
Together with the Navier-Stokes, the code solves the
advection-diffusion equation for a passive scalar 
\begin{equation}
\frac{\p \Theta'}{\p t} + 
\vetU' \bcdot \bnabla \Theta' 
= \frac{1}{Re\,Pr}\nabla^2 \Theta'
\end{equation}
where $\vetU'$ is the dimensionless velocity and
$\Theta'$ the scalar value.
In the $x$- and $z$-direction,
periodic boundary conditions are used for the the scalar and velocity, and
no-slip conditions for the velocity at the walls. Further, $\Theta' = 0$ at one wall
at $y=-1$ and $\Theta' = 1$ at the other wall at $y=1$,
where $y$ is made non-dimensional with $h$. 
The scalar is thus kept at constant but different values at the wall, like
in Johansson \& Wikstr{\"o}m (1999) and Nagano \& Hattori (2003).
In the statistically stationary state the mean scalar fluxes are equal
at both walls.

In the DNS the flow rate and thus $Re$ was kept constant at $20\,000$ and $Pr=0.71$.
The domain size was $8 \pi h$, $2 h$ and $3 \pi$ in the $x$-, $y$- and $z$-direction,
respectively, and the spatial resolution was similar as in other DNS of turbulent
channel flow (Lee \& Moser 2015). 
The rotation number $Ro$ was varied from 0 (no rotation) to a quite high value of 1.2. 
The parameters of the DNS are listed in table \ref{sim_par}.
\begin{table}
\caption{DNS parameters: $N_x$, $N_y$ and $N_z$ are the number of modes in the streamwise,
wall-normal and spanwise direction, respectively.
}
\begin{center}
\begin{tabular}{lrrrr}
$Ro$ & $Re_\tau$ & $Re^u_\tau$ & $Re^s_\tau$ & $N_x \times N_y \times N_z$\\[3pt]
0	& 1000 & 1000 & 1000 & $2560 \times 385 \times 1920$\\	
0.15	& 976 & 1107& 825 & $2304 \times 385 \times 1728$\\	
0.45	& 800 & 964 & 594 & $2048 \times 361 \times 1536$\\	
0.65	& 700 & 851 & 505 & $1920 \times 321 \times 1440$\\	
0.9	& 544 & 677 & 365 & $1536 \times 257 \times 1152$\\	
1.2	& 423 & 501 & 326 & $1152 \times 217 \times 864$
\end{tabular}
\end{center}
\label{sim_par}
\end{table}
The friction velocity is defined as
$u_\tau = [u^2_{\tau u}/2 + u^2_{\tau s}/2 ]^{1/2}$, where
$u_{\tau u}$ and $u_{\tau s}$ are the friction velocity of unstable
and stable channel side, respectively (Grundestam \etal 2008).
$Re^u_\tau$ and $Re^s_\tau$ are
Reynolds numbers based on $u_{\tau u}$ and $u_{\tau s}$, respectively.

Before the statistics were collected, I ran the DNS
for a sufficiently long time to reach a statistically stationary state 
with a constant mean scalar flux.
I experienced that it can take a long time before the scalar field reaches
such a state when the channel is rotating.

\section{3. Flow field}

Before I present and discuss the DNS results on the scalar transport I will briefly
discuss some one-point flow statistics. A more comprehensive study of rotating channel flow
is presented in Brethouwer (2016, 2017) and other publications cited in the Introduction.
Note that in this paper I only show results for $Re=20\,000$ whereas in Brethouwer (2017)
I often show results for other $Re$. Here below, $U$ denotes the mean streamwise velocity
and $u$, $v$ and $w$ the streamwise, wall-normal and spanwise velocity fluctuations, respectively.
An overbar implies averaging over time and homogeneous $x$- and $z$-directions.

\begin{figure}[t]
\centering
(a)\includegraphics[width=6.0cm]{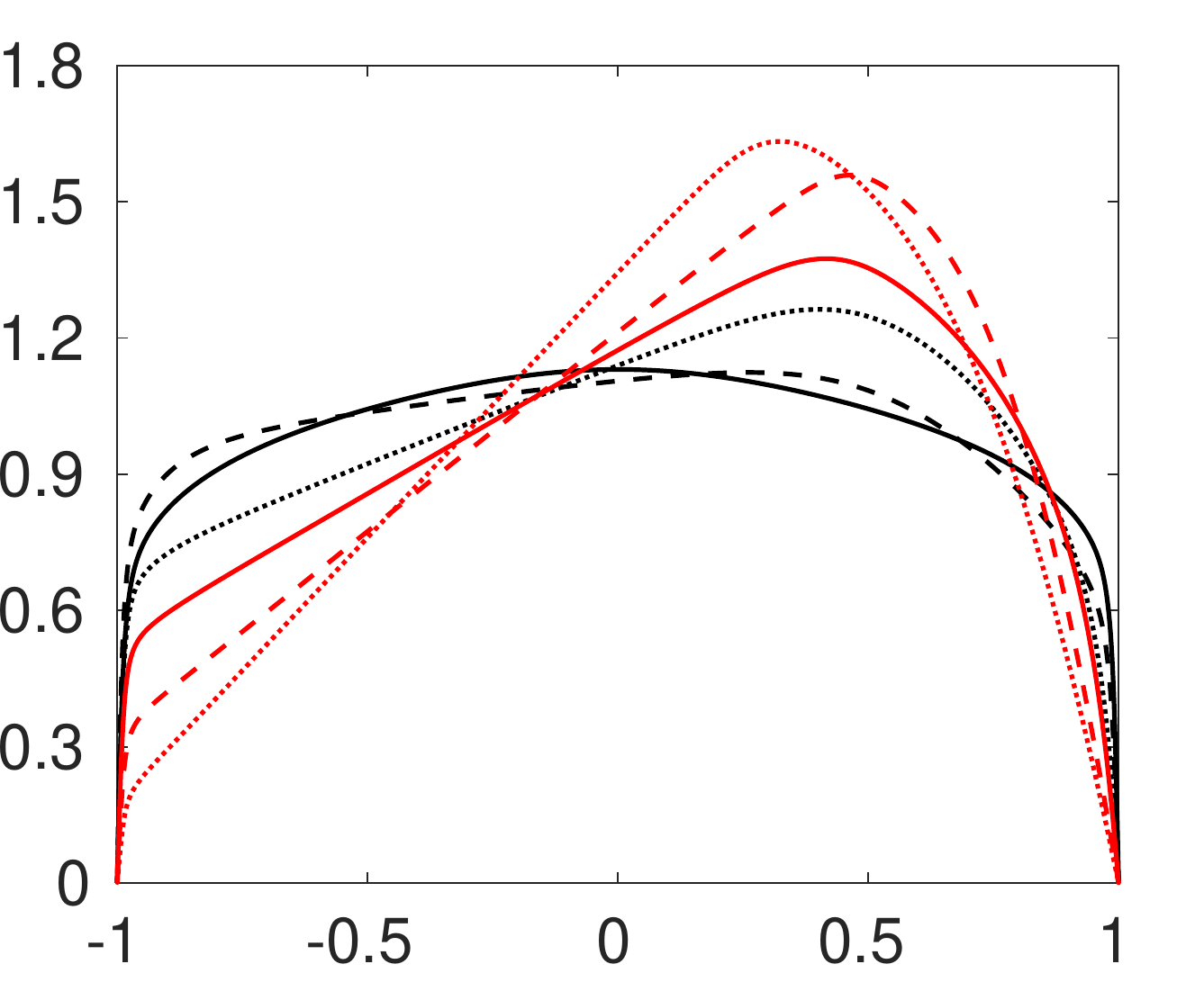}
(b)\includegraphics[width=6.0cm]{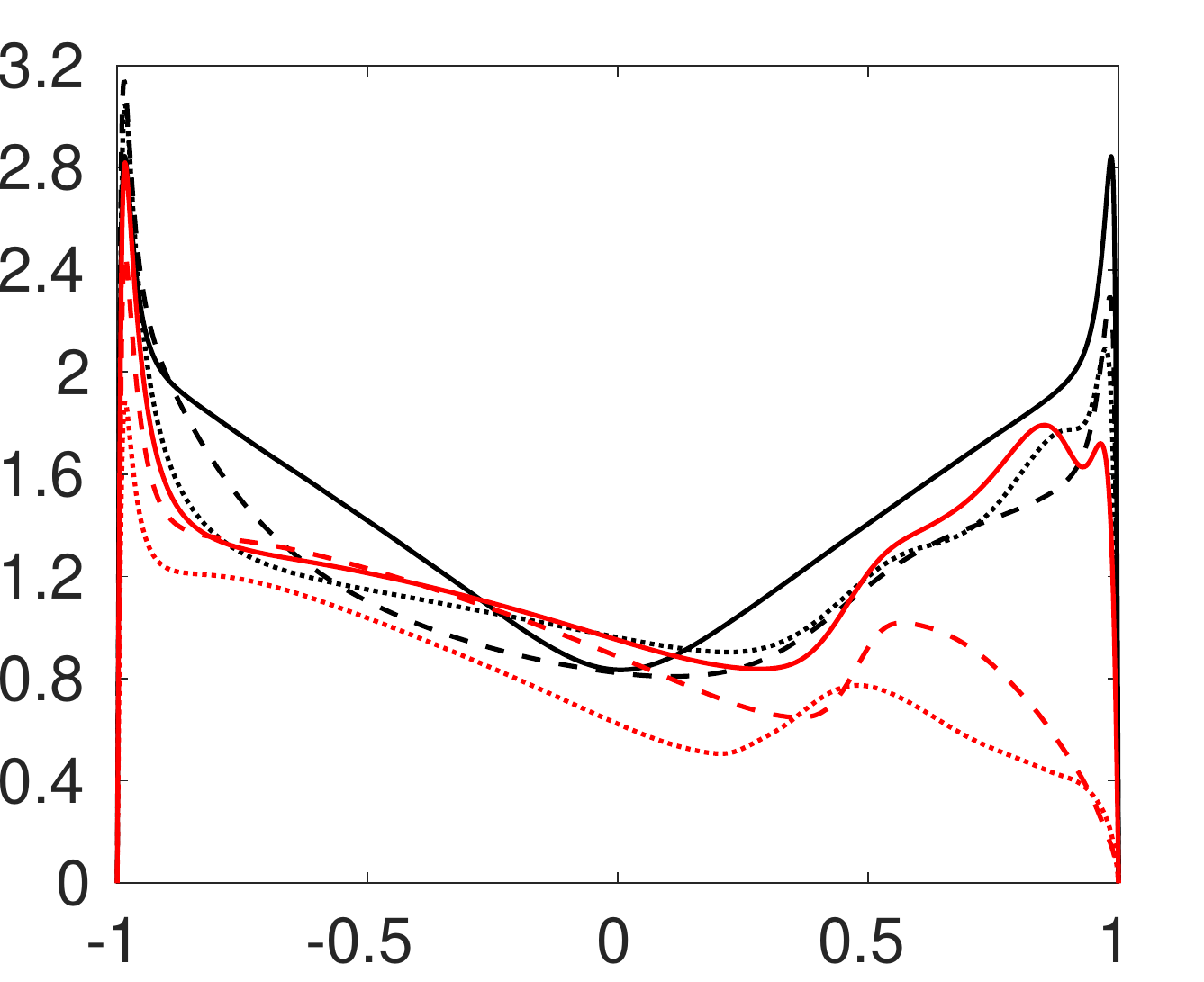}
(c)\includegraphics[width=6.0cm]{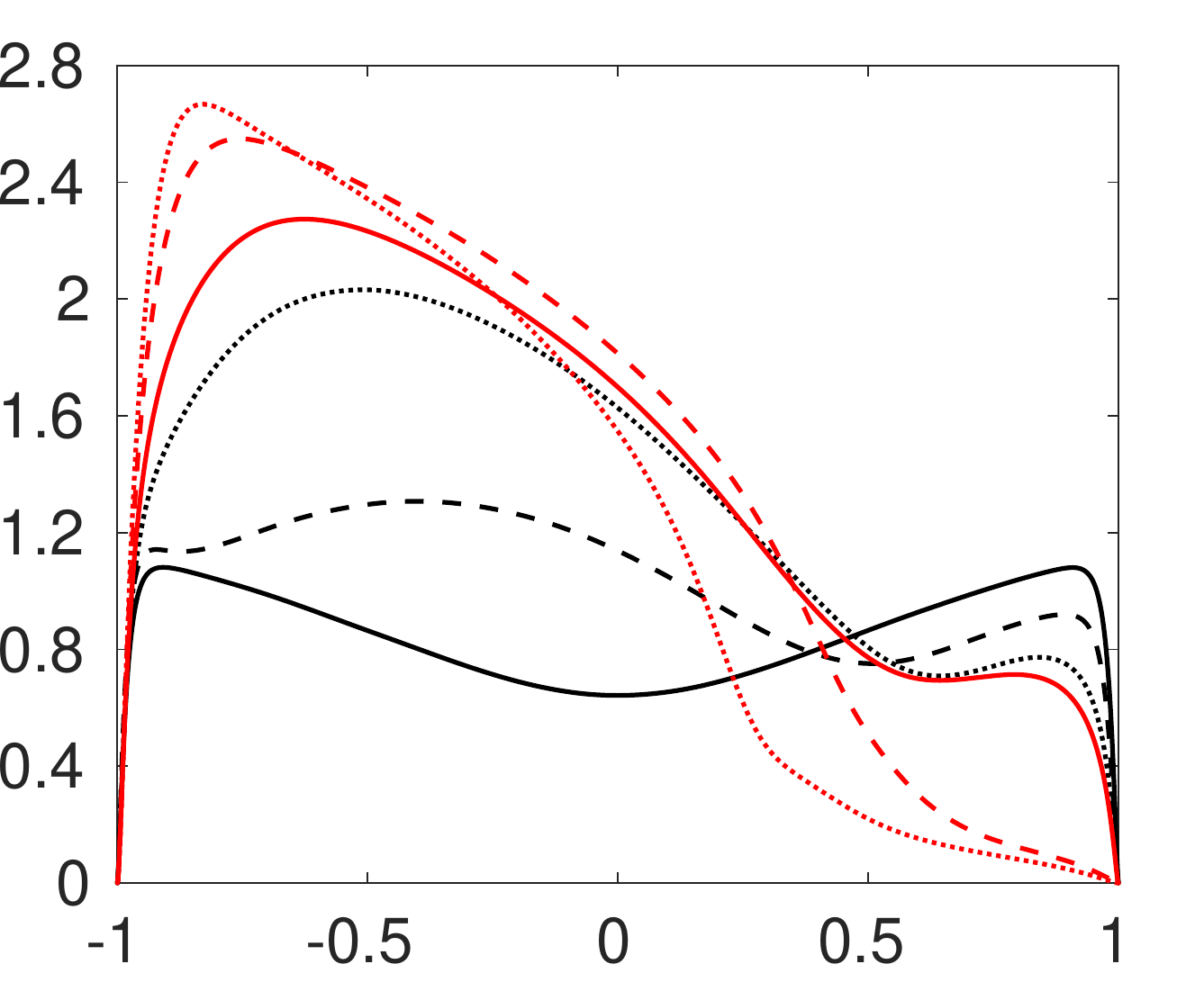}
(d)\includegraphics[width=6.0cm]{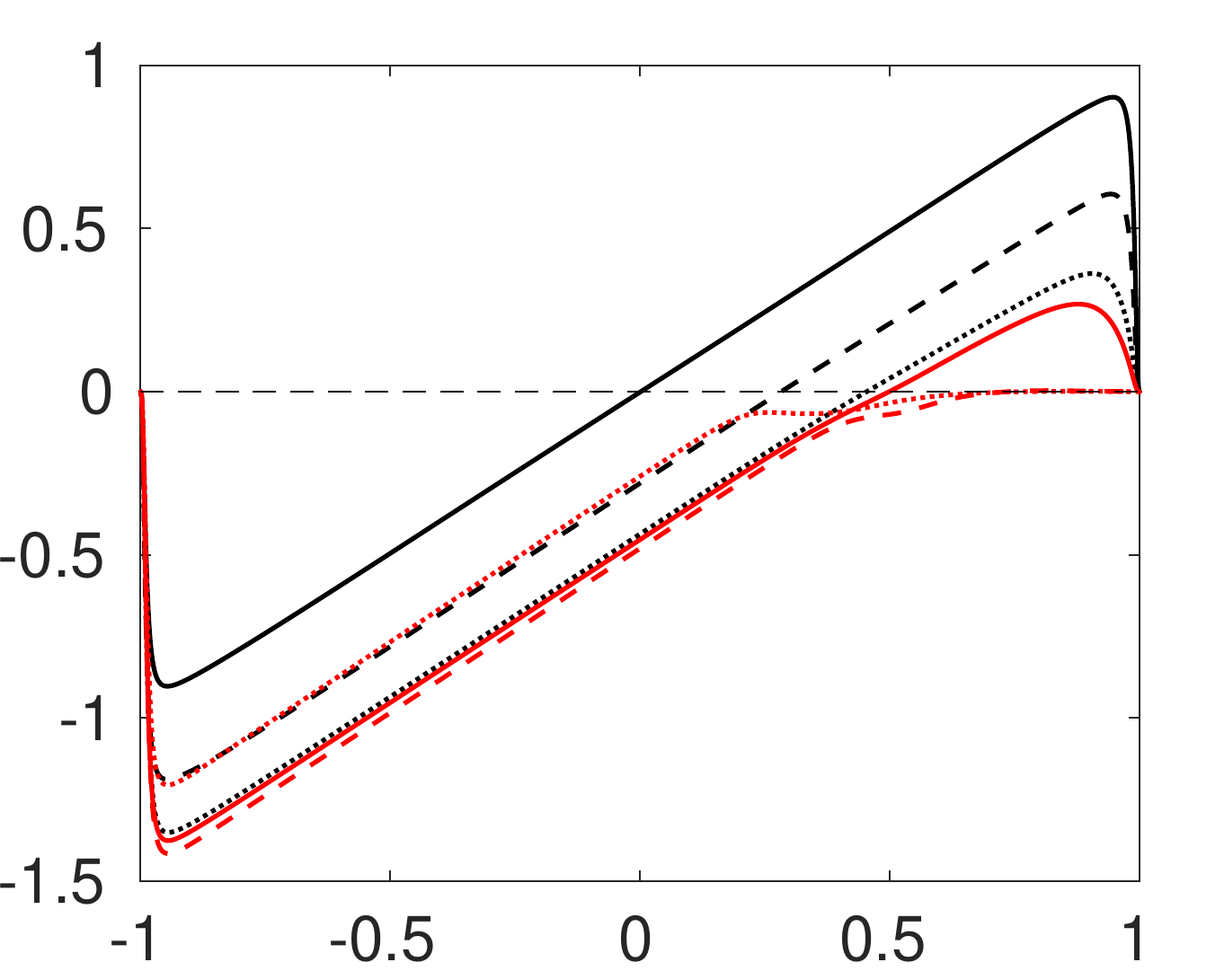}
\caption{Profiles of (a) $U/U_b$,
(b) $u^+$,
(c) $v^+$,
and (d) $uv^+$.
($~^{\line(1,0){20}}$) $Ro=0$,
($---$) $Ro=0.15$,
($\cdot\cdot\cdot$) $Ro=0.45$,
($\textcolor{red}{~^{\line(1,0){20}}}$) $Ro=0.65$,
($\textcolor{red}{---}$) $Ro=0.9$.
($\textcolor{red}{\cdot\cdot\cdot}$) $Ro=1.2$.
}
\label{ustat}
\end{figure}

Figure \ref{ustat}.(a) shows that 
all profiles of $U/U_b$ for $Ro > 0$ are skewed and have a 
linear slope part with $\diff U/\diff y \simeq 2 \Omega$ on the unstable side.
In the DNS $y=-1$ and 1 correspond to the wall 
on the unstable and stable channel sides, respectively.
Figure \ref{ustat}.(b) and (c) show the
profiles of the root-mean-square 
of the streamwise and wall-normal velocity fluctuations
$u^+$ respective $v^+$ scaled by $u_\tau$. 
On the stable side $u^+$ is strongly reduced by rotation
and the sharp peak near the wall disappears.
The maximum of $v^+$ grows, respectively, decays monotonically with $Ro$
on the unstable and stable channel side 
while the normalized Reynolds shear stress, $uv^+ = \overline{uv}/u^2_\tau$,
decays with $Ro$ on the stable side (figure \ref{ustat}.d). 
Both $v^+$ and $uv^+$ are small or nearly zero on the stable channel side
if $Ro \geq 0.9$. 
When $Ro$ is further raised turbulence becomes also weak on the unstable side
and if $Ro \geq 2.4$ turbulent momentum transfer is negligible and the
flow approaches a laminar state (Brethouwer 2017).
In the DNS $Re$ is constant but $Re_\tau$ varies and decreases with $Ro$,
see table \ref{sim_par}, owing to the reduced Reynolds stresses.

\section{4. Visualizations of the scalar field and time series}

\begin{figure}[t]
\centering
(a)\includegraphics[width=8.0cm]{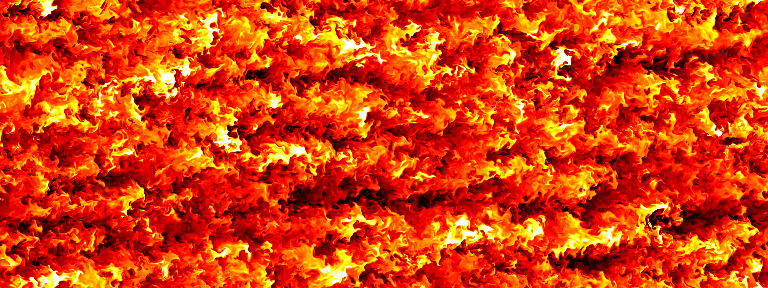}
\vskip2mm
(b)\includegraphics[width=8.0cm]{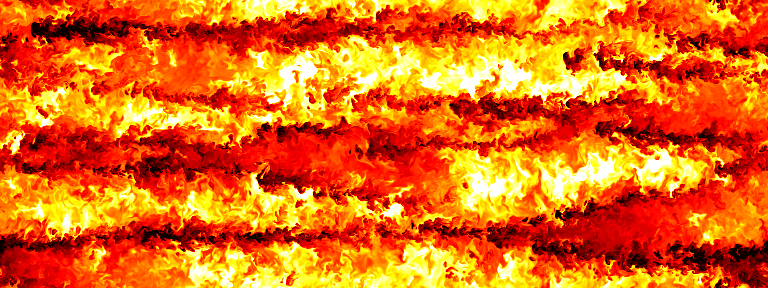}
\vskip2mm
(c)\includegraphics[width=8.0cm]{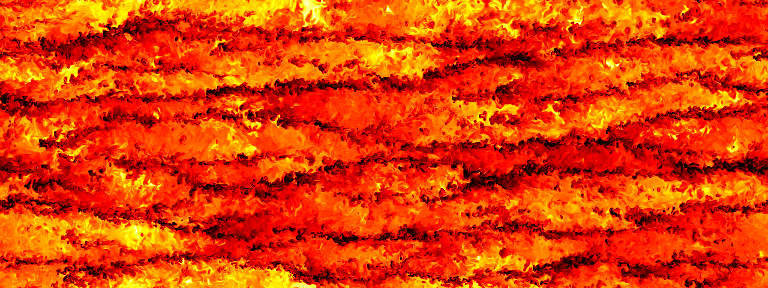}
\vskip2mm
(d)\includegraphics[width=8.0cm]{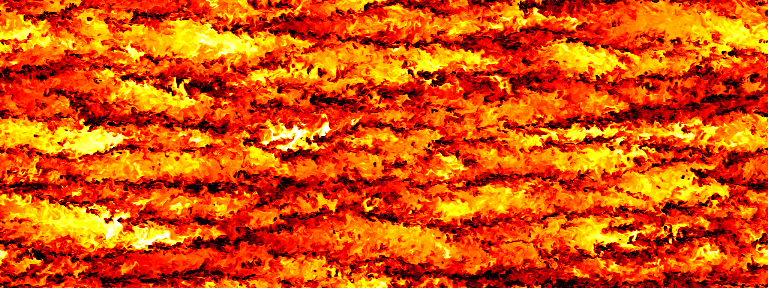}
\vskip2mm
(e)\includegraphics[width=8.0cm]{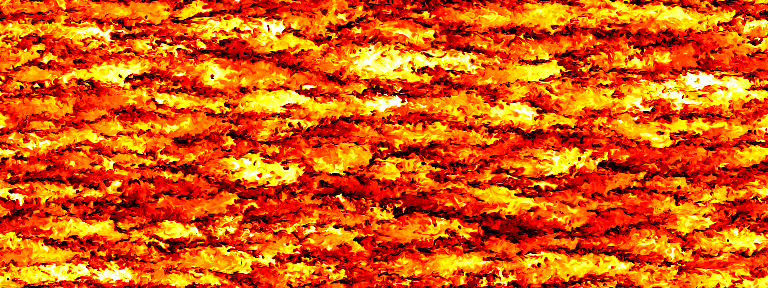}
\caption{
Visualizations of the instantaneous scalar field in an $x$-$z$ plane
at (a) $Ro=0$ and $y=-0.6$,
(b) $Ro=0.15$ and $y=-0.6$,
(c) $Ro=0.45$ and $y=-0.8$,
(d) $Ro=0.65$ and $y=-0.8$ and
(e) $Ro=0.9$ and $y=-0.85$.
Dark colours correspond to low scalar values.
}
\label{visu}
\end{figure}

One-dimensional velocity spectra at $Re=20\,000$ and $Ro=0$, 0.15, 0.45 and 0.9 and 
visualizations of the instantaneous flow field are presented in Brethouwer (2017).
These spectra and visualizations reveal the presence
of large and long streamwise counter-rotating
roll cells on the unstable channel side at
$Ro=0.15$ with a spanwise size of $\pi h/2$.
The roll cells leave an imprint on the scalar field visualized in figure \ref{visu}. 
Whereas no clear
coherent structures can be seen in the scalar field in the non-rotating channel
(figure \ref{visu}.a), at $Ro=0.15$
narrow streaks with low scalar values are observed in the $x$-$z$ plane 
on the unstable channel side in the outer layer (figure \ref{visu}.b).
These are approximately aligned with the flow direction 
and caused by updrafts 
between the counter rotating roll cells 
coming from the lower wall at $y=-1$.
With increasing $Ro$ roll cells tend to become smaller and less obvious,
accordingly, the streaks with low scalar values become less coherent
and the spanwise distance between them diminishes (figure \ref{visu}.c, d) 
and at $Ro=0.9$ roll cells are hardly perceptible (figure \ref{visu}.e).
These observations are consistent with the energy spectra presented in Brethouwer (2017).

\begin{figure}[t]
\centering
(a)\includegraphics[width=8.0cm]{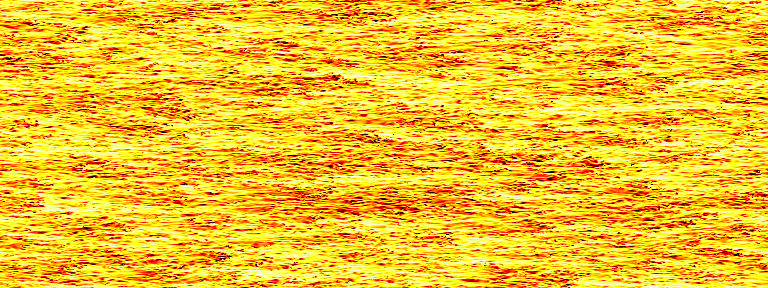}
\vskip2mm
(b)\includegraphics[width=8.0cm]{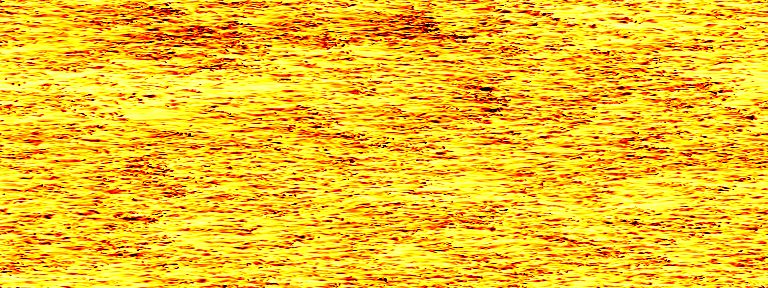}
\vskip2mm
(c)\includegraphics[width=8.0cm]{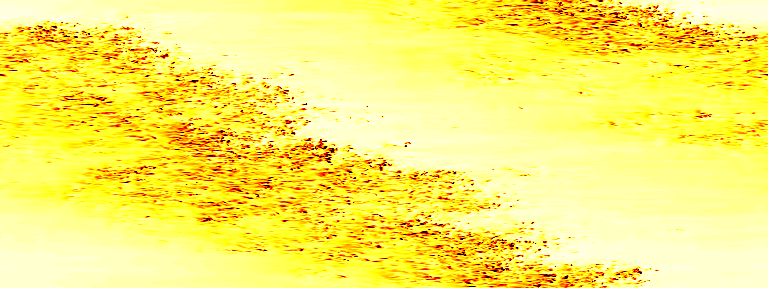}
\vskip2mm
(d)\includegraphics[width=8.0cm]{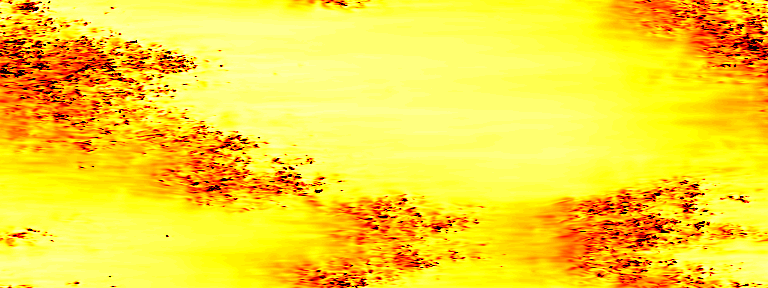}
\vskip2mm
(e)\includegraphics[width=8.0cm]{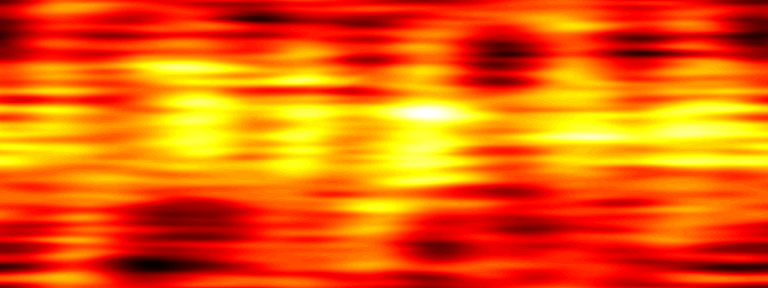}
\caption{
Visualizations of the instantaneous scalar gradient at $y=1$
at (a) $Ro=0$, 
(b) $Ro=0.15$,
(c) $Ro=0.45$,
(d) $Ro=0.65$ and
(e) $Ro=0.9$.
Dark colours correspond to high scalar gradients.
}
\label{viss}
\end{figure}

Figure \ref{viss} shows the instantaneous scalar gradient on the wall
at $y=1$ on the stable channel side.
The fluctuations in the scalar gradient imply that
the flow is fully turbulent here if $Ro \leq 0.15$ (figure \ref{viss}.a, b), 
but at higher $Ro$ it starts to partly relaminarize on the stable channel side. 
At $Ro=0.45$ oblique banded patterns develop with alternating turbulent and
laminar-like flow (Brethouwer 2017). A corresponding pattern emerges
in the scalar gradient field with oblique banded regions with strong scalar 
gradient fluctuations induced by the turbulence and regions where these 
gradient fluctuations are mostly absent since the flow is locally 
laminar-like (figure \ref{viss}.c). Similar oblique turbulent-laminar
patterns have been observed in several flow types, see
for example, Duguet \etal (2009), Brethouwer \etal (2012) and Deusebio \etal (2014).
A distinguishing feature is that the patterns do not span the whole channel
in the present case but are only found on the stable side.
At a higher $Ro=0.65$ the patterns with turbulence and strong scalar gradient fluctuations
become less coherent, as seen in figure \ref{viss}.(d). They also appear
to become more unsteady and their size varies in time, as indicated by 
other visualizations (not shown) and temporal variations in the 
mean scalar gradient and skin friction at $y=1$, shown later.
When $Ro \geq 0.9$ fluctuations are weak on the stable side and 
small-scale fluctuations in the scalar gradient field
cannot be observed
(figure \ref{viss}.e), implying that near-wall turbulence
is suppressed by the system rotation.

\begin{figure}[t]
\centering
(a)\includegraphics[width=8.0cm]{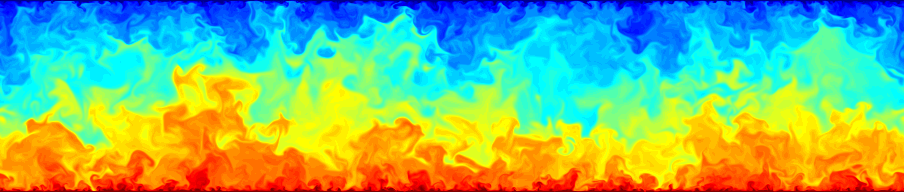}
\vskip2mm
(b)\includegraphics[width=8.0cm]{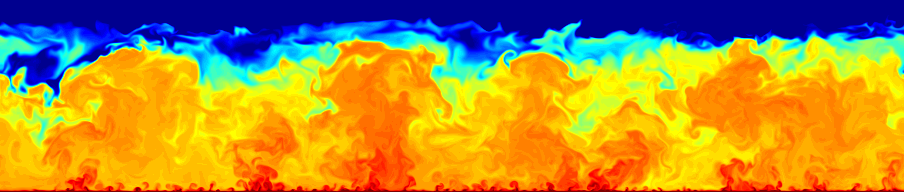}
\vskip2mm
(c)\includegraphics[width=8.0cm]{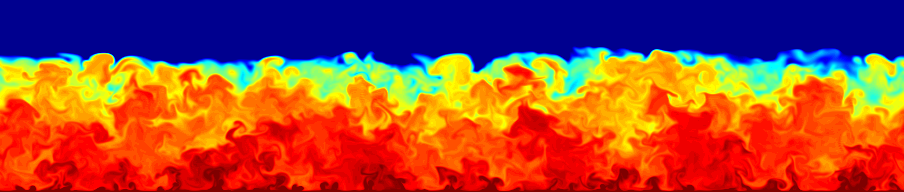}
\caption{
Visualizations of the instantaneous scalar field in a $y$-$z$-plane
at (a) $Ro=0$, 
(b) $Ro=0.45$ and
(c) $Ro=0.9$.
Red and blue colours correspond to low and high scalar values, respectively.
}
\label{visyz}
\end{figure}

Instantaneous plots of the scalar field in a $y$-$z$-plane at 
$Ro=0$, 0.45 and 0.9 are shown in figure \ref{visyz}.
At $Ro=0.45$, the visualization reveals large-scale plume-like structures
in the scalar field (figure \ref{visyz}.b), 
which are likely caused by the large streamwise roll cells.
At higher $Ro$ and $Ro=0$, the roll cells are absent or much smaller (Brethouwer 2017)
and accordingly no or less large plumes are seen in the scalar field
(figure \ref{visyz}.a, c). 
Large jumps in the scalar field are observed when $Ro > 0$ in the core region
near the border of
stable and unstable channel side where the mean scalar gradient is large,
as discussed later.

\begin{figure}[t]
\begin{center}
\setlength{\unitlength}{1cm}
(a)\includegraphics[width=6.0cm]{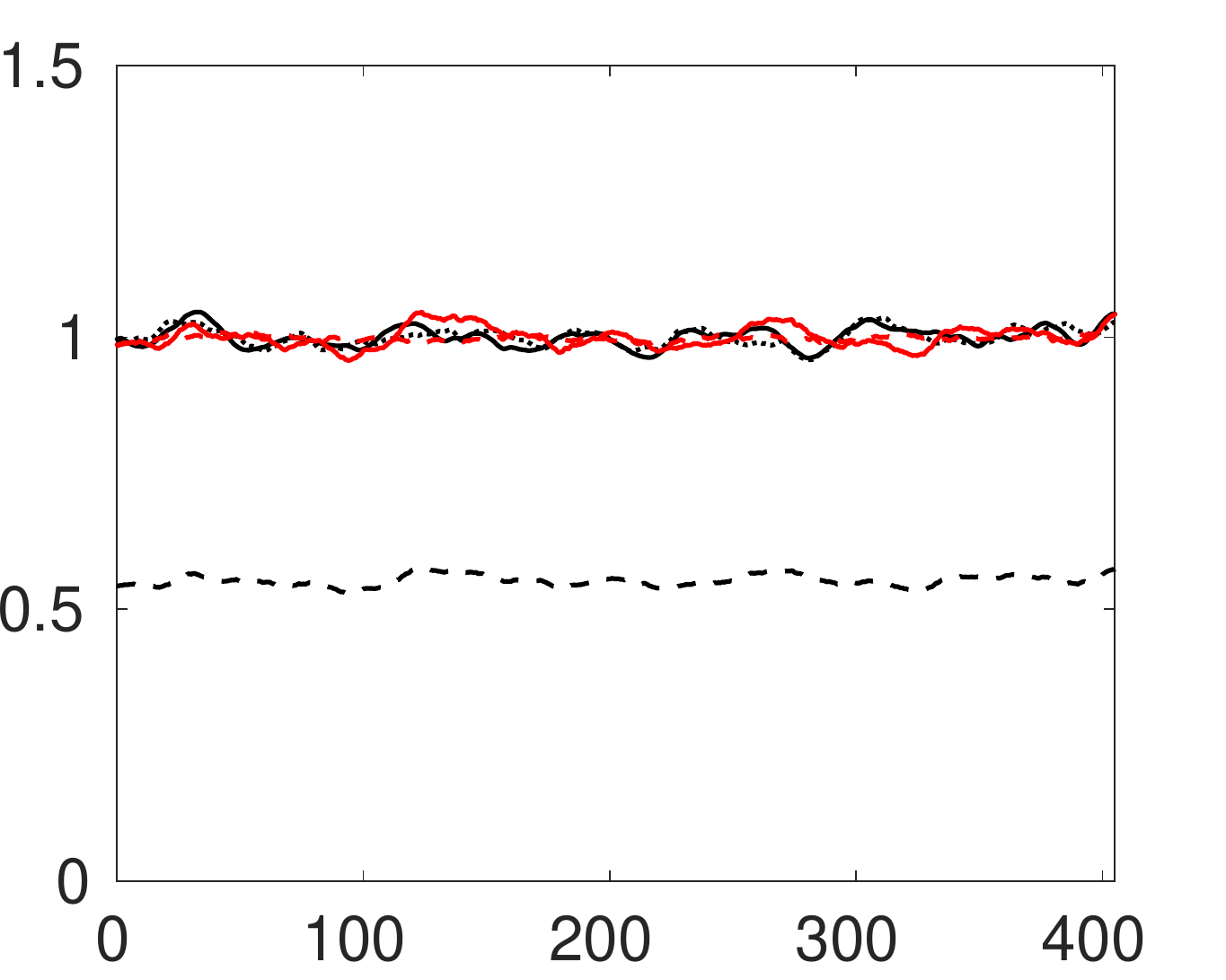}
\put(-3.1,-0.3){$t$}
(b)\includegraphics[width=6.0cm]{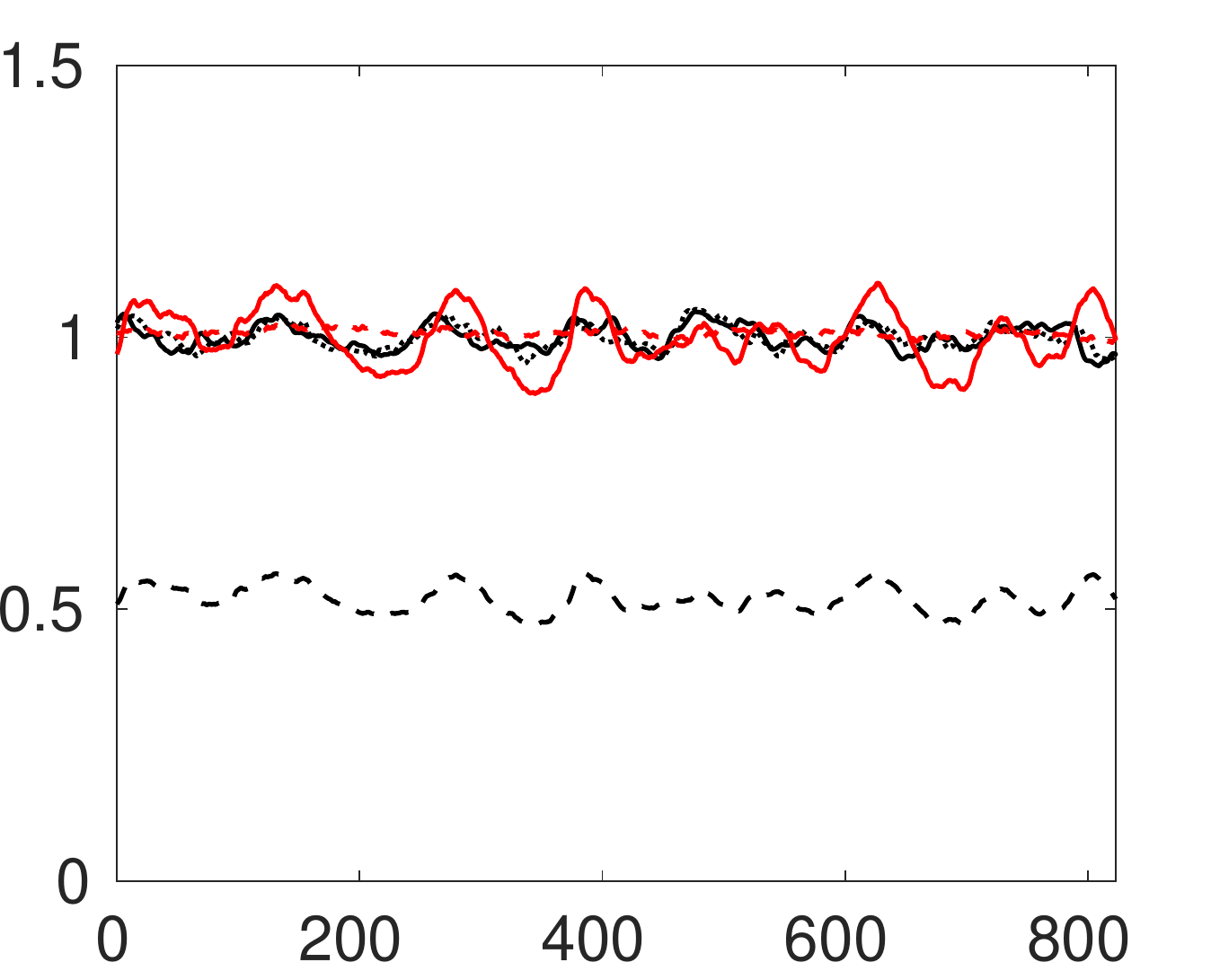}
\put(-3.1,-0.3){$t$}

\vskip2mm
(c)\includegraphics[width=6.0cm]{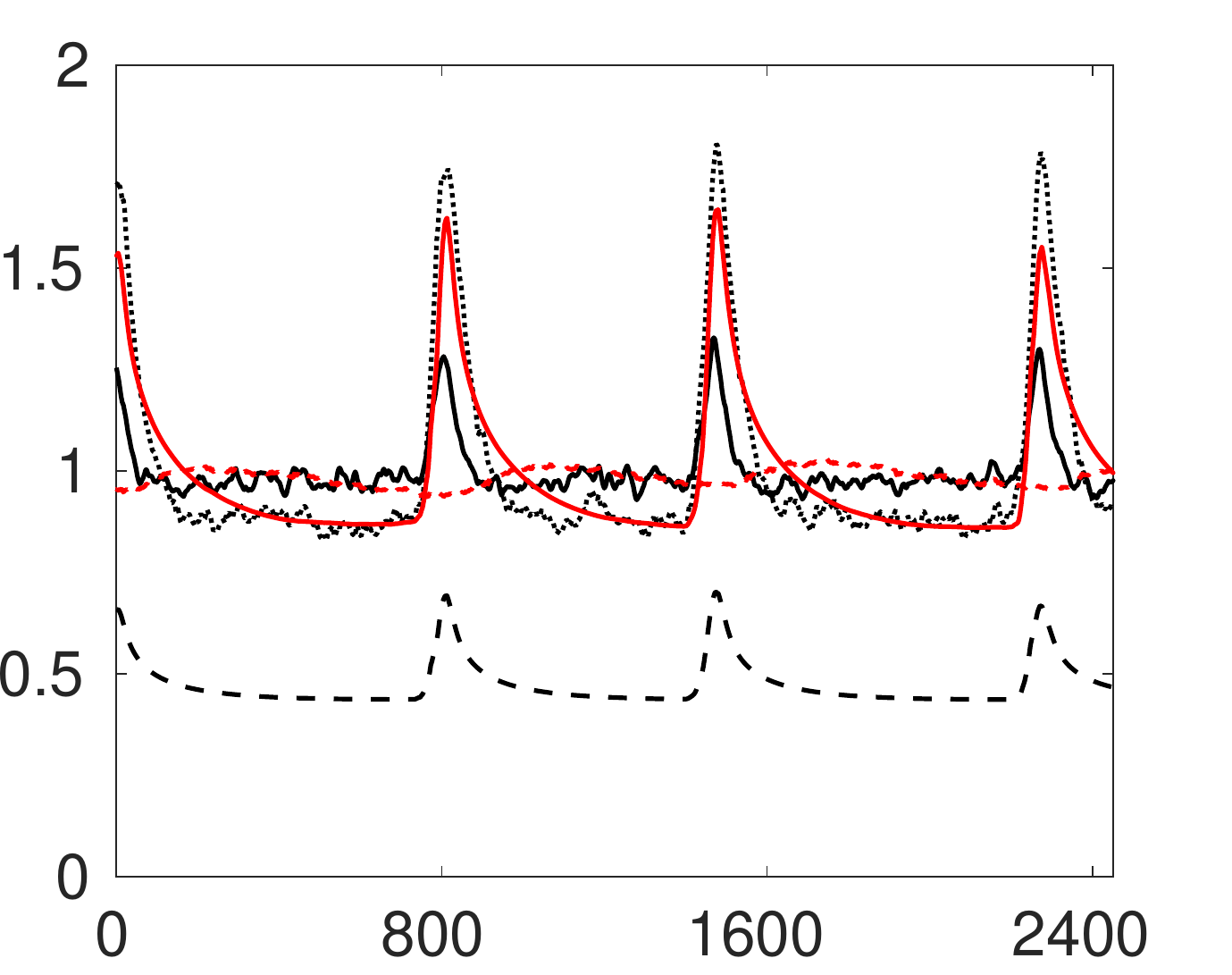}
\put(-3.1,-0.3){$t$}
\hskip2mm
(d)\includegraphics[width=6.0cm]{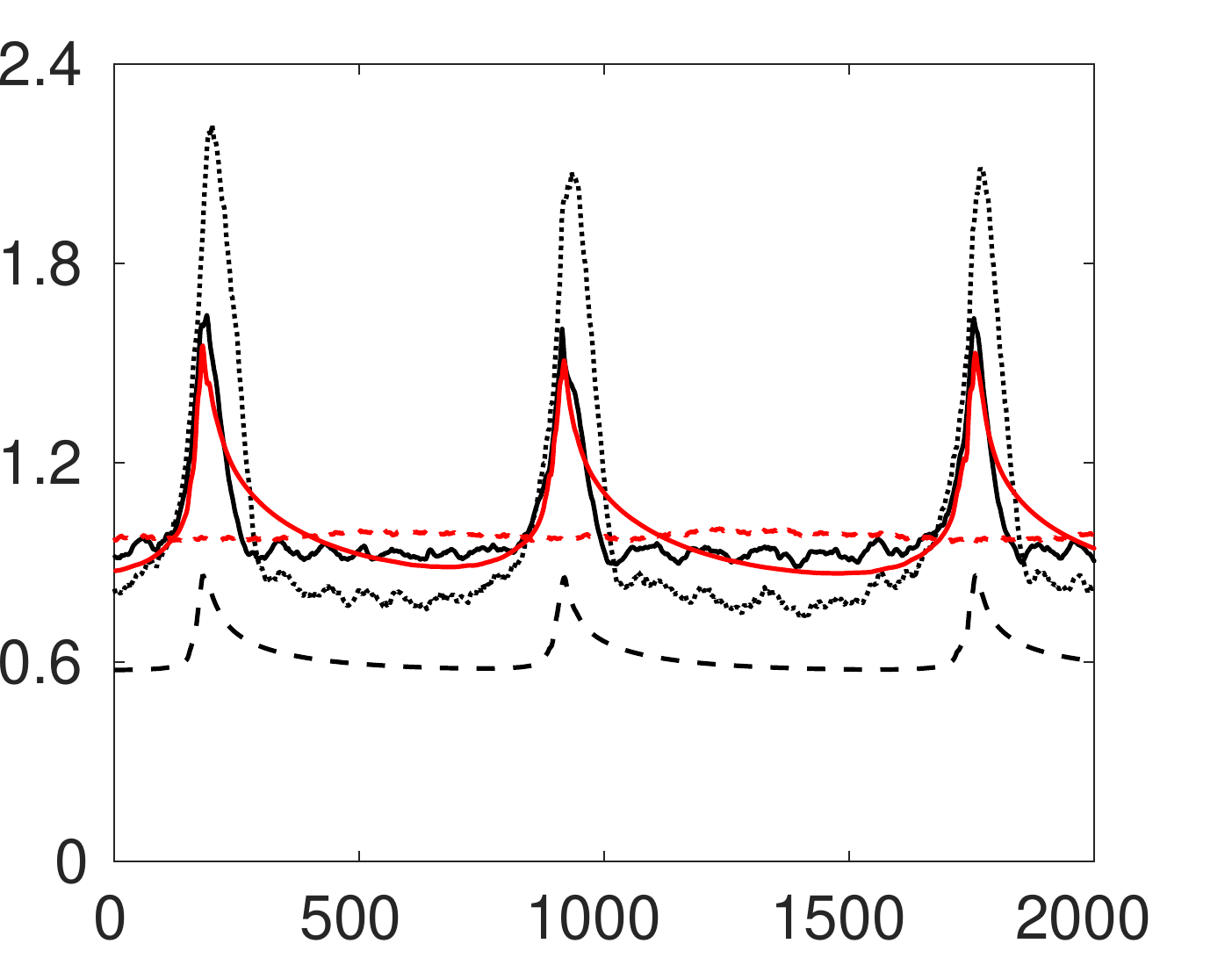}
\put(-3.1,-0.3){$t$}
\end{center}
\caption{
Time series at (a) $R=0.45$,
(b) $R=0.65$,
(c) $R=0.9$ and
(d) $R=1.2$
of the volume averaged $K_m$ 
($~^{\line(1,0){20}}$) 
and $\theta_m$ ($\cdot\cdot\cdot$),
and wall gradients $\tau_{ws}$ ($---$), 
$(\diff \Theta /\diff y)_u$ ($\textcolor{red}{---}$)
and $(\diff \Theta /\diff y)_s$ ($\textcolor{red}{~^{\line(1,0){20}}}$). 
$\tau_{ws}$ is normalized with the time-averaged wall shear 
on both wall whereas the other quantities are normalized
with their respective time-averaged values.
}
\label{times}
\end{figure}

It is important to remark that at 
$Re=20\,000$ a linear instability of Tollmien-Schlichting-like
waves occurs when $Ro \geq 0.9$ (Brethouwer 2016). This instability
causes recurring bursts of turbulence mostly confined to the stable channel side
and its mechanism is examined in Brethouwer \etal (2014)
and Brethouwer (2016).
The instability and strong bursts naturally affect the scalar field, as shown by the time
series in figure \ref{times}.
In this figure, time series of the volume averaged turbulent kinetic energy $K_m$
and root-mean-square of the scalar fluctuations $\theta_m$ are presented as well
as the wall shear stress $\tau_{ws}$ at $y=1$ and wall temperature gradients
$(\diff \Theta /\diff y)_u$ and $(\diff \Theta /\diff y)_s$ at $y=-1$ and $y=1$,
respectively. The stress and gradients are averaged over the wall and
the time $t$ is scaled by $h/U_b$.
When $Ro \leq 0.45$ all these quantities are approximately constant in time
(figures \ref{times}.a) but
at $Ro=0.65$, 
$\tau_{ws}$ and $(\diff \Theta /\diff y)_s$
display noticeable variations caused by the growing and shrinking of the turbulent areas
on the stable channel side, as discussed before.
Dai \etal (2016) have observed similar variations of $\tau_{ws}$
in their DNS and attributed these to the dynamics and changes of the 
streamwise roll cells.
At $Ro=0.9$ and 1.2 the time series of
$\theta_m$ and $(\diff \Theta /\diff y)_s$ show simultaneously with 
$K_m$ and $\tau_{ws}$ recurring sharp peaks with a time interval of about $700 t$,
whereas $(\diff \Theta /\diff y)_u$ and the wall shear stress
at $y=-1$ (not shown) stay approximately constant. 
This shows that the bursts of turbulence triggered by the linear instability
causes sharp and significant but relatively short surges in the scalar fluctuations 
and scalar flux at the wall on the stable channel side.
Such recurring bursts are also observed at higher $Ro$. However, when $Ro \geq 2.1$
and the turbulence is weak in the whole channel (Brethouwer 2017)
the bursts trigger sharp spikes in the scalar flux not only at the wall at 
$y=1$ but also at the other wall (not shown).

Although interesting, I will not further explore these recurring bursts
in the DNS at $Ro \geq 0.9$ but focus in the rest of the paper
in these DNS on the relatively calm periods between the bursts.
From a modelling point of view these calm periods are more relevant
and give a better fundamental insight into the influence of rotation on 
turbulent scalar transport. Including the bursts would add further complexity to
the problem. 
I therefore compute the scalar statistics from the statistics collected
during the calm periods between the bursts.
In Brethouwer (2017) I explain in more detail how I exclude
the periods with the bursts from the statistics.

\section{5. Passive scalar transport}

In this section, I will discuss the basic statistics of the scalar field and scalar
transport. Here below, $\Theta$ is used to denote the mean scalar value, i.e. averaged
over time and $x$- and $z$-directions, and $\theta$ the scalar fluctuation.
A superscript $+$ implies, unless stated otherwise, scaling in terms of 
viscous wall units $\nu$ and $u_\tau$ and
$\theta_\tau = Q_w / u_\tau$ 
where $Q_w =\alpha (\diff \Theta/\diff y)_w$
is the mean scalar flux at the wall, which is equal at both walls in
the statistically stationary state.

\begin{figure}[t]
\begin{center}
\setlength{\unitlength}{1cm}
(a)\includegraphics[height=5.5cm]{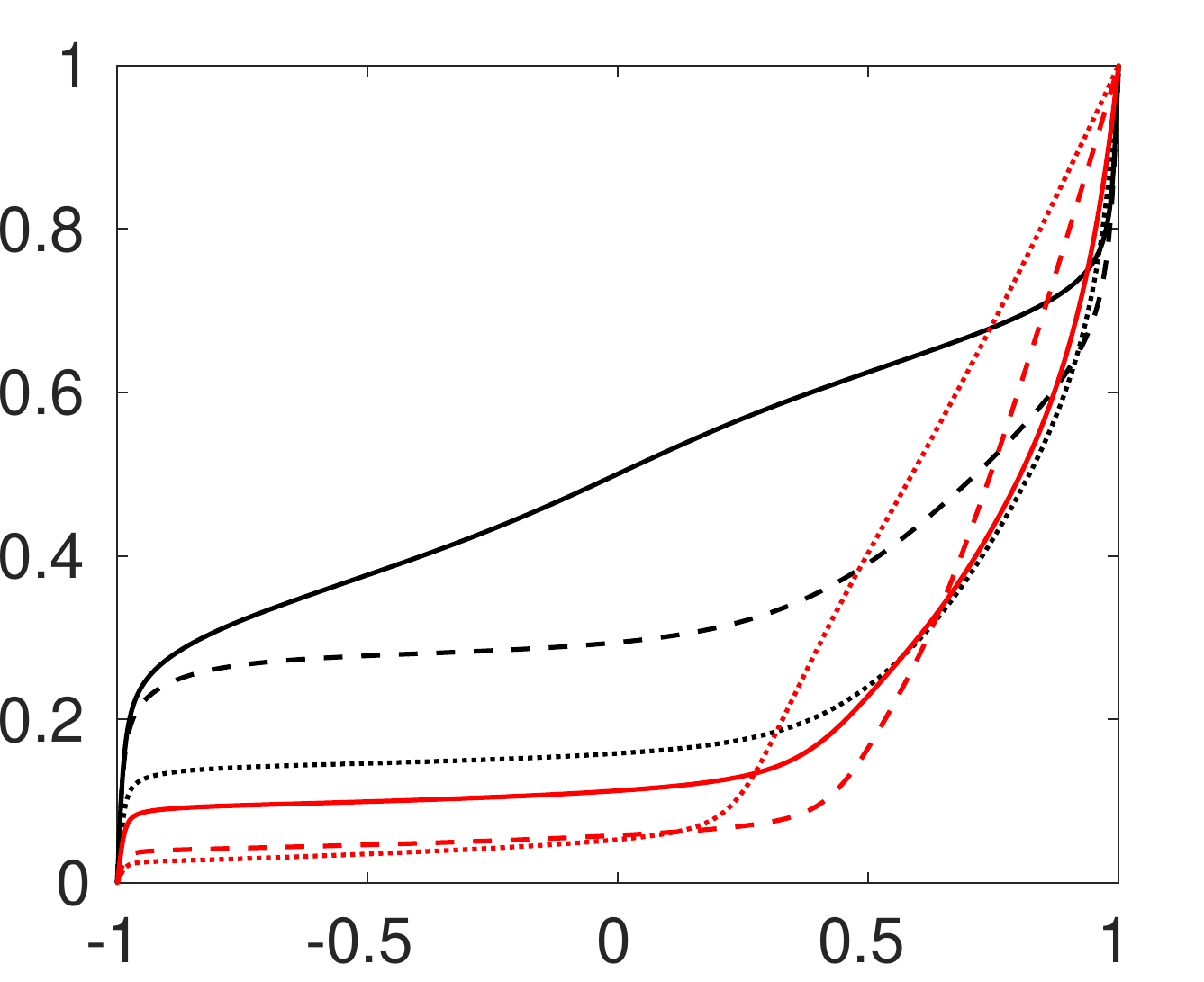}
\put(-3.4,-0.3){$y$}
\put(-7.2,2.7){$\Theta$}
\hskip4mm
(b)\includegraphics[height=5.5cm]{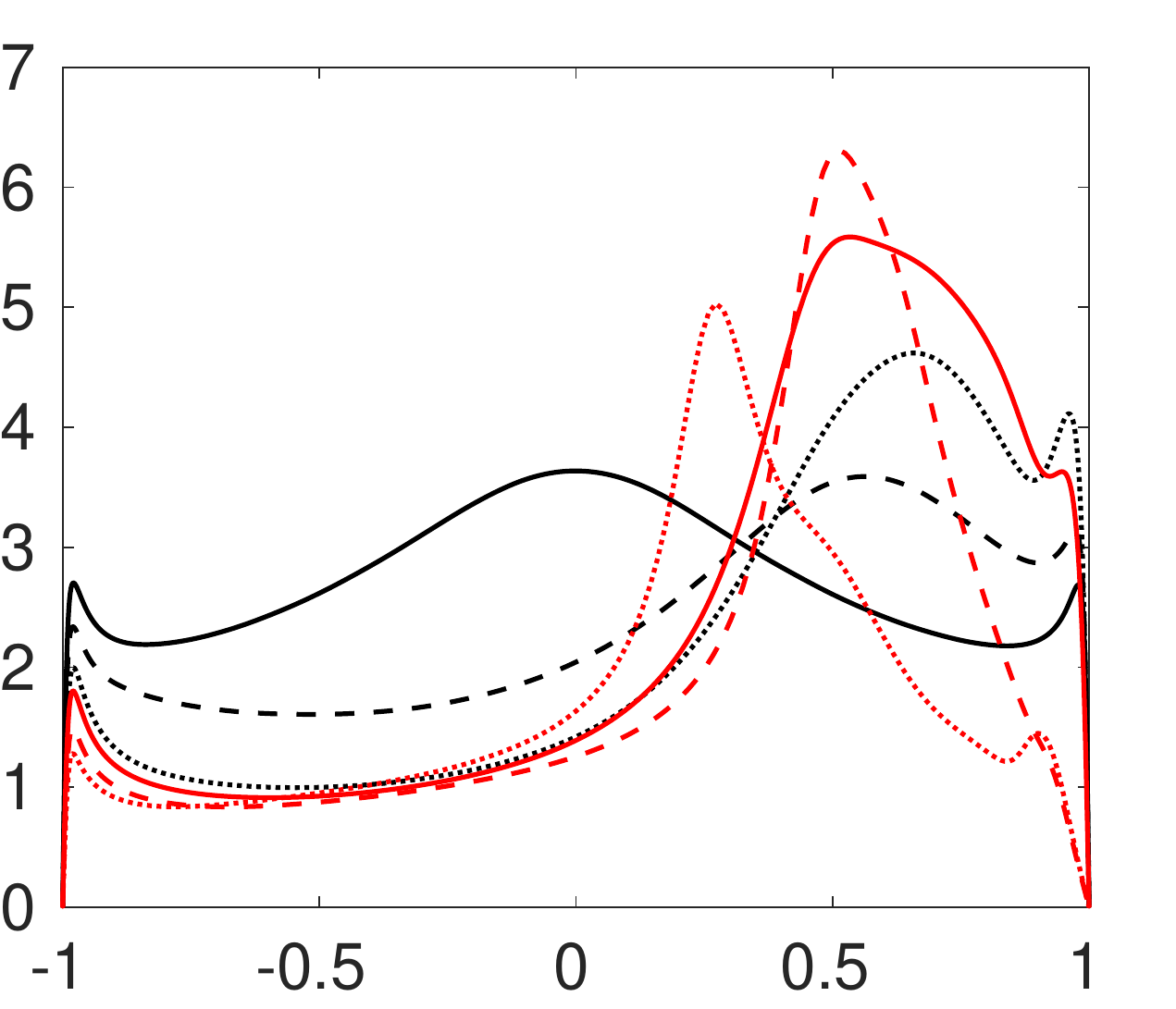}
\put(-3.4,-0.3){$y$}
\put(-7.2,2.6){$\theta^+$}
\end{center}
\caption{
Profiles of (a) $\Theta$ and (b) $\theta^+$.
Lines as in figure \ref{ustat}.
}
\label{tmean}
\end{figure}

Figure \ref{tmean}.(a) shows profiles of $\Theta$ at different $Ro$.
The mean scalar value $\Theta$ goes down monotonically with $Ro$ on the unstable channel side,
whereas on the stable side the mean scalar gradient is steep in rotating channel flow,
much steeper than on the unstable side. This can be understood 
by considering the mean scalar transport across the channel in the steady state
\begin{equation}
\alpha \frac{\diff \Theta}{\diff y} - \overline{v\theta} = Q_w.
\end{equation}
On the stable channel side, $\overline{v\theta}$ is strongly diminished
in rotating channel flow, as shown later, which naturally implies 
a large $\alpha \diff \Theta /\diff y$ to maintain the scalar flux balance.
Profiles of the root-mean-square of the scalar fluctuations $\theta^+$
are shown in figure \ref{tmean}.(b). 
At $Ro=0$ the profile of $\theta^+$ has not only peaks near both walls caused
by intense near-wall turbulence but also a maximum at the centre as a result
of a large $\diff \Theta /\diff y$ and, consequently, a high production
of scalar fluctuations here, as shown later. 
When $0.15 \leq Ro\leq 0.65$, $\theta^+$ including the near-wall peak on the unstable
channel side declines with $Ro$ and grows on the stable channel side.
This growth is caused by the large $\diff \Theta /\diff y$ and related large production
of scalar fluctuations on the stable side. 
Note that scalar fluctuations are strong although turbulence
is weakened by rotation on the stable channel side.
At higher $Ro$,
$\theta^+$ declines in the near-wall region of the stable side. Instead, 
$\theta^+$ has a pronounced maximum near the position where $\diff \Theta /\diff y$
strongly increases going from the unstable towards the stable channel side.
Quite similar trends regarding the profiles of $\Theta$ and $\theta^+$
are also observed at lower Reynolds numbers
(Nagano \& Hattori 2003, Liu \& Lu 2007).

\begin{figure}[t]
\begin{center}
\setlength{\unitlength}{1cm}
(a)\includegraphics[height=5.5cm]{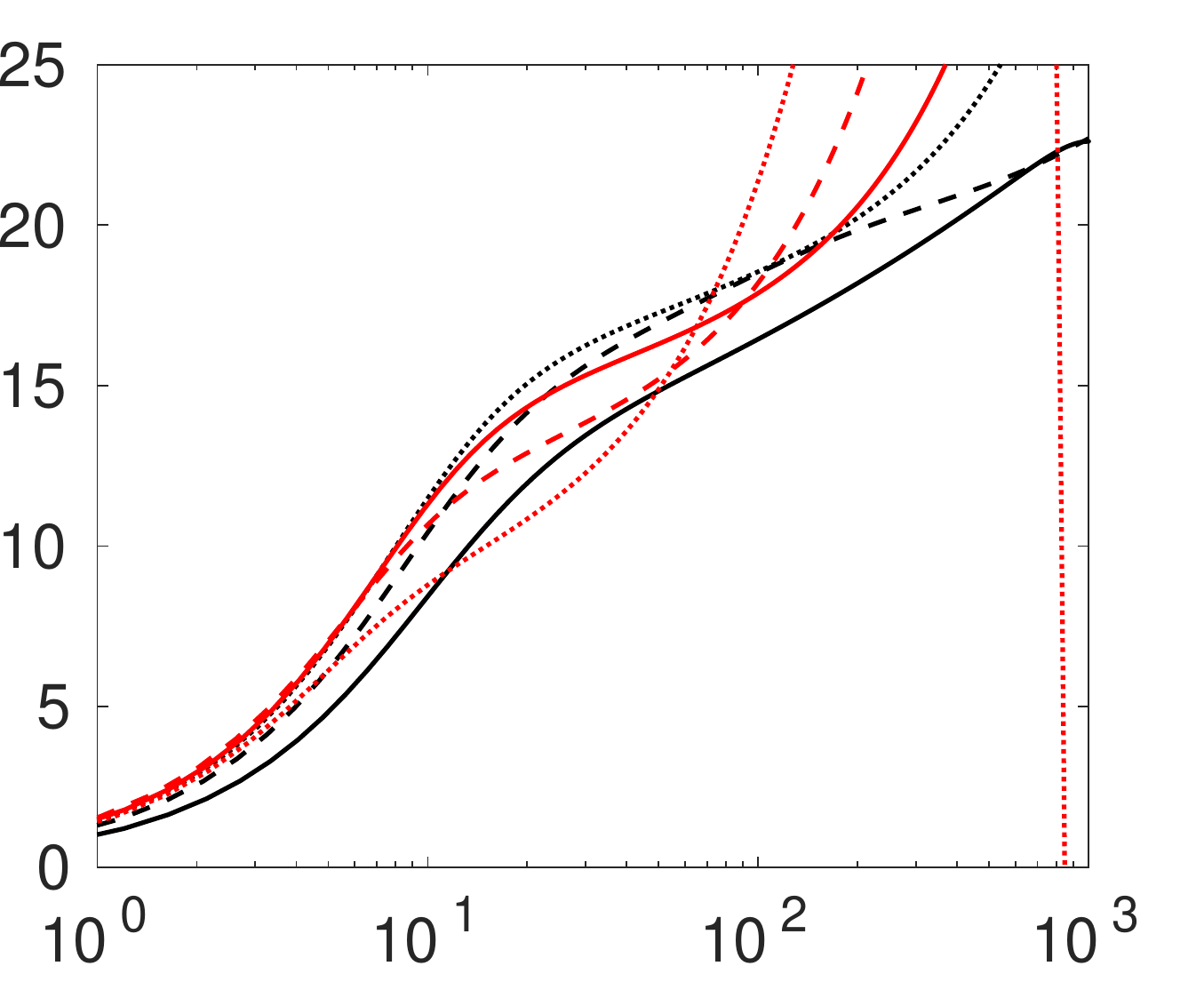}
\put(-3.4,-0.3){$y^+$}
\put(-7.2,2.6){$U^+$}
\hskip4mm
(b)\includegraphics[height=5.5cm]{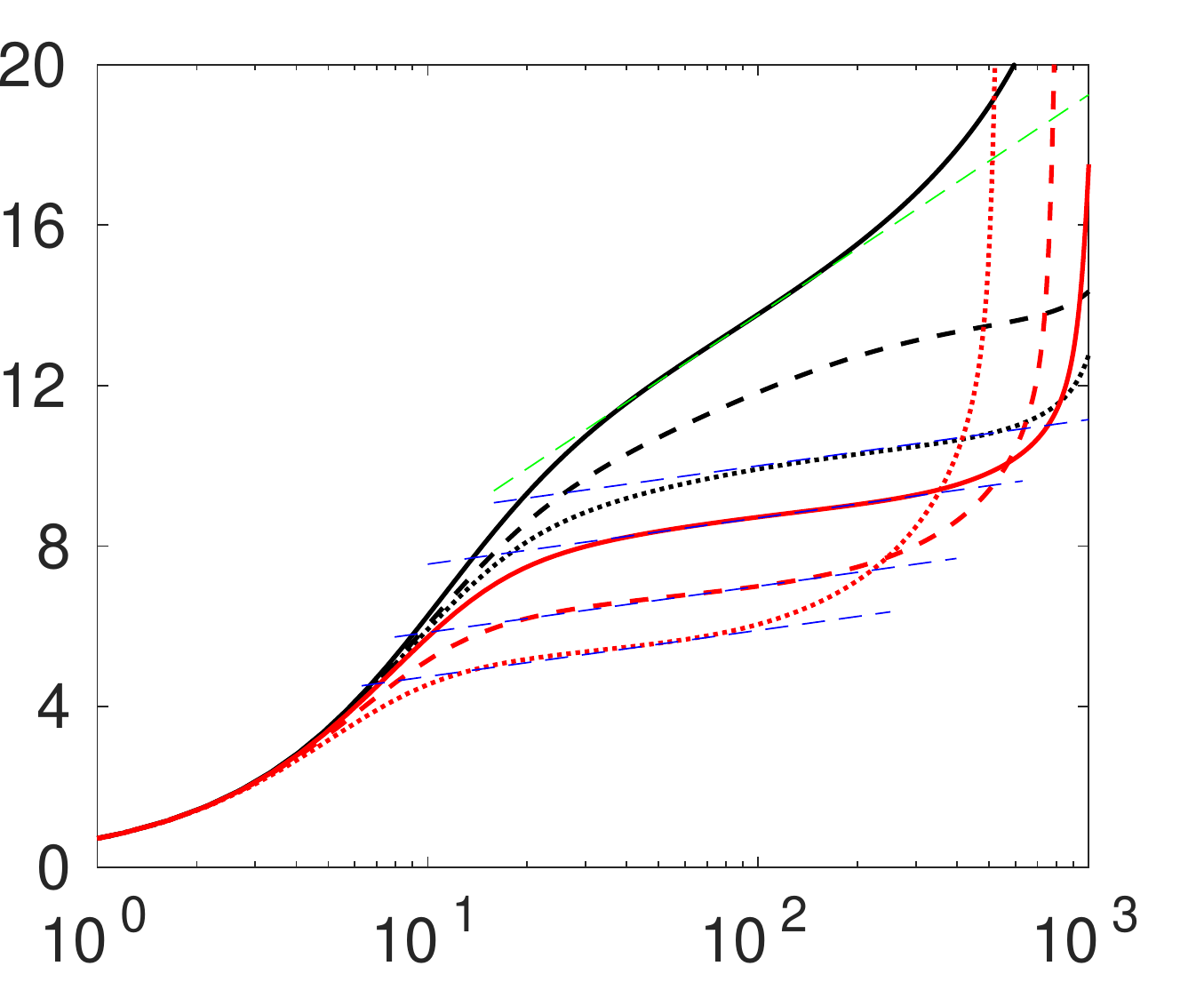}
\put(-3.4,-0.3){$y^+$}
\put(-7.2,2.6){$\Theta^+$}
\end{center}
\caption{
Profiles of (a) $U^+$ and (b) $\Theta^+$ as function of $y^+$.
The straight green and blue dashed lines in (b) are explained in the text.
Other lines as in figure \ref{ustat}.
}
\label{utplus}
\end{figure}

Figure \ref{utplus} shows the mean velocity $U^+$
and scalar $\Theta^+$ profiles
on the unstable channel side as function 
of the distance to the wall in wall units $y^+$.
The mean velocity profile at $Ro=0$ follows approximately a log-law behaviour
between the near-wall and centre region, but a similar log-law behaviour
is not observed when $Ro > 0$ and cannot be expected
from theoretical arguments since the velocity profile depends on $Ro$.
Previous studies have shown that
in the overlap region of non-rotating channel flow 
the $\Theta^+$-profile has a similar log-law behaviour as the velocity
(Kawamura \etal 1999, Pirozzoli \etal 2016).
Accordingly, in the present DNS at $Ro=0$, the mean scalar profile between
the near-wall and centre region approximately matches 
\begin{equation}
\Theta^+= \frac{1}{\kappa_\theta} \log \, y^+ + C_\theta
\label{loglaw}
\end{equation}
with $\kappa_\theta = 0.42$ and $C_\theta = 2.8$ given by the 
straight green dashed line (figure \ref{utplus}.b).
The present value of $\kappa_\theta=0.42$ is similar to the values 
0.43 and 0.46 found by 
Kawamura \etal (1999), respectively, Pirozzoli \etal (2016).
When $Ro \geq 0.45$ the profiles of $\Theta^+$ on the unstable
side away from the wall approximately
also follow the logarithmic profile (\ref{loglaw}) but with
$\kappa_\theta = 2.0$ and decreasing values of $C_\theta$ with $Ro$,
given by the straight blue dashed lines in figure \ref{utplus}.(b),
despite the absence of a similar 
log-law behaviour in the $U^+$-profile.
A closer inspection shows that this log-law region
partly overlaps with the region where the mean velocity profile
is approximately linear and $\diff U / \diff y \approx 2 \Omega$.
Whether this log-law behaviour of the $\Theta^+$-profiles
with the same logarithmic slope
in rotating channel flows is just a coincidence is not yet clear
and there is obvious explanation for this behaviour.

\begin{figure}[t]
\begin{center}
\setlength{\unitlength}{1cm}
(a)\includegraphics[height=5.5cm]{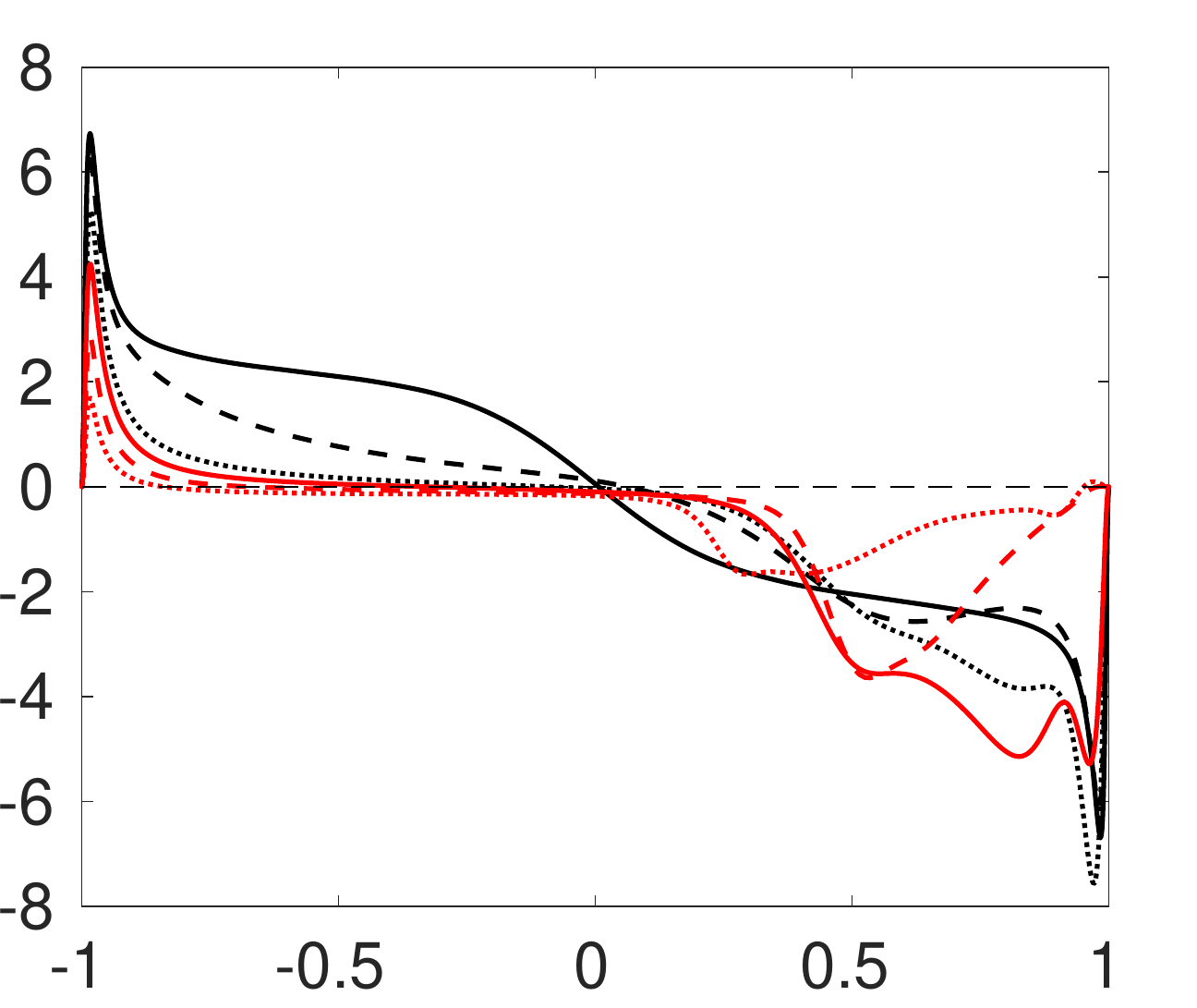}
\put(-3.4,-0.3){$y$}
\put(-7.4,2.6){$\overline{u \theta}^+$}
\hskip4mm
(b)\includegraphics[height=5.5cm]{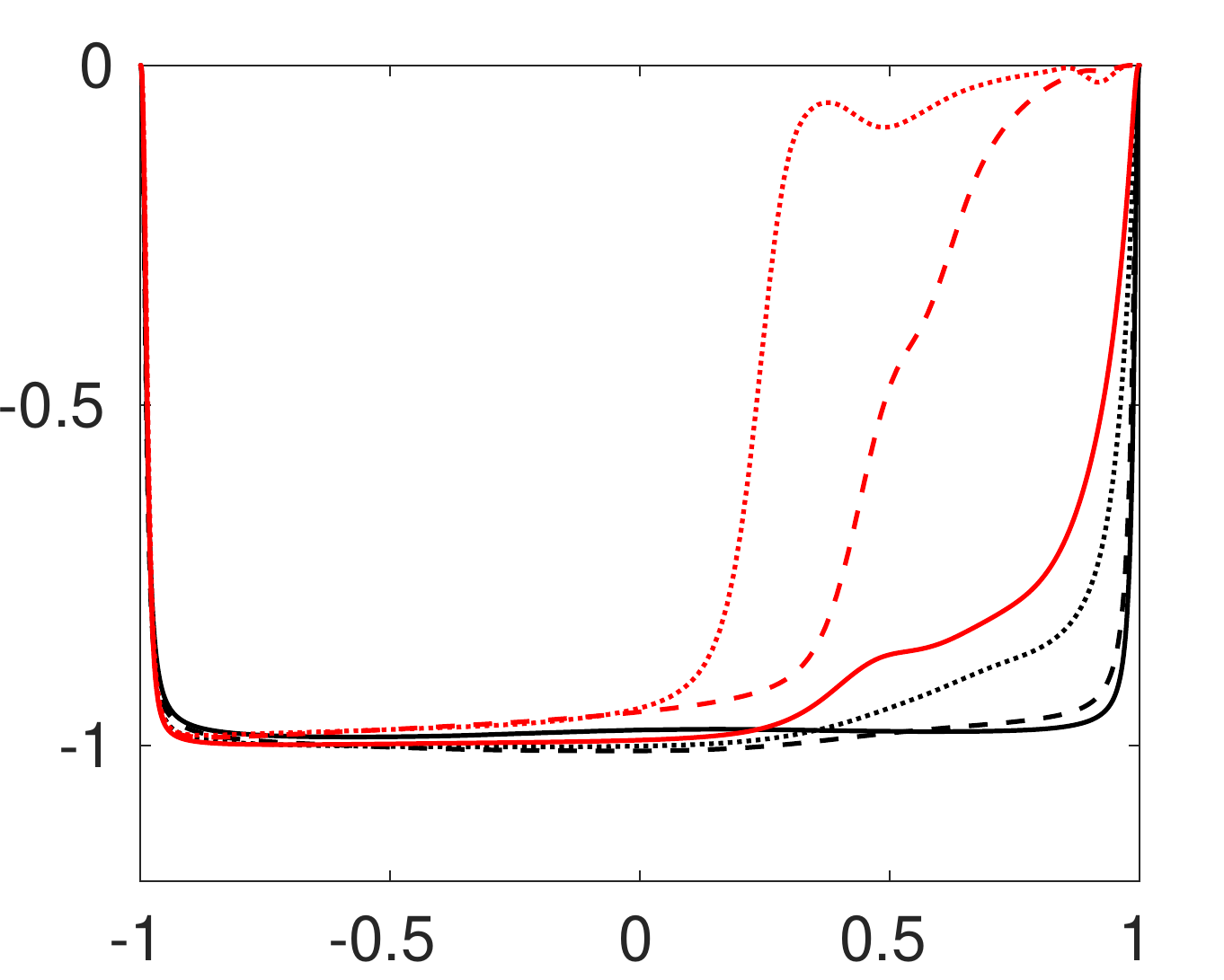}
\put(-3.4,-0.3){$y$}
\put(-7.4,2.6){$\overline{v \theta}^+$}

\vskip2mm
(c)\includegraphics[height=5.5cm]{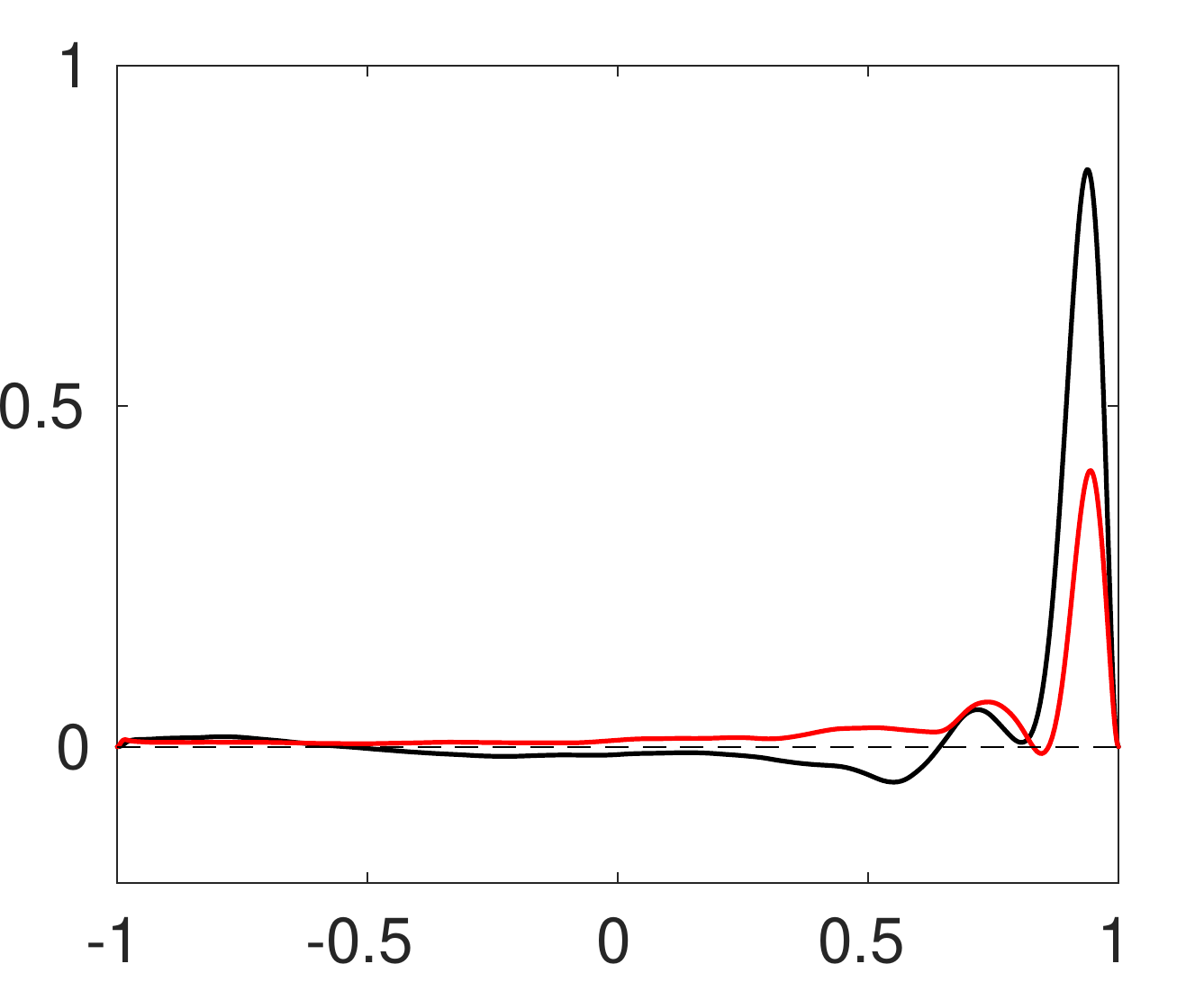}
\put(-3.4,-0.3){$y$}
\put(-7.4,2.6){$\overline{w \theta}^+$}
\end{center}
\caption{
Profiles of (a) $\overline{u \theta}^+$ and (b) $\overline{v \theta}^+$.
Lines as in figure \ref{ustat}.
(c) Profile of $\overline{w \theta}^+$ at 
(${~^{\line(1,0){20}}}$) $Ro=0.45$
$Ro=0.45$ and
($\textcolor{red}{~^{\line(1,0){20}}}$) $Ro=0.65$.
}
\label{tflux}
\end{figure}
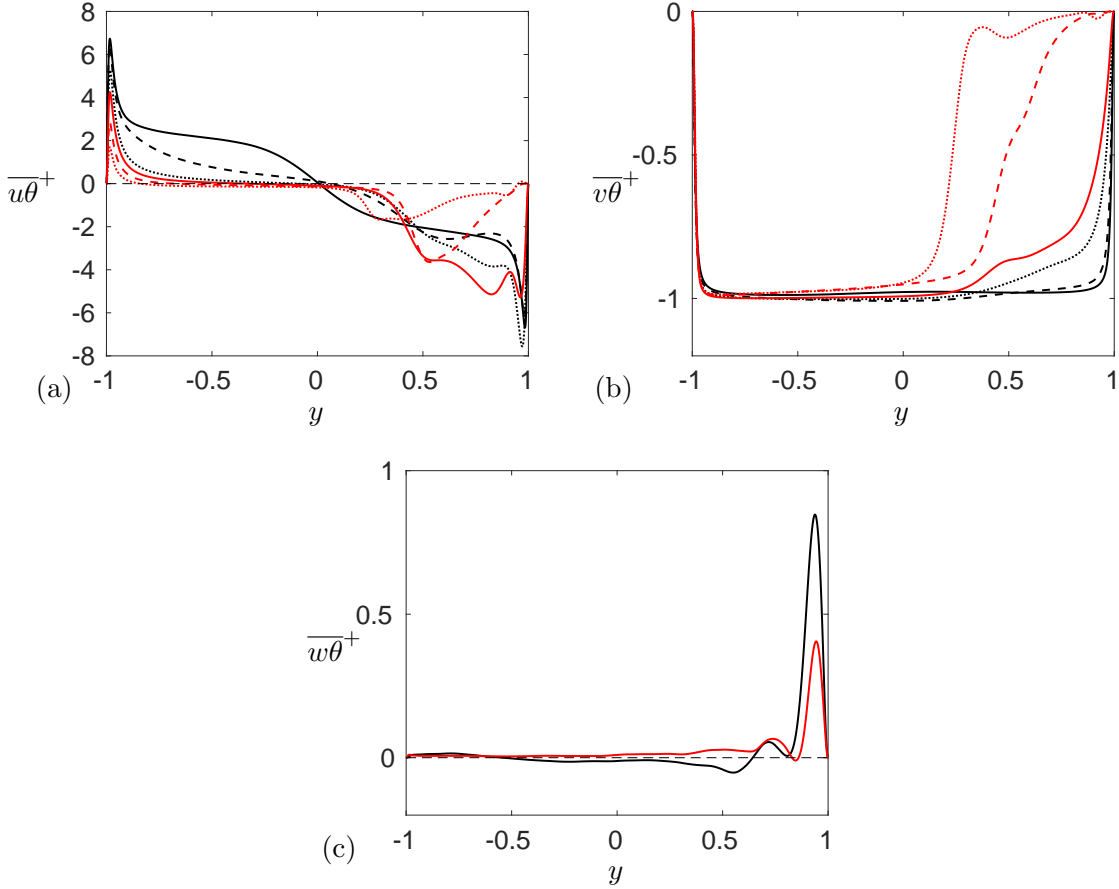

I will now consider turbulent scalar transport. Figure \ref{tflux}.(a)
and (b) show the profiles of the streamwise and wall-normal turbulent scalar fluxes
$\overline{u \theta}^+$ respective $\overline{v \theta}^+$ in wall units.
In non-rotating channel flow 
$\overline{u \theta}^+$ has, except around the centreline, 
a larger magnitude than $\overline{v \theta}^+$ (Johansson \& Wikstr{\"o}m 1999)
meaning that the turbulent scalar flux is mainly aligned with the streamwise direction
and the same applies to the stable channel side in the rotating cases.
When $Ro$ rises, 
$\overline{u \theta}^+$ including its near-wall peak monotonically declines on
the unstable side while 
$-\overline{v \theta}^+$ stays near unity, meaning that the turbulent scalar
flux vector turns towards the wall-normal direction and is 
nearly aligned with the wall-normal direction away from the wall
in the region where
$\diff U / \diff y \approx 2 \Omega$. 
On the stable channel side,
$-\overline{v \theta}^+$ declines with $Ro$ and is very small for $Ro \geq 0.9$
when the flow relaminarizes there, whereas
$\overline{u \theta}^+$ stays large for $Ro \leq 0.65$ but declines at higher $Ro$
and its near-wall peak disappears (figure \ref{tflux}). 
The scalar transport on the stable channel side is thus
mainly diffusive at higher $Ro$.
These results are consistent
with previous DNS of scalar transport in rotating channel flow at
lower $Re$ and with a smaller range of $Ro$ (Nagano \& Hattori 2003, Liu \& Lu 2007).
They are also consistent with
DNS and rapid distortion theory of homogeneous turbulent shear flow subject to spanwise system rotation.
In that case,
the turbulent scalar flux vector is mostly aligned with the streamwise direction
when the rotation is absent, or cyclonic as on the stable channel
side. When the rotation is anti-cyclonic as on the 
unstable channel side the turbulent scalar flux vector becomes more aligned
with the wall-normal direction and is almost fully aligned with the mean scalar
gradient when the mean shear $\diff U / \diff y$ equals $2 \Omega$, where $\Omega$ is the imposed
rotation (Brethouwer 2005, Kassinos \etal 2007), like in the present case.

In non-rotating turbulent channel flow the mean spanwise turbulent scalar flux
$\overline{w \theta}^+$ is naturally zero. But figure \ref{tflux}.(c)
shows that when $Ro=0.45$ or 0.65 and oblique turbulent patterns are
present in the near-wall region of the stable side
(figure \ref{viss}.c, d)
$\overline{w \theta}^+$ is significant in this region.
In these cases, $\overline{uw}$ is also non-zero.
The oblique patterns induce thus a noticeable
spanwise turbulent momentum and scalar transport near the wall.

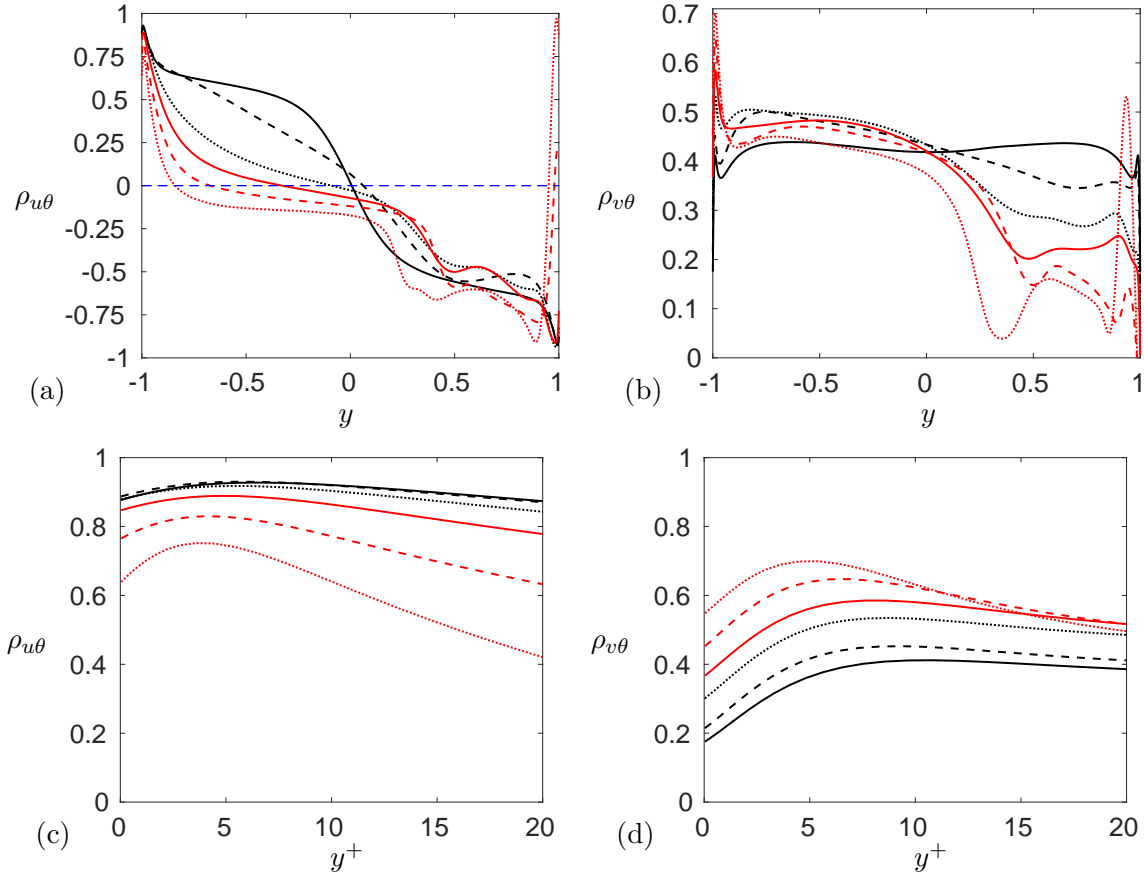
\begin{figure}[t]
\begin{center}
\setlength{\unitlength}{1cm}
(a)\includegraphics[height=5.5cm]{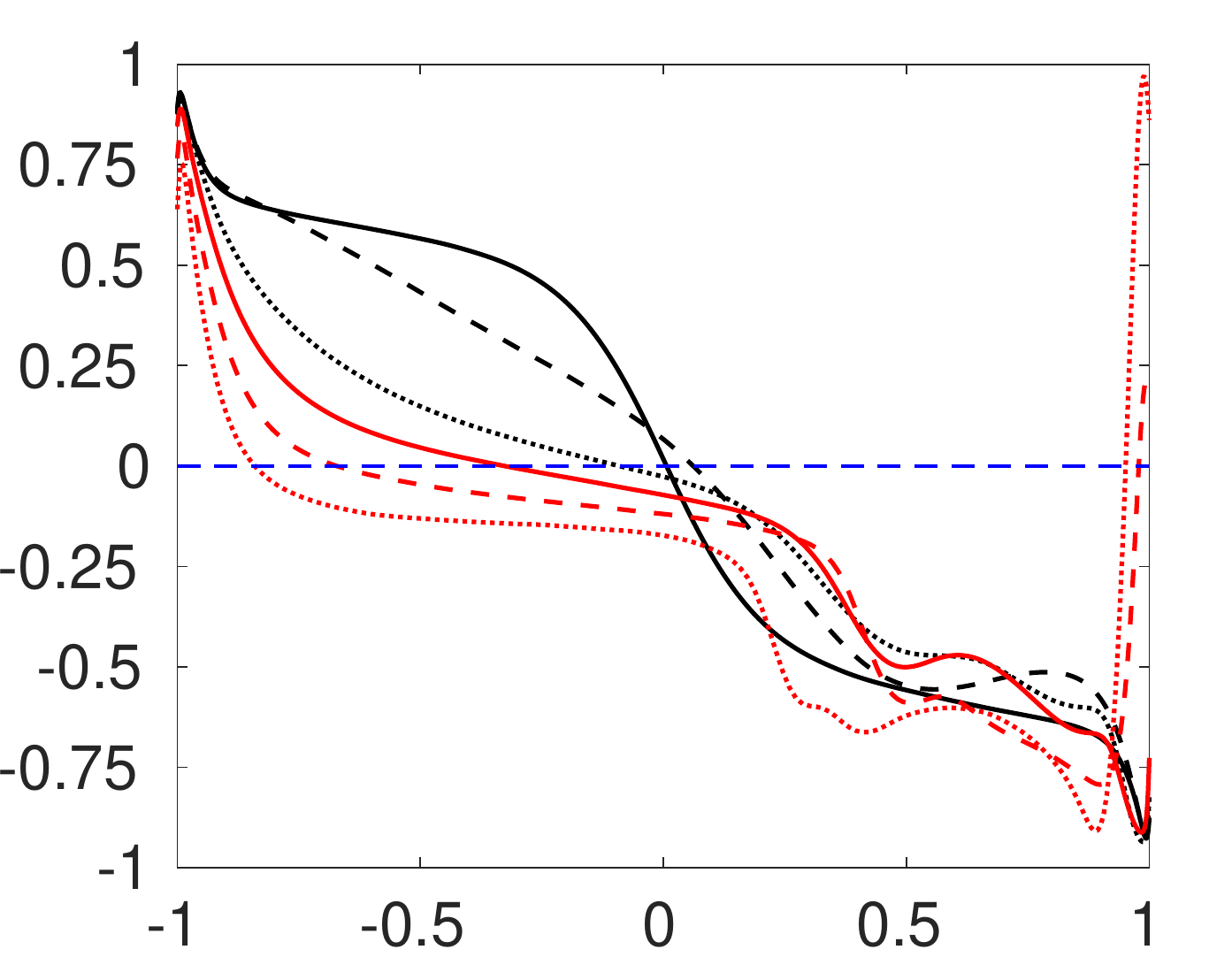}
\put(-3.4,-0.3){$y$}
\put(-7.7,2.6){$\rho_{u \theta}$}
\hskip4mm
(b)\includegraphics[height=5.5cm]{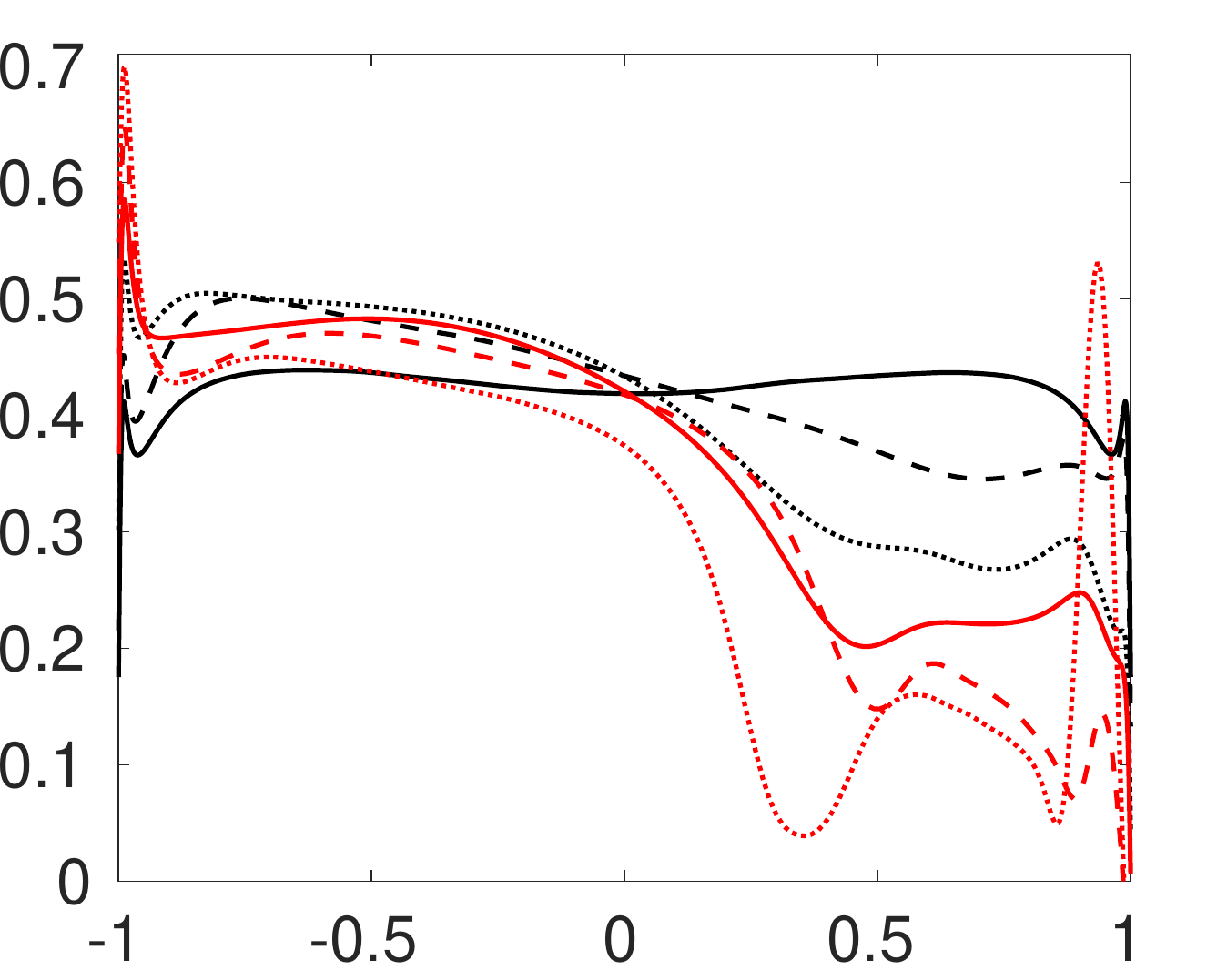}
\put(-3.4,-0.3){$y$}
\put(-7.7,2.6){$\rho_{v \theta}$}

(c)\includegraphics[height=5.5cm]{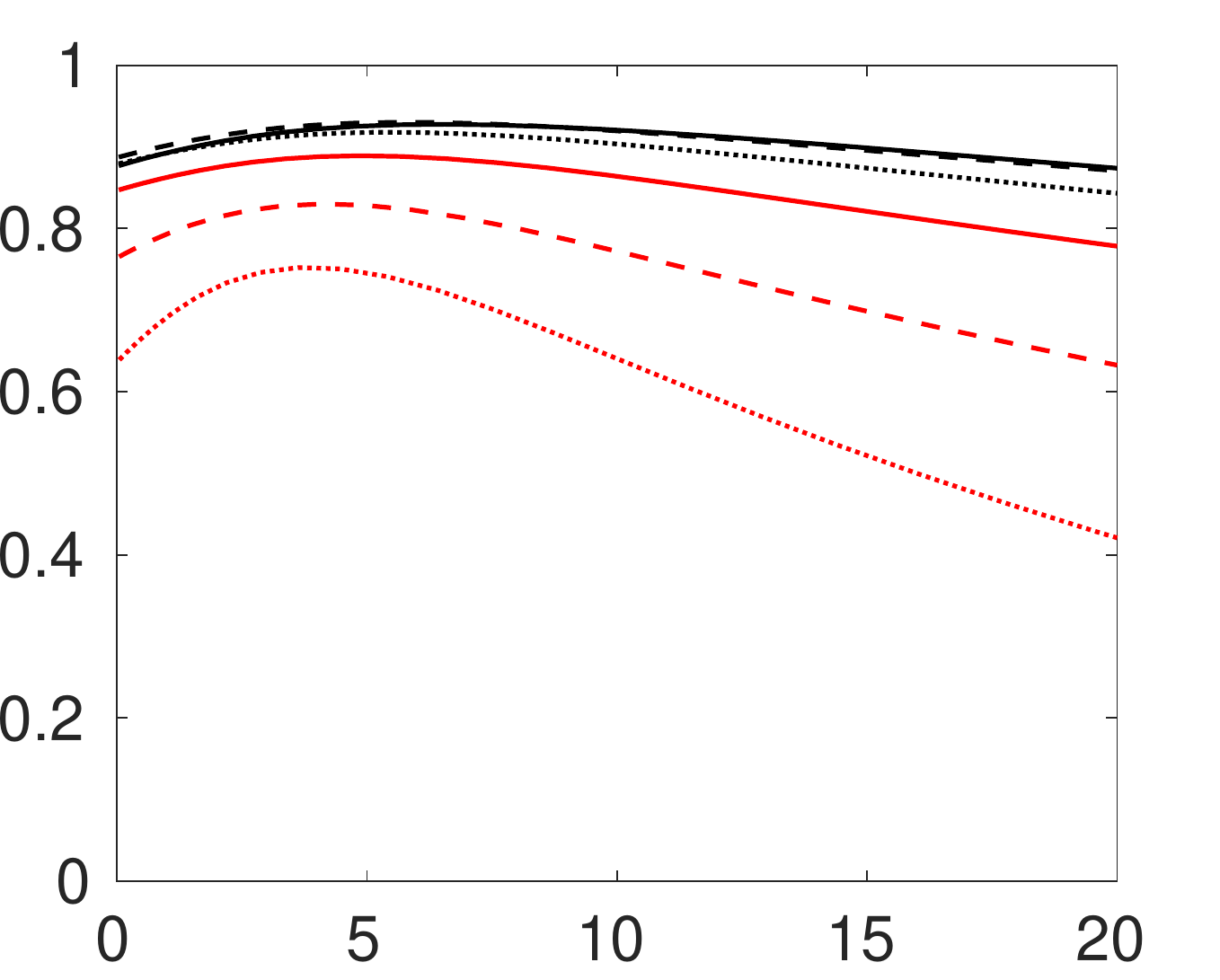}
\put(-3.4,-0.3){$y^+$}
\put(-7.7,2.6){$\rho_{u \theta}$}
\hskip4mm
(d)\includegraphics[height=5.5cm]{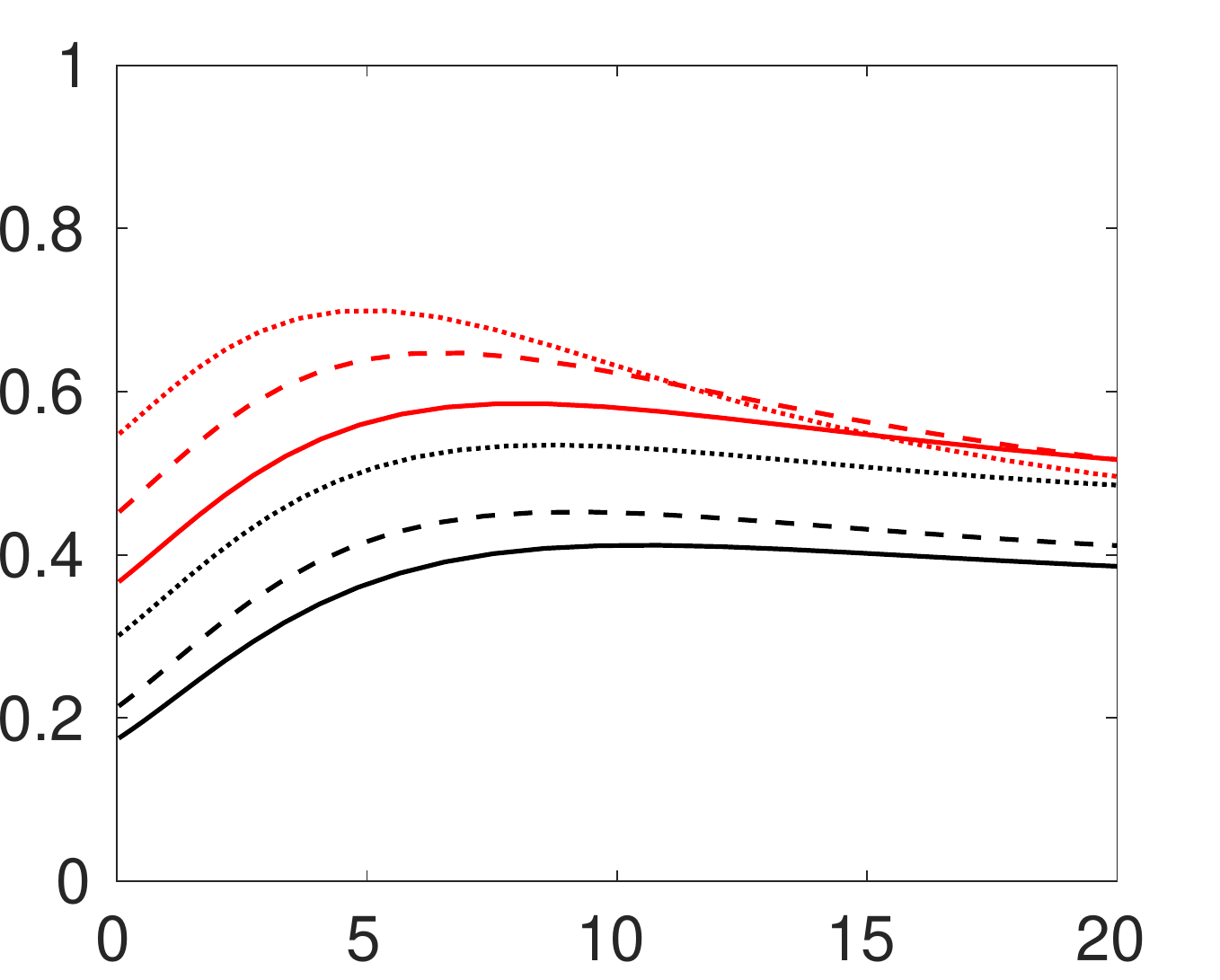}
\put(-3.4,-0.3){$y^+$}
\put(-7.7,2.6){$\rho_{v \theta}$}
\end{center}
\caption{
Profiles of (a,c) $\rho_{u \theta}$ and (b,d) $\rho_{v \theta}$,
(c) and (d) show a close-up near the wall on the unstable side.
Lines as in figure \ref{ustat}.
}
\label{correl}
\end{figure}

Figure \ref{correl} shows the correlation coefficients 
$\rho_{u\theta} = \overline{u\theta}/ (u'\theta')$ and
$\rho_{v\theta} = -\overline{v\theta}/ (v'\theta')$, where
a prime $'$ denotes root-mean-square values.
Consistent with Abe \& Antonia (2009) and Pirozzoli \etal (2016),  
very high streamwise flux correlations $\rho_{u\theta}$ are found at $Ro=0$
near the wall owing the very high similarity between the streamwise
velocity and scalar field (figure \ref{correl}.c), 
while in the outer layer, except around the centreline, 
$\rho_{u\theta}$ is lower yet still large (figure \ref{correl}.a).
A large correlation means here large positive or large negative values of $\rho_{u\theta}$.
In the rotating cases, the magnitude of
$\rho_{u\theta}$ remains large on the stable channel side,
but on the unstable
side it declines rapidly in the outer layer with $Ro$
and even becomes negative at high $Ro$ (figure \ref{correl}.a). 
Close to the wall on the unstable side,
$\rho_{u\theta}$ declines when $Ro \geq 0.65$ but remains quite large (figure \ref{correl}.c).
Thus, rotation reduces the similarity between the $u$- and 
$\theta$-field on the unstable side, especially in the outer layer. Yang \etal (2011) came
to a similar conclusion for DNS of rotating channel flow at much lower $Re$.
The high near-wall correlation at $Ro=0$ is motivated by
the similarity between the governing equations for $u$ and $\theta$
with the main difference being the pressure gradient term (Abe \& Antonia 2009).
However, if $Ro > 0$ an additional Coriolis term appears in the former,
which diminishes the similarity between the governing equations and, accordingly,
the correlation between $u$ and $\theta$.

By contrast,
$\rho_{v\theta}$ is quite insensitive to $Ro$ on the unstable channel side
away from the wall with values between 0.4 and 0.5 
whereas it declines with $Ro$ on the stable side (figure \ref{correl}.b). 
On the other hand, near the wall on the unstable side,
$\rho_{v\theta}$ increases considerably with $Ro$ and becomes quite large 
(figure \ref{correl}.d). 
The behaviour and magnitude of $\rho_{v\theta}$ with $Ro$ 
follows here closely the correlation coefficient for the Reynolds shear stress
$\overline{uv}/ (u'v')$ (not shown) which also becomes large near the wall at higher $Ro$.
These high correlations might be related to the 
formation of elongated streamwise near-wall vortices in rotating channel flow
on the unstable side (Yang \& Wu 2012).

To conclude,
the alignment of the scalar flux
with the mean scalar gradient with $Ro$ on the unstable side is 
related to the reduced correlation between streamwise velocity and
scalar fluctuations whereas the reduced turbulent scalar fluxes on the
stable side are caused by both declining $v$-$\theta$ correlations and Reynolds stresses.

\begin{figure}[t]
\setlength{\unitlength}{1cm}
\begin{center}
\includegraphics[height=5.5cm]{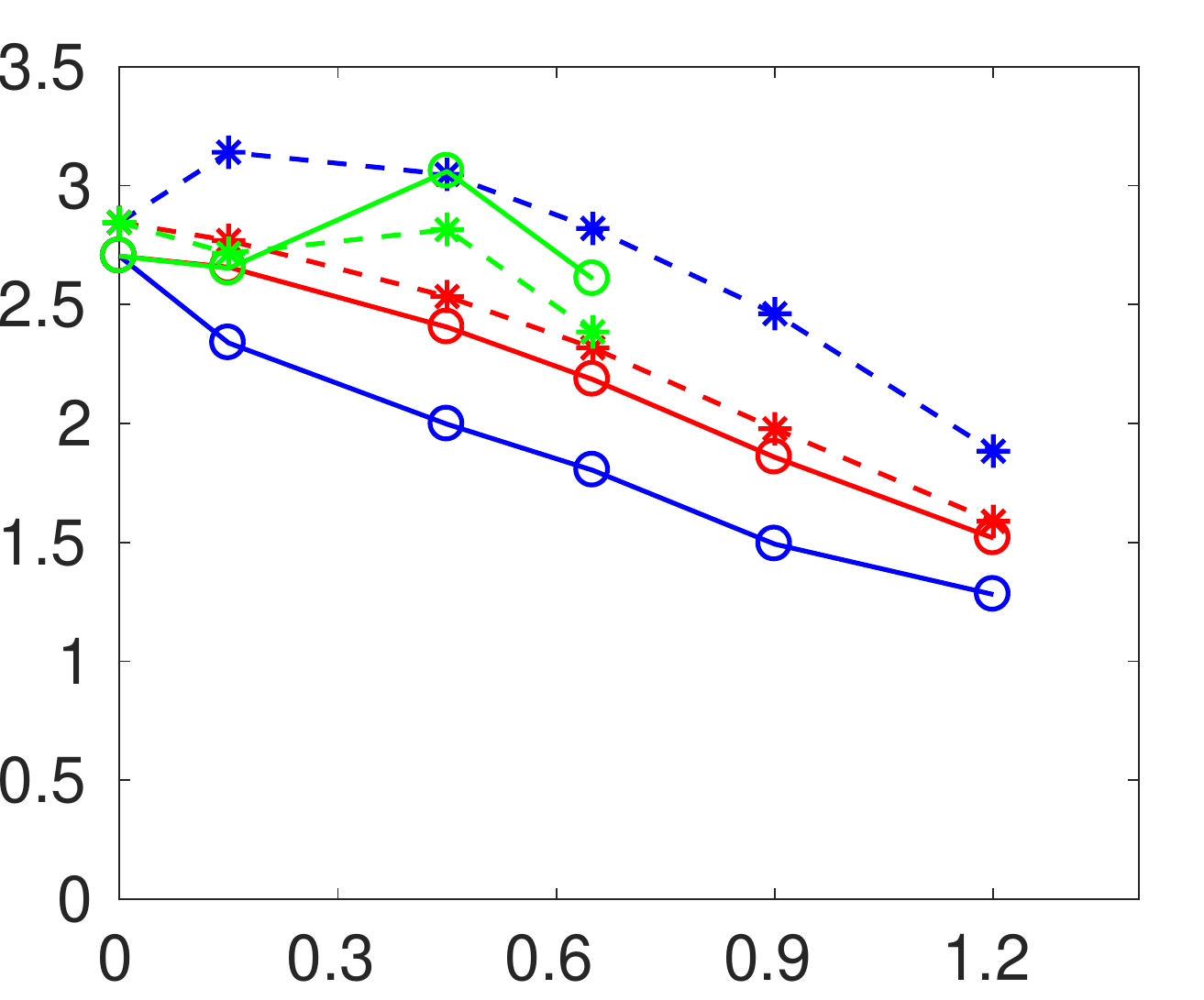}
\put(-3.4,-0.3){$Ro$}
\put(-7.5,2.9){$u^+_{max}$}
\put(-7.5,2.3){$\theta^+_{rms}$}
\end{center}
\caption{The maximum near-wall value of $u^+$ ($\ast---\ast$)
and $\theta^+$ ($\circ^{\line(1,0){1}}\circ$)
on the unstable (blue and red lines/symbols)
and stable channel side (green lines/symbols) as function of $Ro$.
Blue lines and symbols: using wall units based on $u_\tau$.
Red lines and symbols: using wall units based on $u_{\tau u}$.
Green lines and symbols: using wall units based on $u_{\tau s}$.
}
\label{peak}
\end{figure}

Figure \ref{peak} shows the near-wall peak values of 
$u^+$ and $\theta^+$ as function of $Ro$.
The velocity and scalar fluctuations on the unstable side are either scaled
by $u_\tau$ and $\theta_\tau = Q_w/u_\tau$, respectively, or
by $u_{\tau u}$ and $\theta_{\tau u} = Q_w/u_{\tau u}$, respectively,
while on the stable channel side they are scaled
by $u_{\tau s}$ and $\theta_{\tau s} = Q_w/u_{\tau s}$, respectively.
The high similarity between the $u$ and $\theta$ field in the near-wall region
in non-rotating channel flow 
(Abe \& Antonia 2009) 
is reflected by the closeness of the peak values of 
$u^+$ and $\theta^+$, 
although the peak value of
$u^+$ is slightly higher than that of $\theta^+$, as a consequence of a 
smaller than unity $Pr$ (Pirozzoli \etal 2016).  
In rotating channel flow the difference between the peak values of
$u^+$ and $\theta^+$ 
is considerable if the scaling is based on $u_\tau$ and $\theta_\tau$,
but if the scaling of the peak values on the unstable and stable channel side
is based on
$u_{\tau u}$, $\theta_{\tau u}$ and
$u_{\tau s}$, $\theta_{\tau s}$, respectively, the differences are much smaller.
In the latter case, the peak values of 
$u^+$ and $\theta^+$ 
on the unstable channel side
show a very similar downward trend with $Ro$.
The trend on the stable channel side is non-monotonic, which may be related to the
appearance of turbulent-laminar patterns at $Ro = 0.45$ and 0.65.
Rotation reduces thus the similarity between the $u$- and $\theta$-field
on the unstable side, but their near-wall peak values are still highly correlated
and display a similar scaling in terms of wall units.

\section{6. Budget equations}

In this section, the budgets of the governing equations 
of the scalar energy $K_\theta = \frac{1}{2}\overline{\theta \theta}$ 
and $\overline{u\theta}$ and $\overline{v\theta}$ are considered
to obtain insights into the generation of
scalar fluctuations and fluxes.
Budgets for scalar transport in non-rotating channel have been studied 
by e.g. Johansson \& Wikstr{\"o}m (1999)
and in spanwise rotating channel flow by Nagano \& Hattori (2003) and Liu \& Lu (2007),
albeit at low Reynolds numbers.

\begin{figure}
\begin{center}
\setlength{\unitlength}{1cm}
(a)\includegraphics[height=5.5cm]{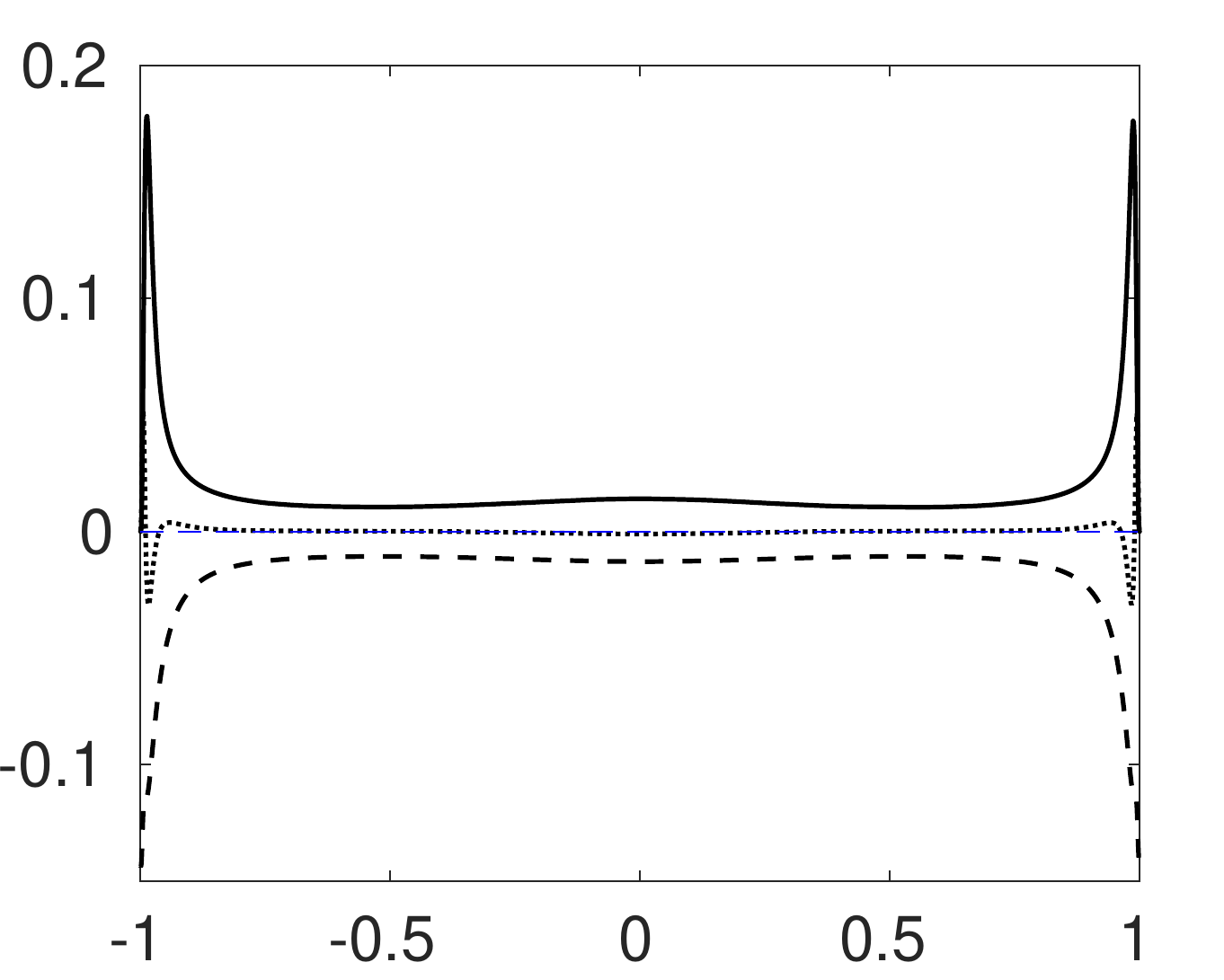}
\put(-3.4,-0.3){$y$}
\hskip4mm
(b)\includegraphics[height=5.5cm]{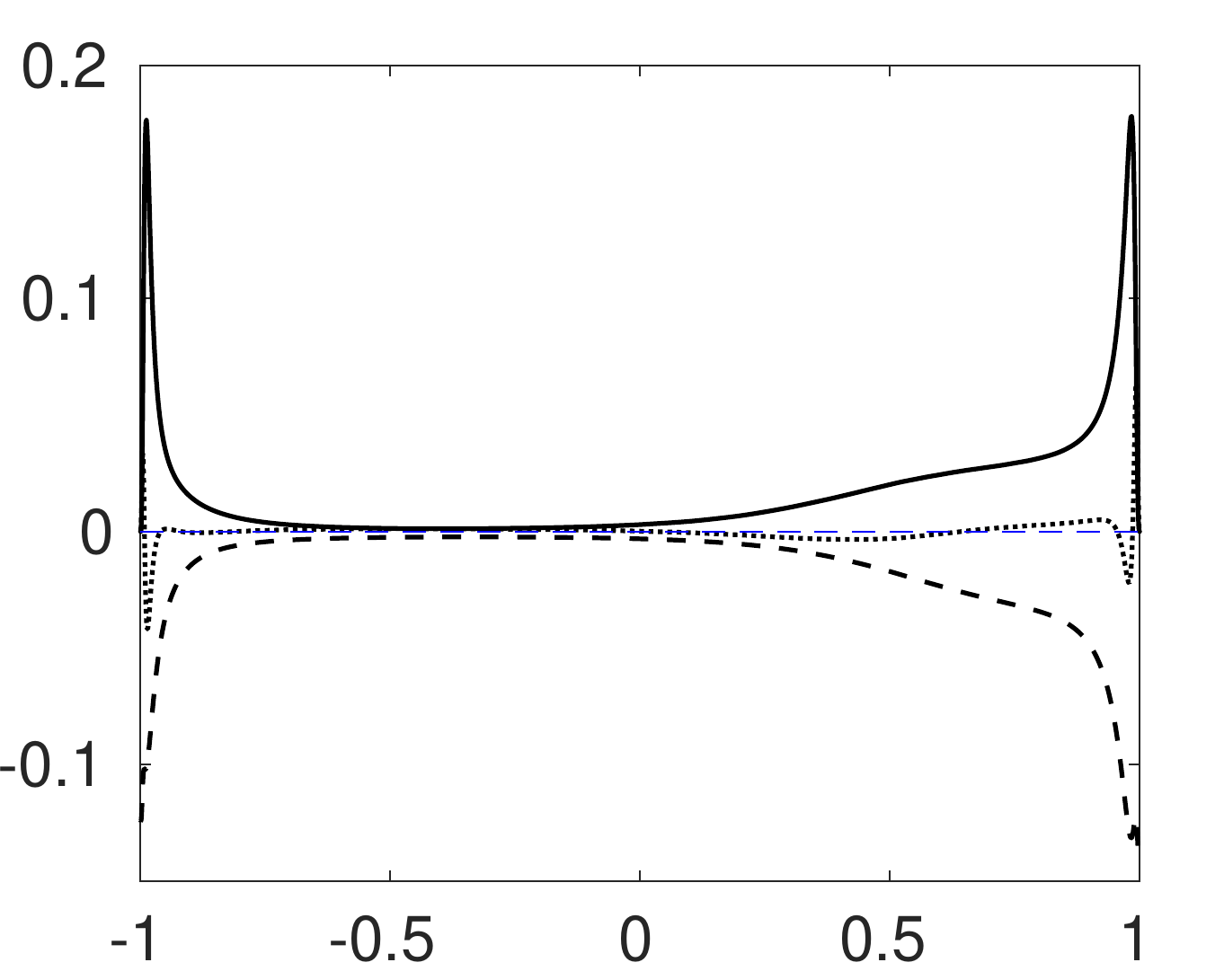}
\put(-3.4,-0.3){$y$}

\vskip2mm
(c)\includegraphics[height=5.5cm]{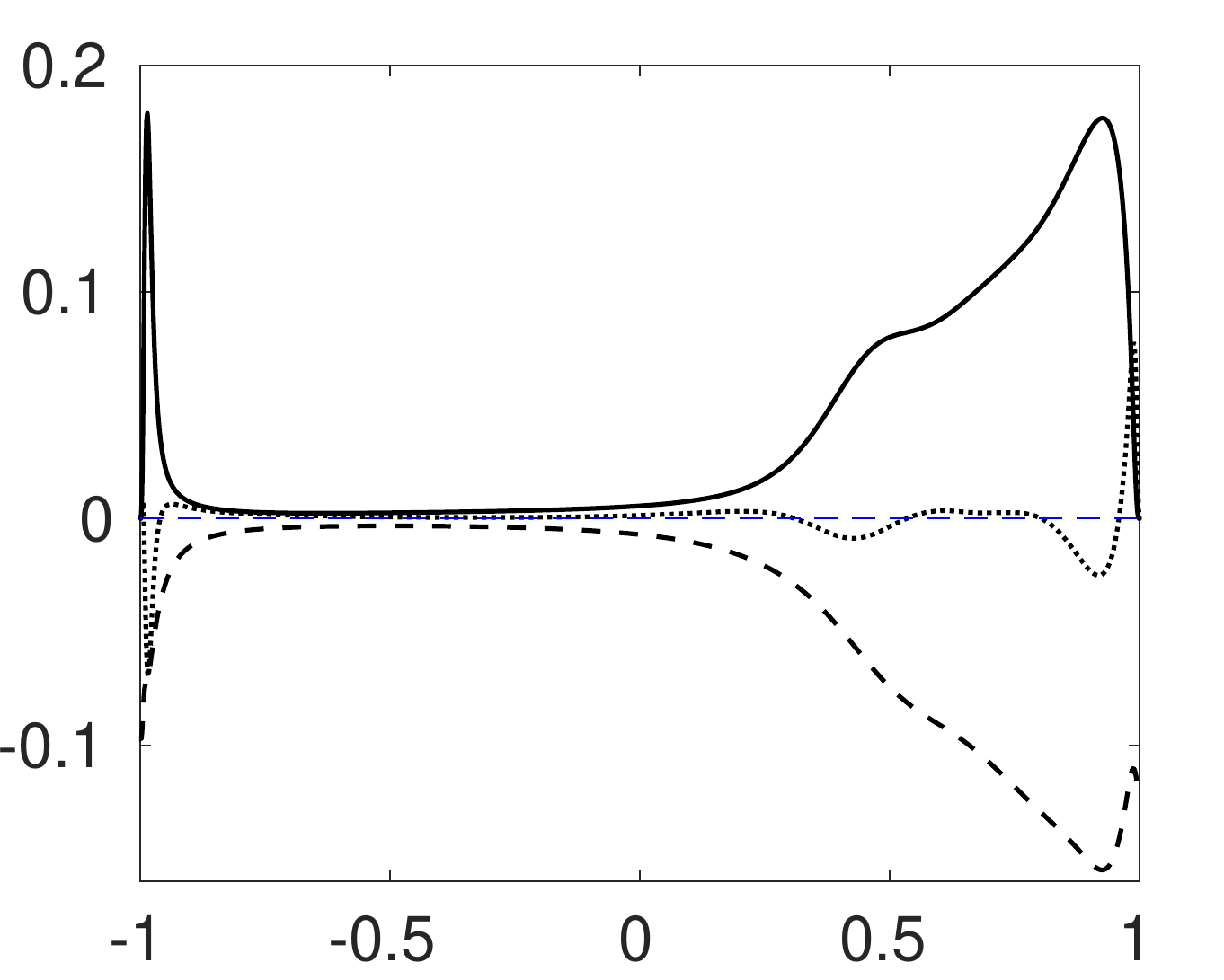}
\put(-3.4,-0.3){$y$}
\hskip4mm
(d)\includegraphics[height=5.5cm]{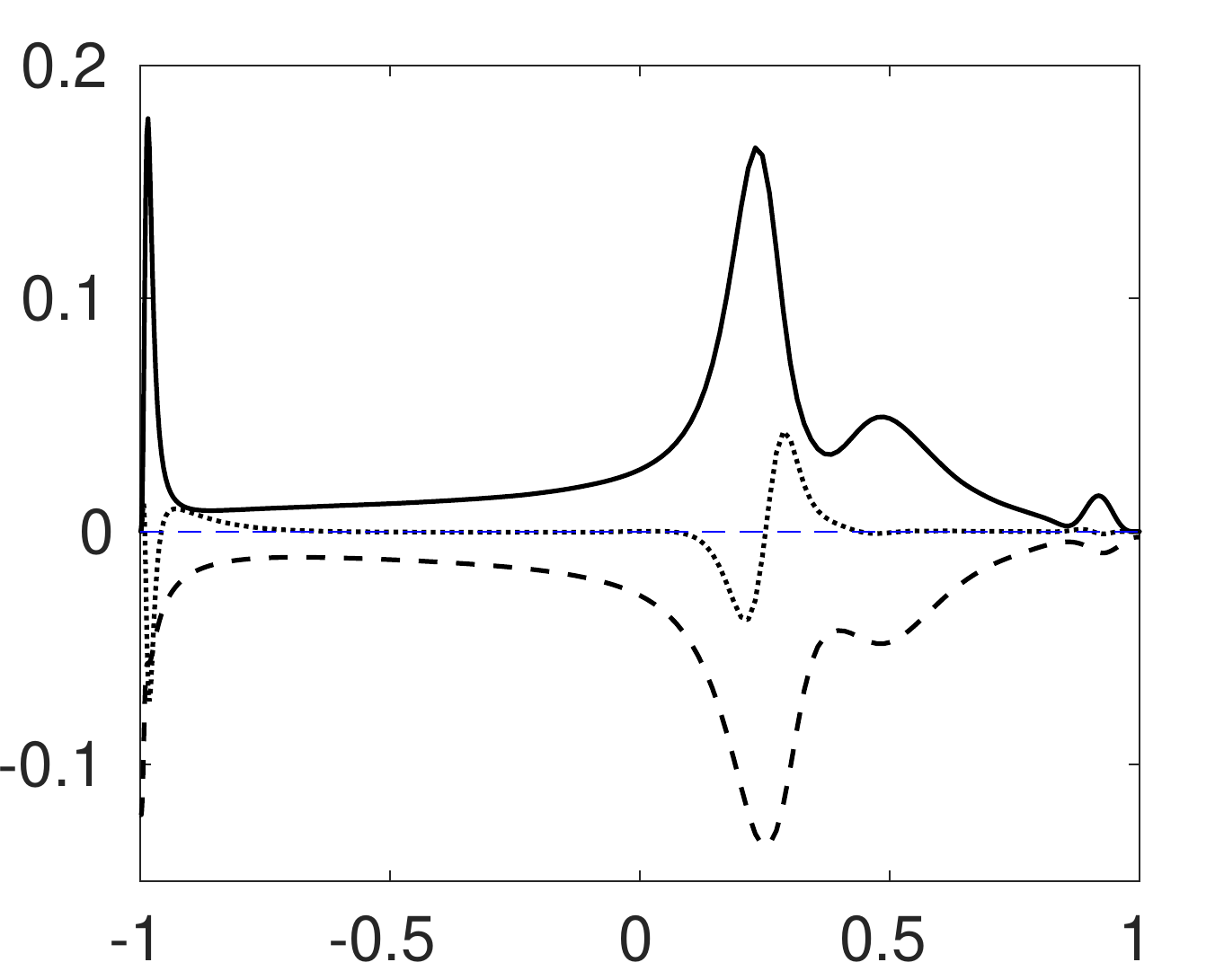}
\put(-3.4,-0.3){$y$}
\end{center}
\caption{Budgets of the $K_\theta$-balance equation in wall units
at (a) $Ro=0$, (b) $Ro=0.15$, (c) $Ro=0.65$ and (d) $Ro=1.2$.
($~^{\line(1,0){20}}$) $P^+_\theta$,
($---$) $\varepsilon^+_\theta$,
($\cdot\cdot\cdot$) $D^{t+}_\theta$.
}
\label{ktbud}
\end{figure}

\begin{figure}
\begin{center}
\setlength{\unitlength}{1cm}
(a)\includegraphics[height=5.5cm]{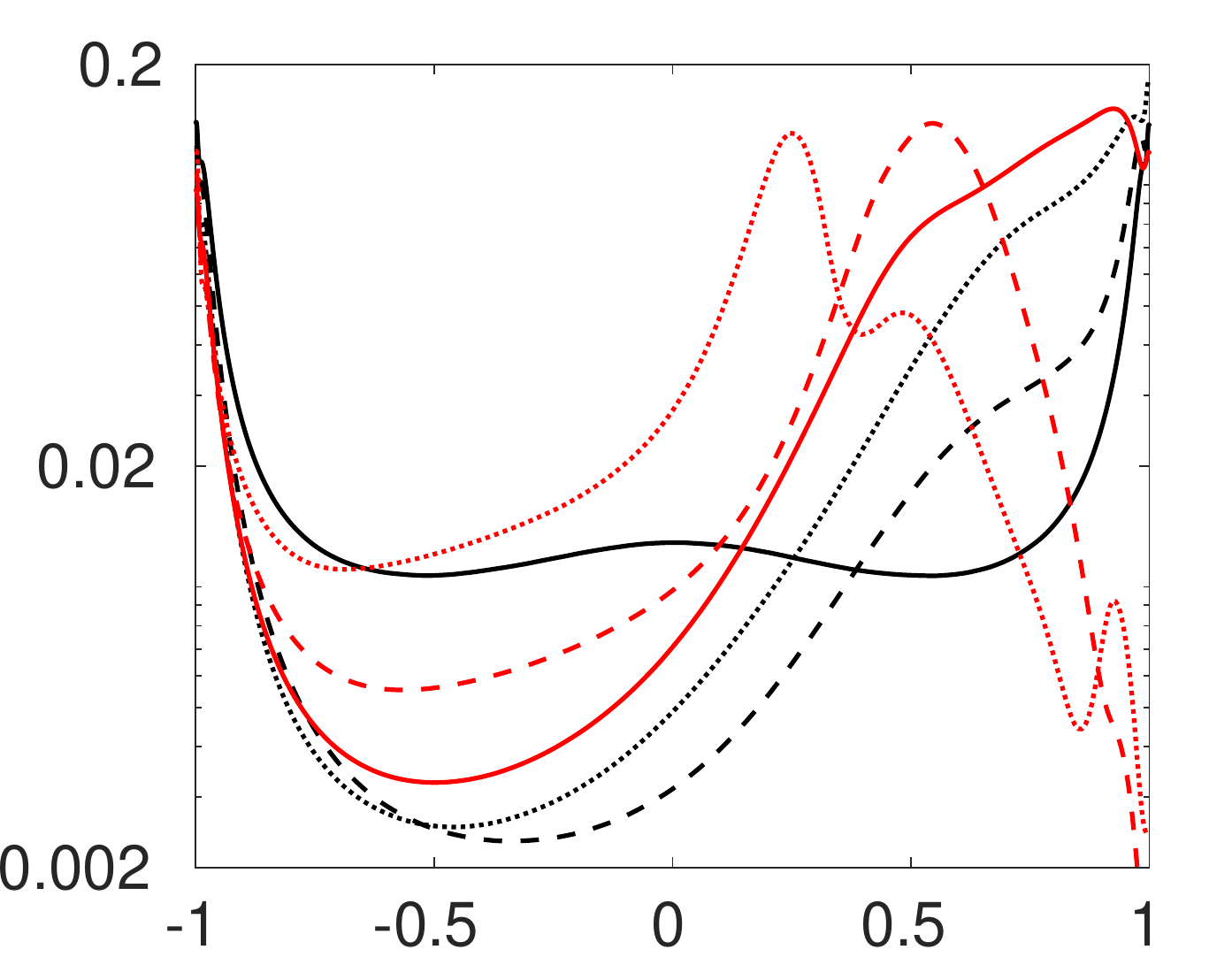}
\put(-3.4,-0.3){$y$}
\put(-7.4,2.6){$\varepsilon^+_\theta$}
\hskip4mm
(b)\includegraphics[height=5.5cm]{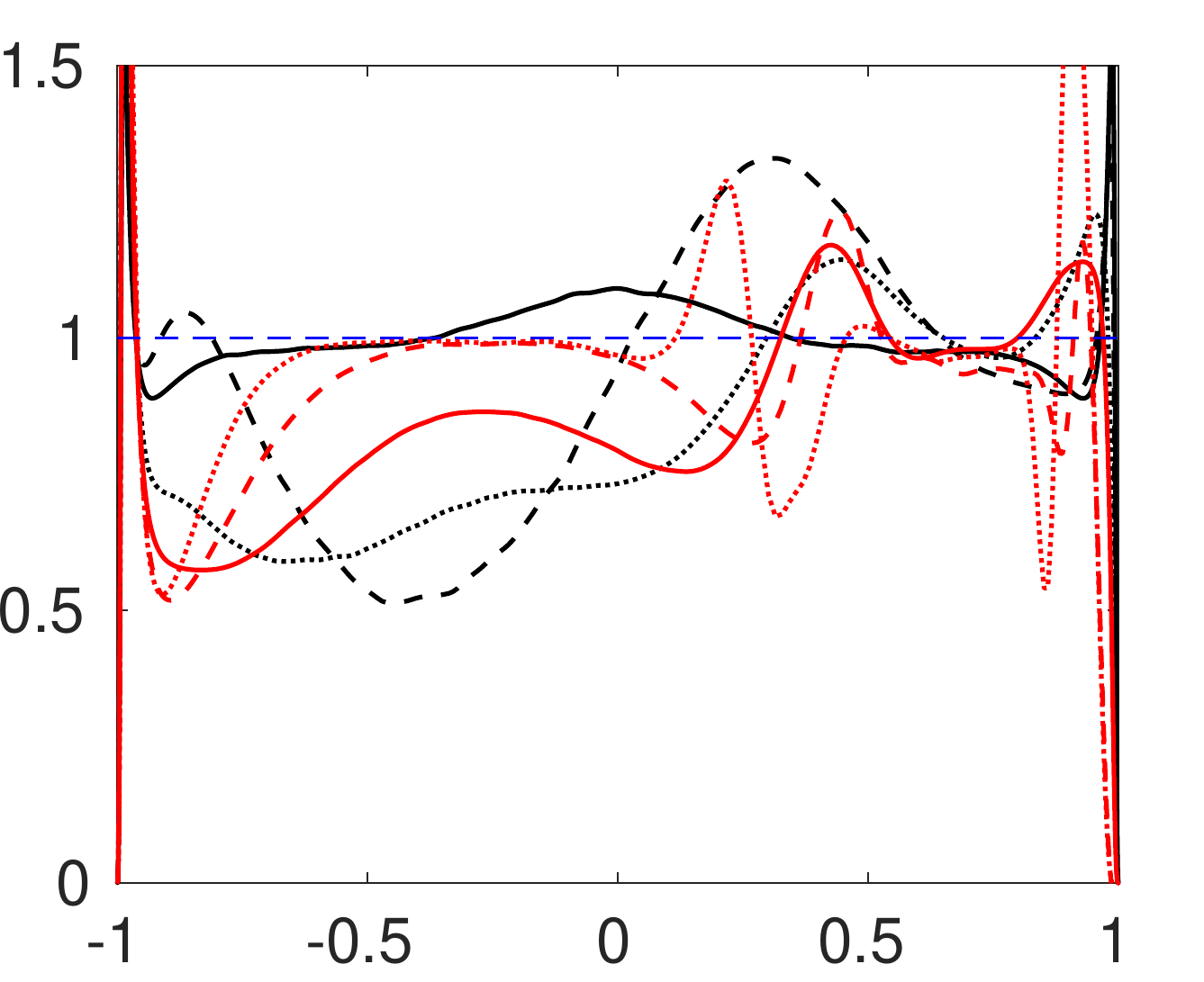}
\put(-3.4,-0.3){$y$}
\put(-7.4,2.6){$\frac{P_\theta}{\varepsilon_\theta}$}
\end{center}
\caption{Profiles of (a) $\varepsilon^+_\theta$ 
and (b) $P_\theta / \varepsilon_\theta$.
Lines as in figure \ref{ustat}.
}
\label{diss}
\end{figure}

In the present steady-state case,
the governing equation of the scalar energy $K_\theta$ reads
\begin{equation}
%
0 =
\underbrace{-\overline{v \theta}\frac{\p \Theta}{\p y}}_{P_\theta}
\underbrace{- \frac{1}{2} \frac{\p \overline{v \theta^2}}{\p y}}_{D^t_{\theta}}
+\underbrace{\alpha \frac{\p^2 K_\theta}{\p y^2}}_{D^m_{\theta}}
\underbrace{-\alpha \overline{\frac{\p \theta}{\p x_k}\frac{\p \theta}{\p x_k}}}_{\varepsilon_\theta},
\end{equation}
where $P_\theta$ represents production, 
$D^t_\theta$ and $D^m_\theta$ turbulent and molecular diffusion, respectively,
and $\varepsilon_\theta$ dissipation.
Figure \ref{ktbud} shows the budgets $P^+_\theta$, 
$D^{t+}_\theta$ and $\varepsilon^+_\theta$ in wall units at $Ro=0$, 0.15, 0.65 and 1.2.
The molecular diffusion $D^m_\theta$ is not shown since it is small.
The turbulent diffusion $D^t_\theta$ appears only significant near the wall and
at $Ro=1.2$ in the region away from the wall where $K_\theta$ and $P_\theta$ are large,
implying that there is primarily a balance between $P_\theta$ and $\varepsilon_\theta$.
Accordingly, at $Ro=0$ and $Ro \geq 0.9$ the ratio $P_\theta/\varepsilon_\theta$ is 
near unity in the outer region away from the border between the stable and unstable
channel side (figure \ref{diss}.b).
However, $P^+_\theta$ and $\varepsilon^+_\theta$ are both very small
on the unstable side away from the wall when $0.15 \lesssim Ro \lesssim 0.65$
(figure \ref{diss}.a) since the mean scalar gradient is small.
In these cases the ratio $P_\theta/\varepsilon_\theta$ deviates
significantly from unity (figure \ref{diss}.b) implying that $D^{t}_\theta$ is significant.
This disbalance between $P_\theta$ and $\varepsilon_\theta$ and the considerable
diffusion of $K_\theta$ are likely caused by the large-scale streamwise roll cells
being present at these rotation rates but which are much less dominant at higher $Ro$
(Brethouwer 2017). Liu \& Lu (2007) show that these roll cells indeed have a
large impact on scalar transport in rotating channel flows.
The maximum value of $P^+_\theta$ close to the walls is independent of $Ro$ and close
to its theoretical value $Pr/4$ (Johansson \& Wikstr{\"o}m 1999).
At $Ro=0.15$ and even more so at $Ro=0.65$, both 
$\varepsilon^+_\theta$ and $P^+_\theta$ are large on the stable channel side (figure \ref{ktbud}.b, c)
where also the mean scalar gradient and $K_\theta$ are large (figure \ref{tmean}).
At higher $Ro$, the peaks of
$\varepsilon^+_\theta$ and $P^+_\theta$ on the stable side move away from the wall to
the position, going from the unstable towards the stable side, 
where $\diff \Theta / \diff y$ steepens
while $\overline{v\theta}$ is still large (figure \ref{ktbud}.d).
Accordingly, the maximum of $\theta^+$ at high $Ro$
moves away from the wall at high $Ro$ (figure \ref{tmean}.b).
Moving closer to the wall
on the stable side, $\overline{v\theta}$ and consequently
$P^+_\theta$ and $\varepsilon^+_\theta$ diminish rapidly.

The respective governing equations of $\overline{u\theta}$
and $\overline{v\theta}$ read in the present case
\begin{align}
%
0 &=&
\underbrace{-\overline{uv}\frac{\p \Theta}{\p y}-\overline{v\theta}\frac{\p U}{\p y}}_{P_1}
\underbrace{- \frac{\p \overline{uv\theta}}{\p y}}_{D^t_{1}}
+\underbrace{\frac{\p}{\p y}\left(
\alpha \overline{u \frac{\p \theta}{\p y}}
+\nu \overline{\theta \frac{\p u}{\p y}}
\right)}_{D^m_{1}}
+\underbrace{\overline{\frac{p}{\rho} \frac{\p\theta}{\p x}}-(\alpha+\nu) \overline{\frac{\p u}{\p x_k}\frac{\p \theta}{\p x_k}}}_{\Pi_1}
+\underbrace{2 \Omega \overline{v\theta}}_{C_1}\label{uteq}\\
%
%
0 &=& 
\underbrace{-\overline{v^2}\frac{\p \Theta}{\p y}}_{P_2}
\underbrace{- \frac{\p \overline{v^2\theta}}{\p y}}_{D^t_{2}}
\underbrace{- \frac{\p \overline{p\theta}}{\p y}}_{D^p_{2}}
+\underbrace{\frac{\p}{\p y}\left(
\alpha \overline{v \frac{\p \theta}{\p y}}
+\nu \overline{\theta \frac{\p v}{\p y}}
\right)}_{D^m_{2}}
+\underbrace{\overline{\frac{p}{\rho} \frac{\p\theta}{\p y}}-(\alpha+\nu) \overline{\frac{\p v}{\p x_k}\frac{\p \theta}{\p x_k}}}_{\Pi_2}
\underbrace{-2 \Omega \overline{u\theta}}_{C_2}
\label{vteq}
%
\end{align}
where $P_1$, $P_2$ represent production, 
$D^t_1$, $D^t_2$ turbulent diffusion,
$D^p_2$ pressure diffusion,
and $D^m_1$, $D^m_2$ molecular diffusion.
The sum of pressure scalar-gradient correlation and diffusive and viscous dissipation,
$\Pi_1$ and $\Pi_2$, is considered because 
diffusive and viscous dissipation are generally considered to be small and therefore
often added to the pressure scalar-gradient correlation term in turbulence modelling
(Wikstr{\"o}m \etal 2000).
The Coriolis force leads to the additional terms $C_1$ and $C_2$
in the governing equations of the scalar fluxes.

\begin{figure}
\begin{center}
\setlength{\unitlength}{1cm}
(a)\includegraphics[height=5.5cm]{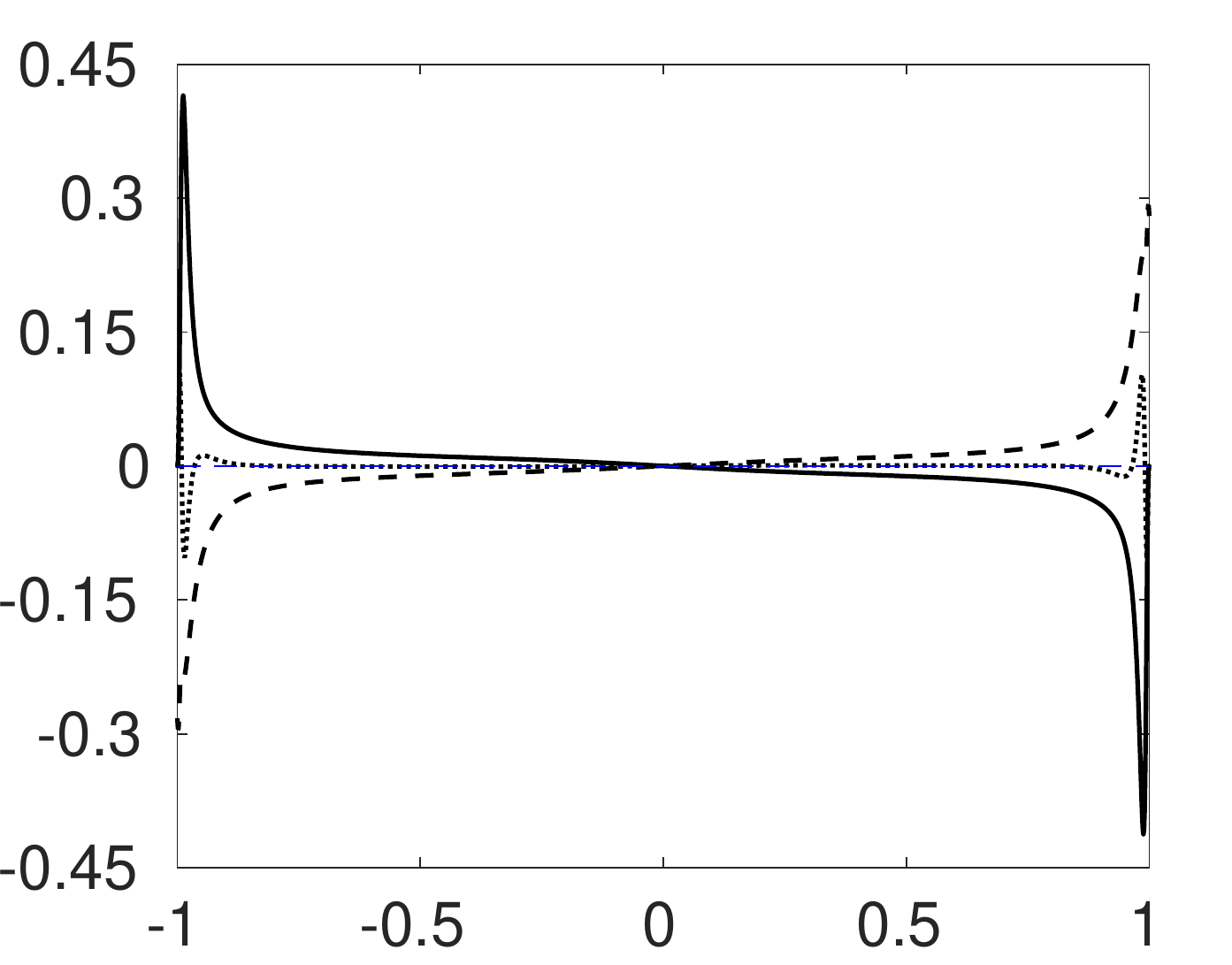}
\put(-3.4,-0.3){$y$}
\hskip4mm
(b)\includegraphics[height=5.5cm]{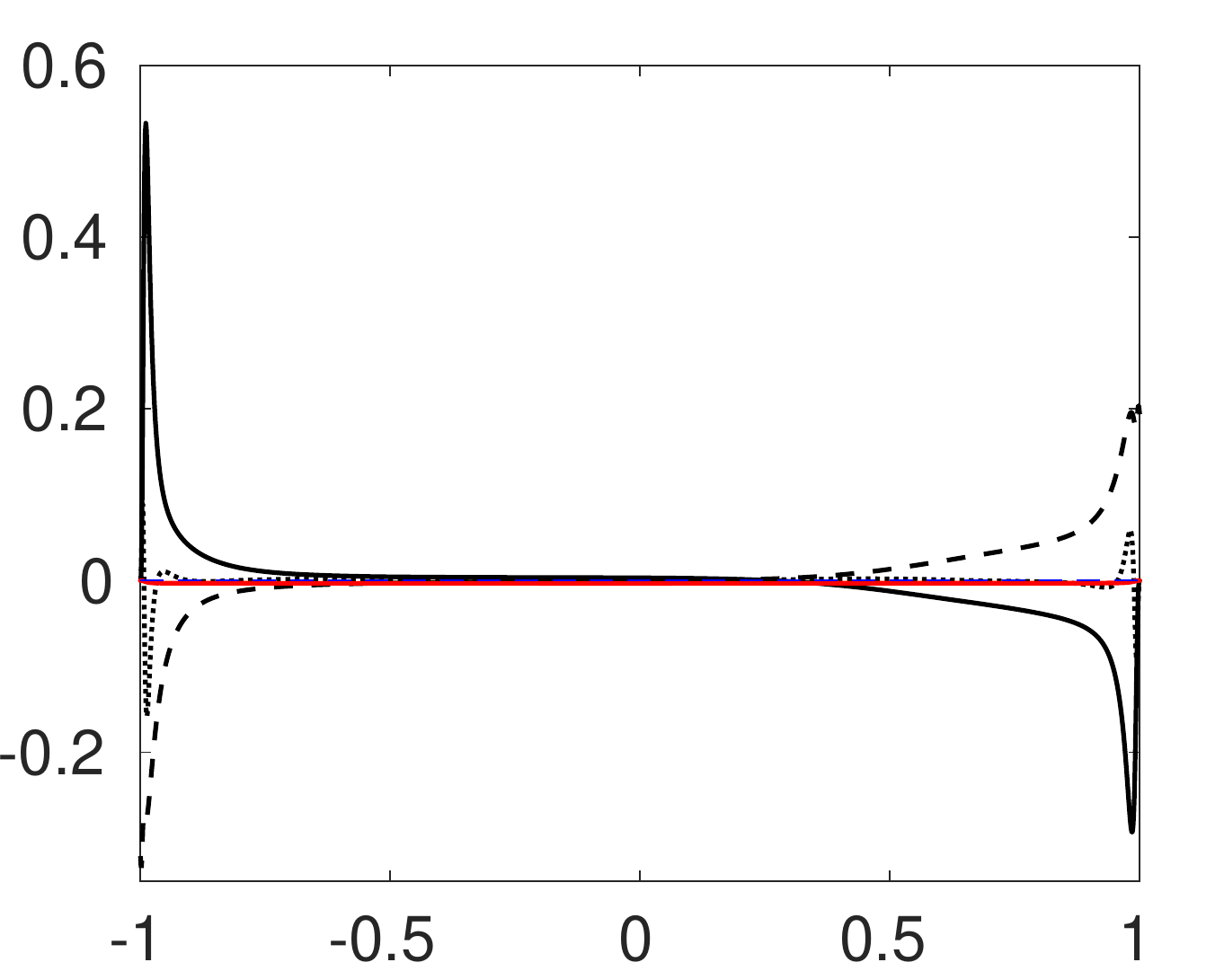}
\put(-3.4,-0.3){$y$}

\vskip2mm
(c)\includegraphics[height=5.5cm]{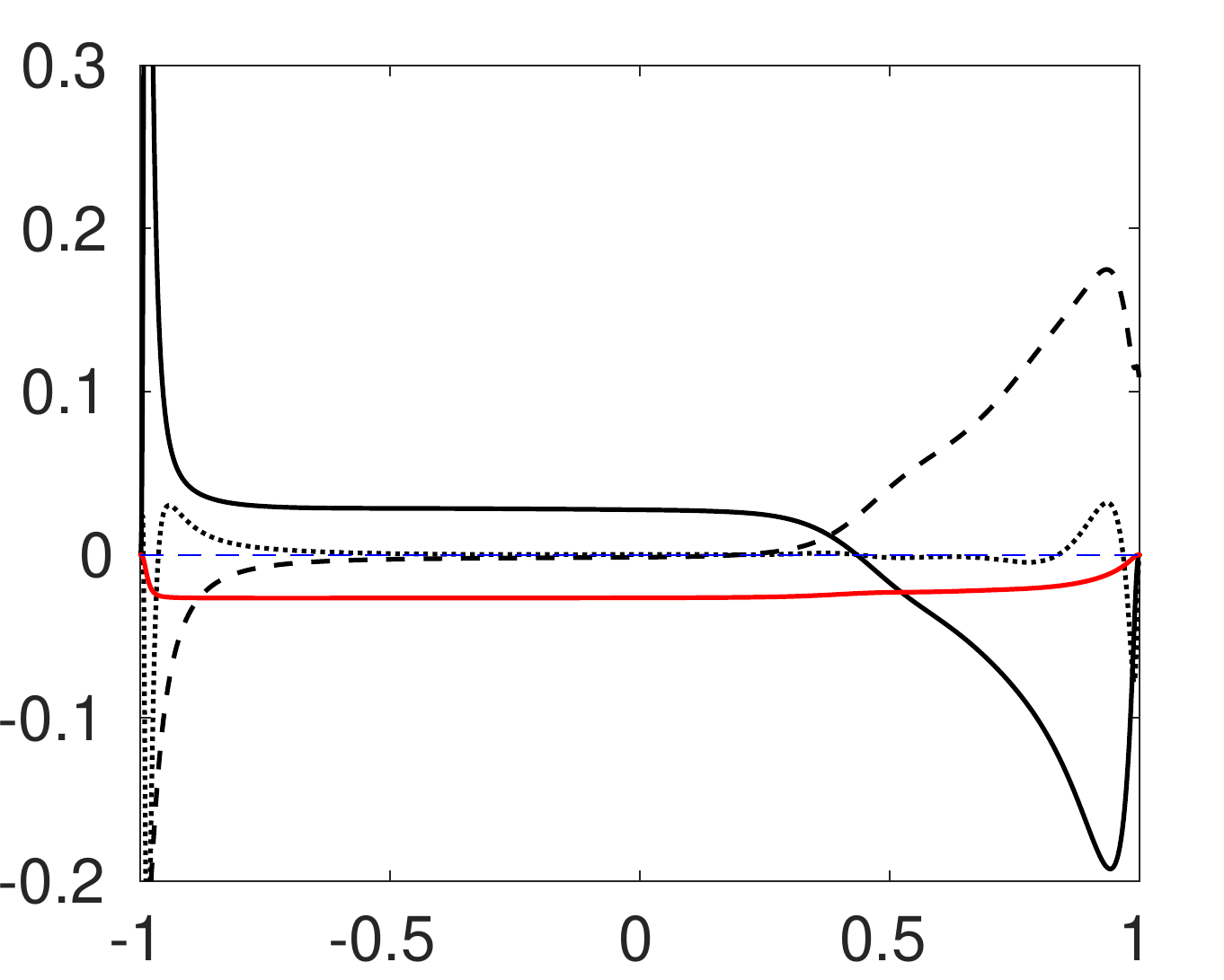}
\put(-3.4,-0.3){$y$}
\hskip4mm
(d)\includegraphics[height=5.5cm]{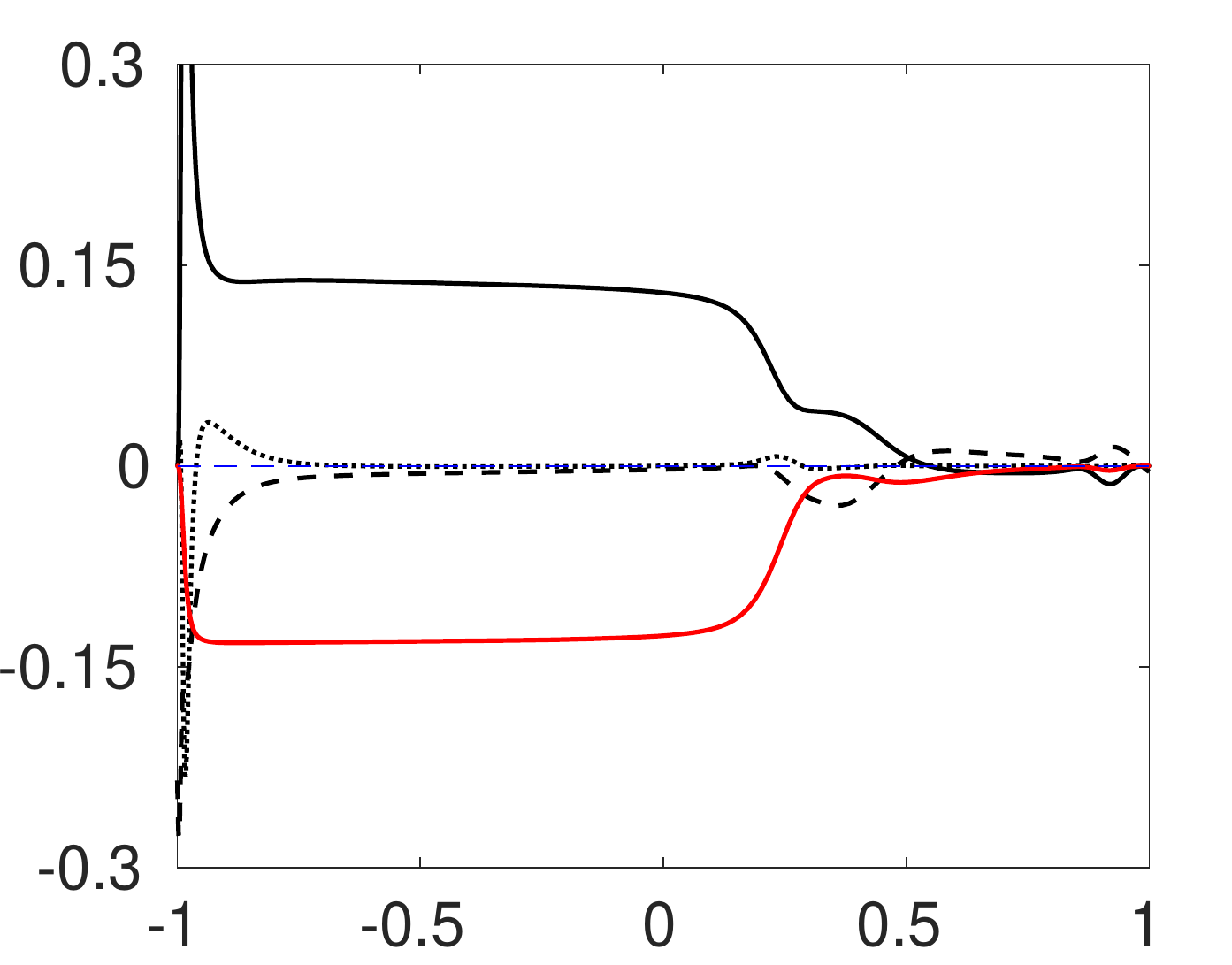}
\put(-3.4,-0.3){$y$}
\end{center}
\caption{Budgets of the $\overline{u\theta}$-balance equation in wall units
at (a) $Ro=0$, (b) $Ro=0.15$, (c) $Ro=0.65$ and (d) $Ro=1.2$.
($~^{\line(1,0){20}}$) $P^+_1$,
($---$) $\Pi^+_1$,
($\cdot\cdot\cdot$) $D^{t+}_1$,
($\textcolor{red}{~^{\line(1,0){20}}}$) $C^+_1$.
}
\label{utbud}
\end{figure}

Figure \ref{utbud} shows the budgets $P^+_1$, $\Pi^+_1$,
$D^{t+}_1$ and $C^+_1$ of the governing equation (\ref{uteq}) of $\overline{u\theta}$
in wall units at $Ro=0$, 0.15, 0.65 and 1.2.
The molecular diffusion $D^m_1$ is again not shown because of its smallness.
At $Ro=0$, $P^+_1$ is in the whole channel mainly balanced by $\Pi^+_1$ and the same applies
to the near wall region at $Ro=0.15$ (figure \ref{utbud}.a, b).
When $-0.6 \lesssim y \lesssim 0.4$ all budget terms are small in the latter
case, but a closer inspection reveals that the Coriolis term
$C^+_1$ is of the same order as $P^+_1$ and $\Pi^+_1$ and negative since $\overline{v\theta}$
is negative. At a higher $Ro=0.65$, $P^+_1$ and $\Pi^+_1$ are large and
approximately balance each other in the near-wall region
on the unstable side and a large part of the stable side (figure \ref{utbud}.c). On the other hand,
in the outer region of the unstable side,
$C^+_1$ is significant and balanced by $P^+_1$ 
and contributes to a small $\overline{u\theta}$ in this
part of the channel (figure \ref{tflux}.a).
When $Ro$ is raised to 1.2 turbulence disappears on the stable side
and all budget terms are small there (figure \ref{utbud}.d).
In the major part of the unstable channel side $D^{t+}_\theta$ and $\Pi^+_1$ are small but
$P^+_1$ and $C^+_1$ are both large and balance each other.
The balance of $P^+_1$ and $C^+_1$ in rotating channel flow on the unstable channel side
can be understood by considering the production term more closely. This term has two
contributions, namely $-\overline{uv}(\p \Theta/\p y)$ and $-\overline{v\theta}(\p U/\p y)$,
see equation (\ref{uteq}). The first part is small on the unstable side
away from the wall because $\p \Theta/\p y$ is small. The second part
$-\overline{v\theta}(\p U/\p y)$ is large but is nearly balanced by 
$C_1 = \overline{v\theta}2 \Omega$ since $\p U / \p y \approx 2 \Omega$ on the unstable side.
The term induced by the Coriolis force has thus a major impact on the streamwise scalar flux
and contributes to the alignment between the turbulent scalar flux vector
and mean scalar gradient in rotating channel flow.

\begin{figure}
\begin{center}
\setlength{\unitlength}{1cm}
(a)\includegraphics[height=5.5cm]{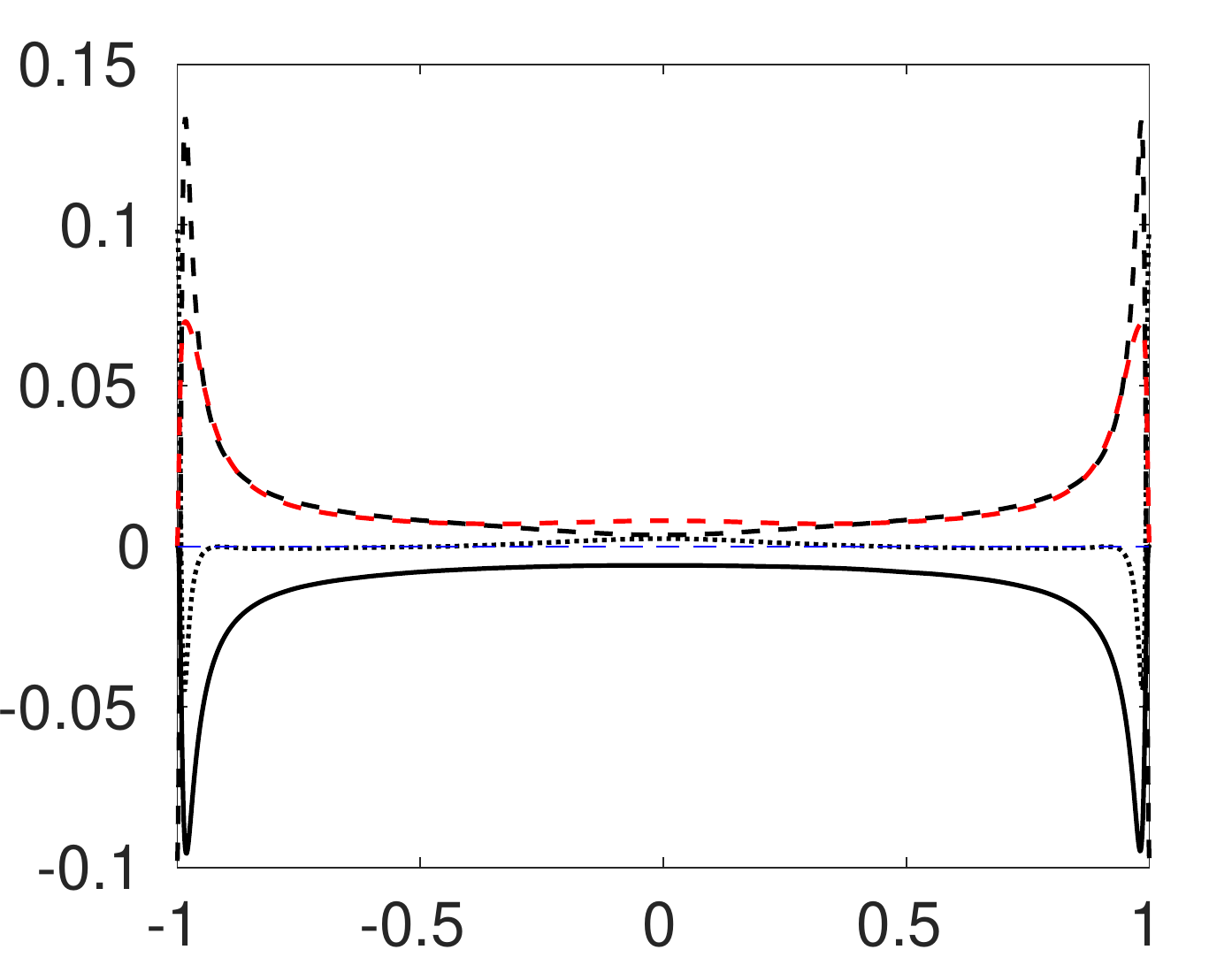}
\put(-3.4,-0.3){$y$}
\hskip4mm
(b)\includegraphics[height=5.5cm]{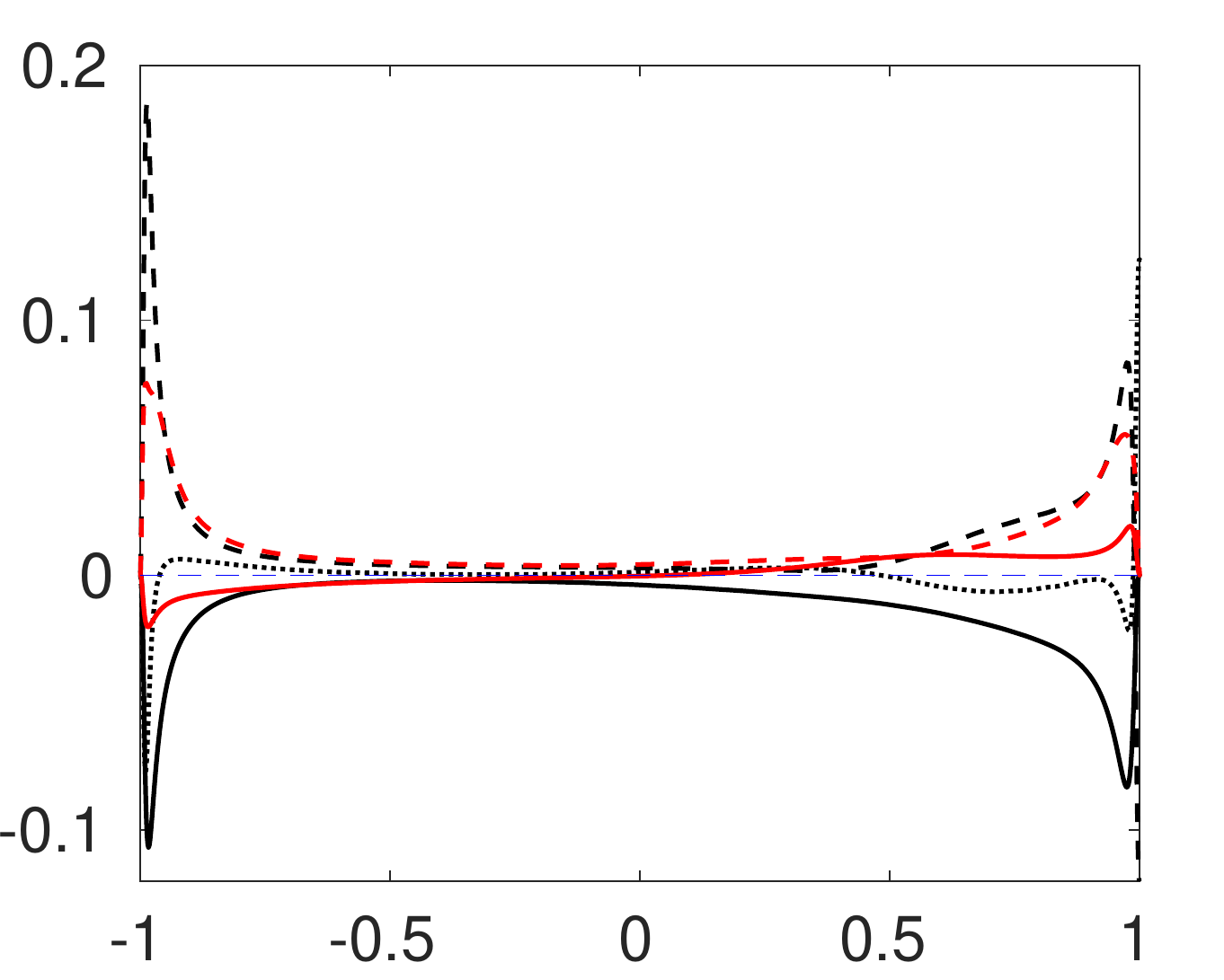}
\put(-3.4,-0.3){$y$}

\vskip2mm
(c)\includegraphics[height=5.5cm]{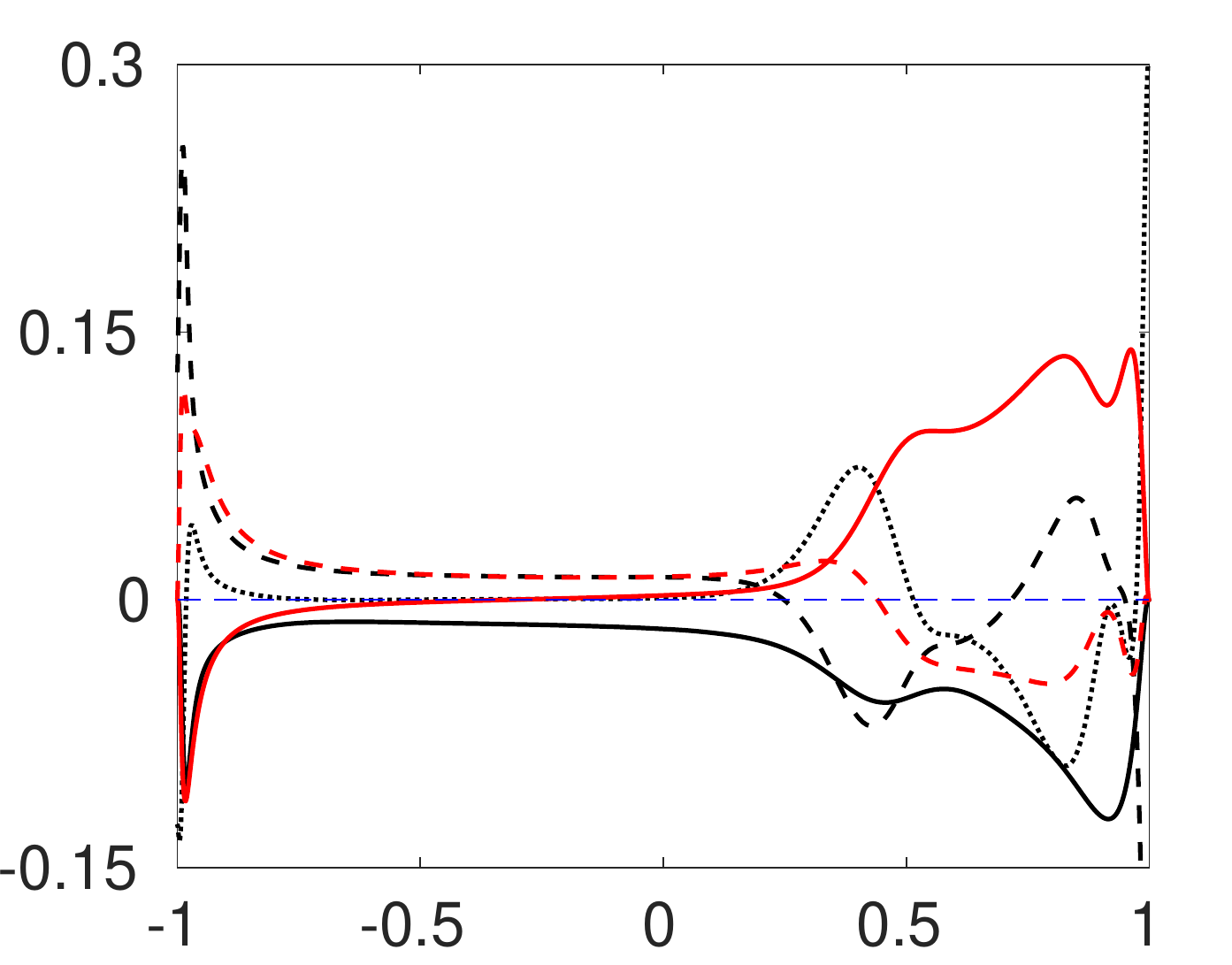}
\put(-3.4,-0.3){$y$}
\hskip4mm
(d)\includegraphics[height=5.5cm]{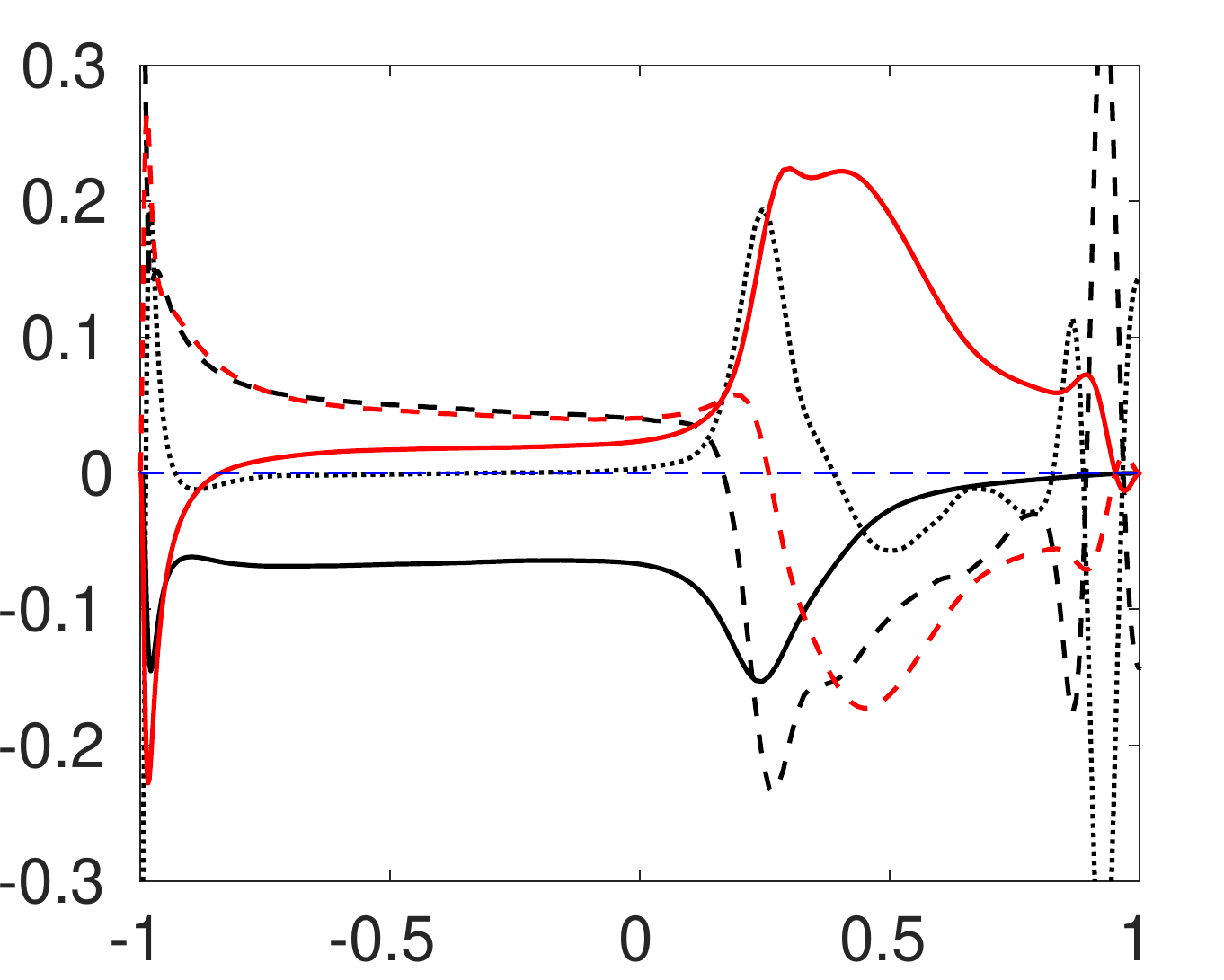}
\put(-3.4,-0.3){$y$}
\end{center}
\caption{Budgets of the $\overline{v\theta}$-balance equation in wall units
at (a) $Ro=0$, (b) $Ro=0.15$, (c) $Ro=0.65$ and (d) $Ro=1.2$.
($~^{\line(1,0){20}}$) $P^+_2$,
($---$) $\Pi^+_2$,
($\cdot\cdot\cdot$) $D^{t+}_2+D^{p+}_2$,
($\textcolor{red}{---}$) $\Pi^+_2+D^{p+}_2$,
($\textcolor{red}{~^{\line(1,0){20}}}$) $C^+_2$.
}
\label{vtbud}
\end{figure}

Figure \ref{vtbud} shows the budgets $P^+_2$, $\Pi^+_2$,
the sum $D^{t+}_2 + D^{p+}_2$ and $C^+_2$ in the governing equation 
(\ref{vteq}) of $\overline{v\theta}$
in wall units at $Ro=0$, 0.15, 0.65 and 1.2.
The molecular diffusion $D^m_2$ is again small.
At $Ro=0$ the dominant contributions 
come from $P^+_2$ and $\Pi^+_2$, except very near the walls and near the centreline
where diffusion 
$D^{t+}_2 + D^{p+}_2$ is significant. At $Ro=0.15$ diffusion and $C^+_2$ are
significant as well as $P^+_2$ and $\Pi^+_2$ in the main part of the channel.
The Coriolis term $C^+_2$ contributes to the production of negative 
$\overline{v\theta}$ on the unstable side but counteracts it on the stable channel side
since $\overline{u\theta}$ changes sign. At a higher $Ro=0.65$, $C^+_2$ is important
and large on the stable side and near the wall on the unstable side but small
away from the wall since $\overline{u\theta}$ is small.
In a large part of the unstable channel side there is thus primarily a balance between
$P^+_2$ and $\Pi^+_2$. At a higher $Ro=1.2$, $P^+_2$ is nearly constant away from the wall
on the unstable side and balanced by $\Pi^+_2$ and $C^+_2$ while it rapidly declines approaching
the wall on the stable channel side 
where both $\Pi^+_2$ and $C^+_2$ are large
(figure \ref{vtbud}.d). 
Note that $C_2$ is positive on the stable side in rotating channel flow
and thus contributes to the reduced wall-normal turbulent scalar transport on that side.
On the stable channel side, both $\Pi^+_2$ and diffusion dominated by
$D^{p+}_2$ display large variations at $Ro=0.65$ and 1.2. By contrast, the profile of the sum 
$\Pi^+_2+D^{p+}_2 = -((\overline{\theta \p p}/\p y)/\rho)^+$, 
which is also displayed in figure \ref{vtbud},
is much smoother and does not show extreme values near the wall in rotating channel flow. 
This suggests that it might be easier to model the latter term instead of 
$\Pi_2$ and $D^p_2$ separately.

\begin{figure}
\begin{center}
\setlength{\unitlength}{1cm}
(a)\includegraphics[height=5.5cm]{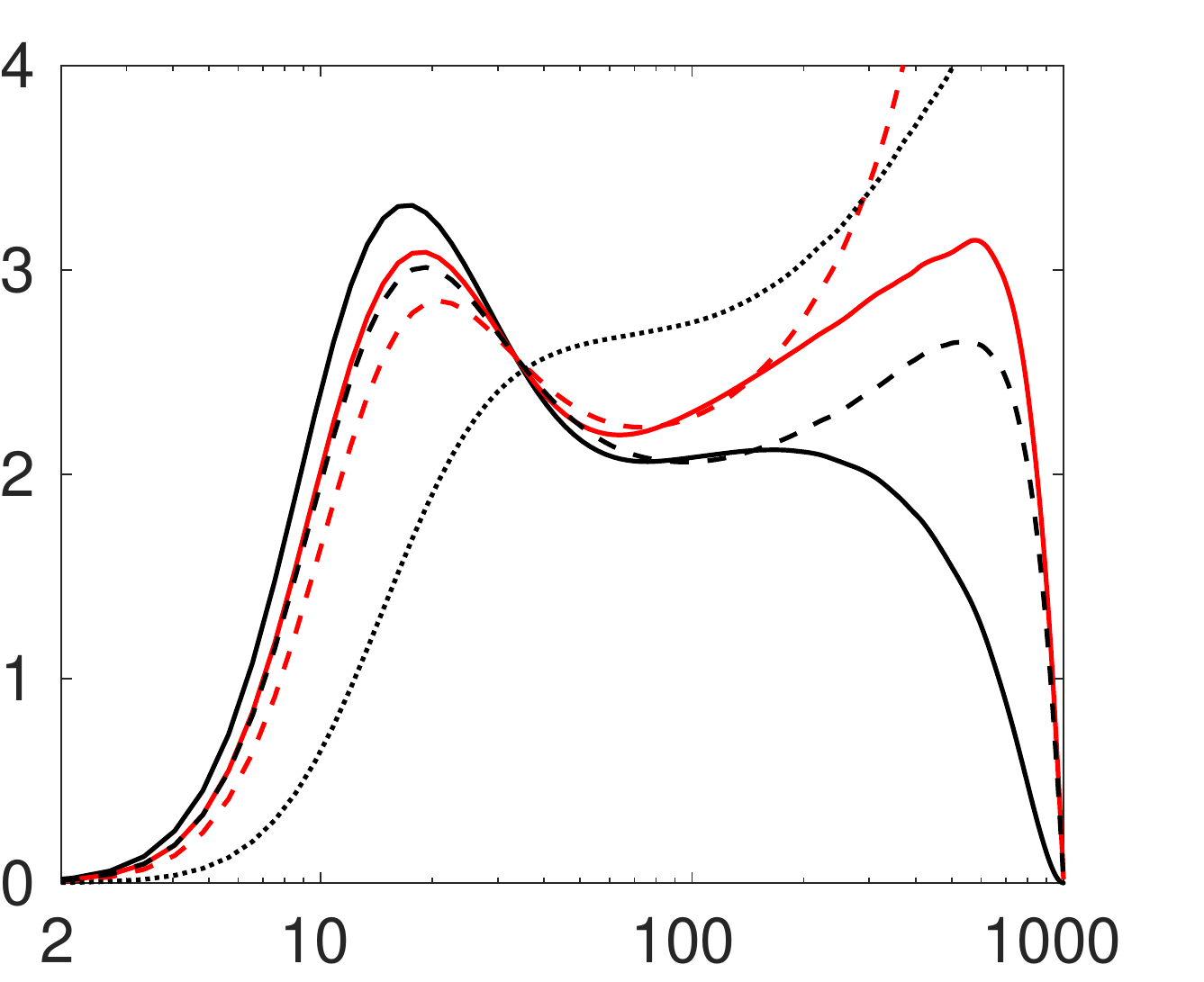}
\put(-3.4,-0.3){$y^*$}
\hskip4mm
(b)\includegraphics[height=5.5cm]{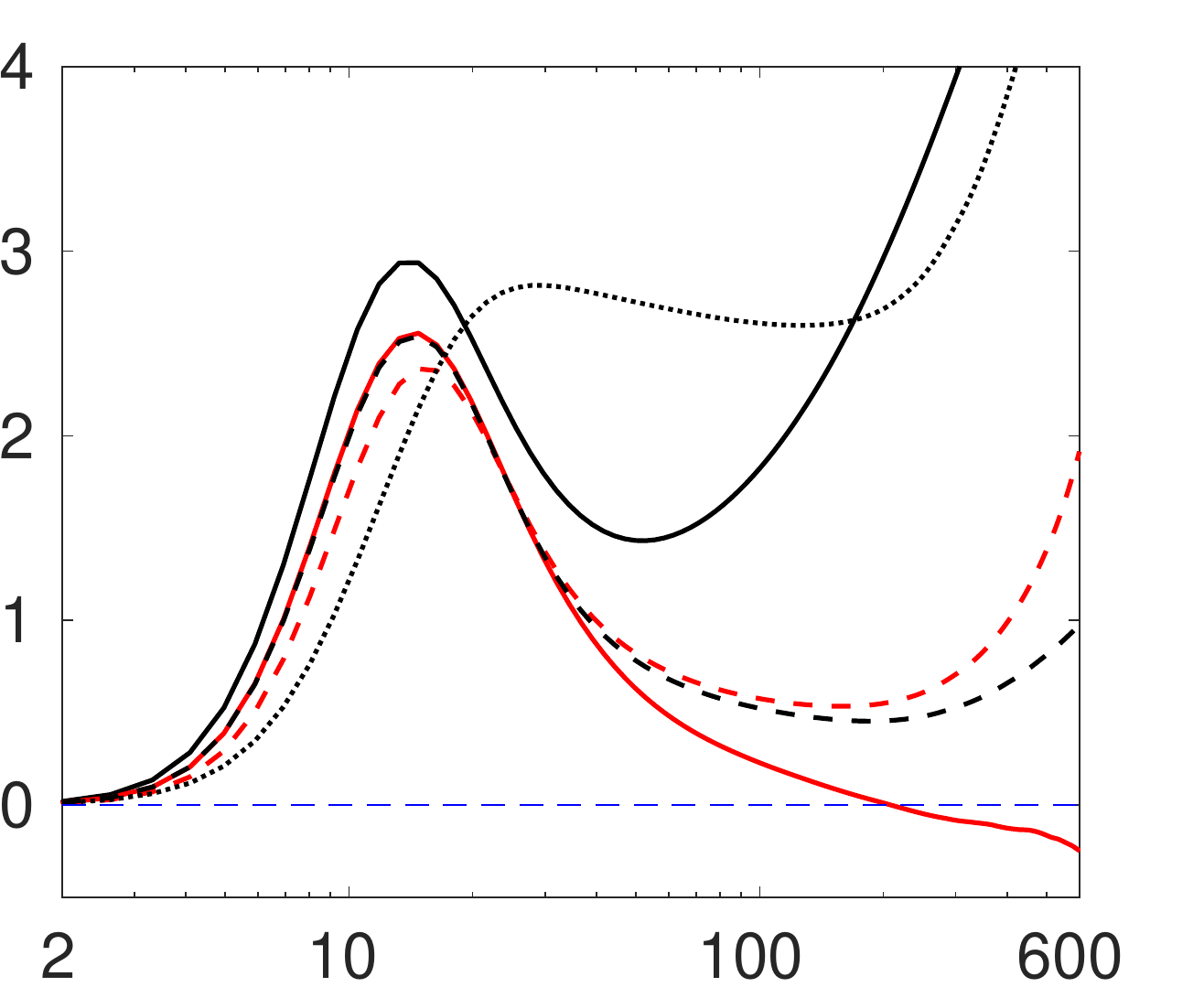}
\put(-3.4,-0.3){$y^*$}
\end{center}
\caption{Premultiplied production terms in wall units on the unstable side
at (a) $Ro=0$ and (b) $Ro=0.65$.
($~^{\line(1,0){20}}$) $y^* P^*_K$,
($\textcolor{red}{---}$) $y^* P^*_\theta$,
($---$) $y^* P^*_{1\theta}$,
($\textcolor{red}{~^{\line(1,0){20}}}$) $y^* P^*_{1u}$,
($\cdot\cdot\cdot$) $y^*(P^*_2+C^*_2 )$.
}
\label{prod}
\end{figure}

Figure \ref{prod} shows in more detail the production terms of turbulent 
kinetic energy $P_K$ and $P_\theta$ near the wall on the unstable side at $Ro=0$ and 0.65.
Furthermore, the figure shows the sum $P_2 + C_2$, 
which can be regarded as a total production of $\overline{v\theta}$,
and $P_{1\theta}=-\overline{uv}(\partial \Theta/\partial y)$ and
$P_{1u}=\overline{v\theta}(2\Omega\overline{v\theta} -\partial U/\partial y)$, 
which are the production owing to the mean scalar gradient, respective,
the production by mean shear together with
Coriolis term contribution in the governing equation (\ref{uteq}) of $\overline{u\theta}$.
The latter term, $P_{1u}$, is small away from the wall on the unstable side of a rotating channel when
$\p U / \p y \approx 2 \Omega$, as explained before.
All terms and the distance to the wall $y^*$ are in wall units. 
The production terms are premultiplied by $y^*$ to accentuate the behaviour away from
the wall.
Here, $u_{u\tau}$ is used instead of $u_\tau$ for the scaling at $Ro=0.65$
because this velocity scale is more appropriate on the unstable side.
A DNS of scalar transport with $Pr=1$ 
in a non-rotating channel flow (not presented here) shows that the profiles 
of $P^*_K$, $P^*_\theta$, $P^*_{1\theta}$ and $P^*_{1u}$ including the maxima
all collapse near the wall, see also Pirozzoli \etal (2016). 
Figure \ref{prod}.(a) shows that in 
the present case with $Pr=0.71$ the profiles
do not collapse when $Ro=0$ since the maximum of $P^*_K$ approaches $1/4$ whereas 
the maximum of $P^*_\theta$ is $Pr/4$ as explained before.
The maxima of $P^*_{1\theta}$ and $P^*_{1u}$ are in between these two previous maxima
and approximately equal to each other, showing that production owing to
the mean scalar gradient and mean shear are equally important. 
All these maxima are nearly independent
of $Ro$ (compare figure \ref{prod}.a and b). The profiles of
$P^*_{1\theta}$ and $P^*_{1u}$ approximately collapse for $y^* \lesssim 20$
(figure \ref{prod}.b),
also at other $Ro$, while they do not collapse if the Coriolis term
is not included in $P^*_{1u}$, which is somewhat remarkable.

\section{7. Efficiency of scalar transport}

In this section, some quantities characterising the efficiency of scalar transport
are discussed;
some of them are directly relevant for engineering.

An important quantity in engineering is the Nusselt number.
It is here defined as the ratio of wall-normal scalar flux
in the present turbulent channel flows and laminar channel flow,
i.e.
\begin{equation}
Nu = \frac{2 h}{\alpha \Delta \Theta} Q_w,
\end{equation}
where $\Delta \Theta$ is the imposed scalar difference at the walls.
With this definition, $Nu=1$ for laminar channel flow.

However, $Nu$ does not distinguish between the scalar transport on
the stable and unstable channel sides. To examine the effectivity of scalar
transport on both channel sides, I have also computed a Nusselt number
defined similarly as in some previous studies, e.g. Pirozzoli \etal (2016),
to account for difference in the mean scalar gradients on both sides.
For the unstable channel side it is defined as
\begin{equation}
Nu^*_u = 
\frac{5}{8} Q_w \frac{\delta}{\alpha |\Theta_w - \Theta_m |}.
\label{nu_eq}
\end{equation}
Here, $\delta = y_0-y_w$ where $y_0$ is the position where the total shear stress, i.e. the sum of
the viscous and turbulent shear stress, is zero, and $y_w$ the wall position 
and $\Theta_w$ the scalar value at the wall on the unstable channel side.
I consider $y_0$ the place separating the stable and unstable channel sides.
Further
\begin{equation}
\Theta_m = \frac{1}{\delta}\int^{y_0}_{y_w} \frac{U \Theta}{U_m} \diff y,~~~~~
U_m = \frac{1}{\delta}\int^{y_0}_{y_w} U \diff y.
\label{int_eq}
\end{equation}
The Nusselt number for the stable channel side, $Nu^*_s$, is defined
in the same way with the integrals in (\ref{int_eq}) from $y_0$ to $y_w$
where $y_w$ and $\Theta_w$ are now the wall
position and the scalar value at the wall on the stable channel side.
The factor $5/8$ in (\ref{nu_eq}) and in the similar expression for $Nu^*_s$ ensures that 
$Nu^*_u=Nu^*_s=1$ if the flow is laminar.
The main difference between the expressions for $Nu$
and $Nu^*_u$, $Nu^*_s$ is that $\Delta\Theta/2h$ is replaced by $|\Theta_w - \Theta_m |/\delta$
as an effective mean scalar gradient to characterize both channel sides.

\begin{figure}[t]
\begin{center}
\setlength{\unitlength}{1cm}
(a)\includegraphics[height=5.5cm]{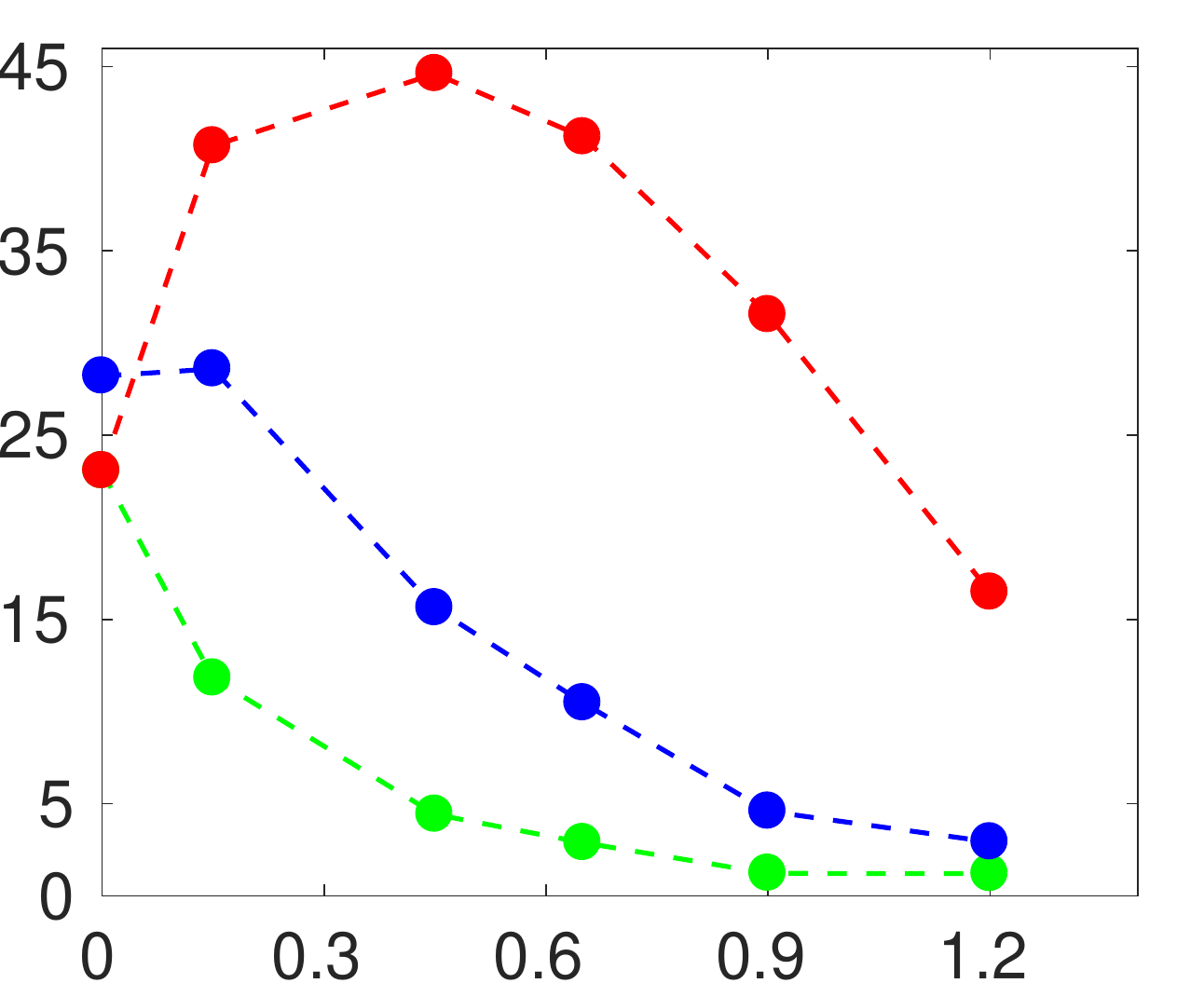}
\put(-3.4,-0.3){$Ro$}
\put(-7.2,2.5){$Nu$}
\hskip12mm
(b)\includegraphics[height=5.5cm]{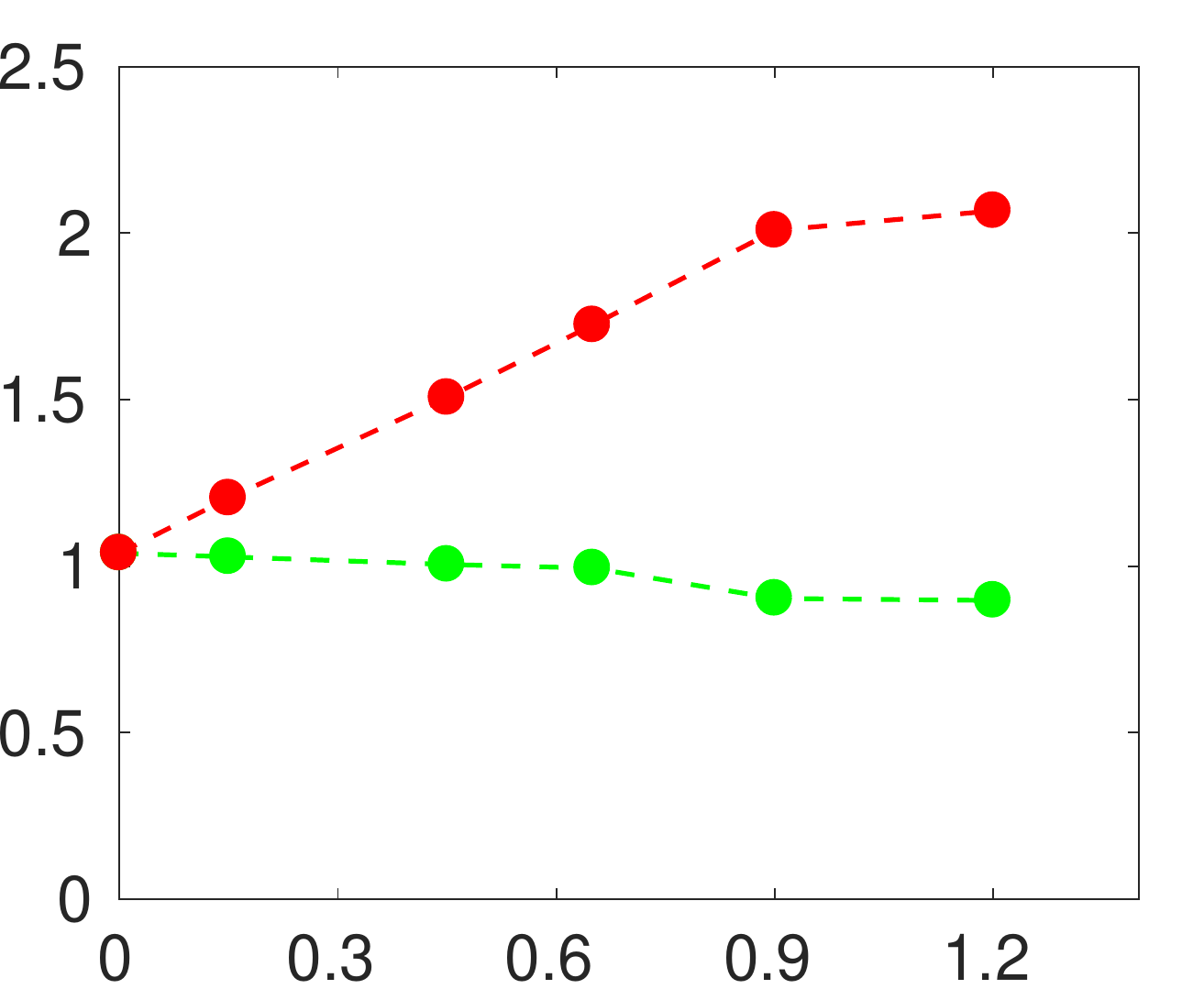}
\put(-3.4,-0.3){$Ro$}
\put(-7.8,2.7){$2St/C_f$}
\end{center}
\caption{(a) $Nu$ (blue line/symbols), $Nu^*_u$ (red line/symbols) and
$Nu^*_s$ (green line/symbols). (b) 
$2St/C_f$ on the unstable (red line/symbols) 
and stable channel side (green line/symbols). 
}
\label{nus}
\end{figure}
Figure \ref{nus}(a) shows $Nu$
together with $Nu^*_u$ and $Nu^*_s$ as function of $Ro$.
$Nu$ slightly grows at first but then declines rapidly with $Ro$,
implying that rotation drastically inhibits the cross channel scalar transport.
The behaviour of $Nu$ in the present case is different from that in 
a rotating channel flow at a lower $Re_\tau = 194$ when $Nu$
monotonically decreases with $Ro$ (Liu \& Lu 2007).
The high values of 
$Nu^*_u$ for $0.15 \leq Ro \leq 0.9$ 
and rapid decline of $Nu^*_s$ with $Ro$
reflect the rapid and slow turbulent
scalar transport on the unstable and stable channel sides, respectively.
Both $Nu$ and $Nu^*_s$ approach unity at high $Ro$ meaning that diffusive scalar transport
becomes significant.

Figure \ref{nus}(b) shows the ratio of the Stanton number to
skin friction coefficient $2St/C_f$ where 
\begin{equation}
St = \frac{Q_w}{U_m T_m},~~~~~
C_f = \frac{2u^2_\tau}{U_m}.
\label{st_eq}
\end{equation}
Here, $U_m$, $T_m$ and $u_\tau$ are either computed for the unstable or
the stable channel side according to equation (\ref{int_eq}). 
If the Reynolds analogy for the momentum and scalar transport is valid
$2St/C_f$ should be near unity (Abe \& Antonia 2017),
which indeed it is in the non-rotating channel (figure \ref{nus}.b).
On the stable channel side
$2St/C_f$ stays near unity while on the unstable channel side
it grows with $Ro$ and clearly deviates from unity, 
suggesting that the Reynolds analogy is valid
for the stable channel side but not for the unstable side
where scalar transport is relatively rapid.

\begin{figure}[t]
\begin{center}
\setlength{\unitlength}{1cm}
(a)\includegraphics[height=5.5cm]{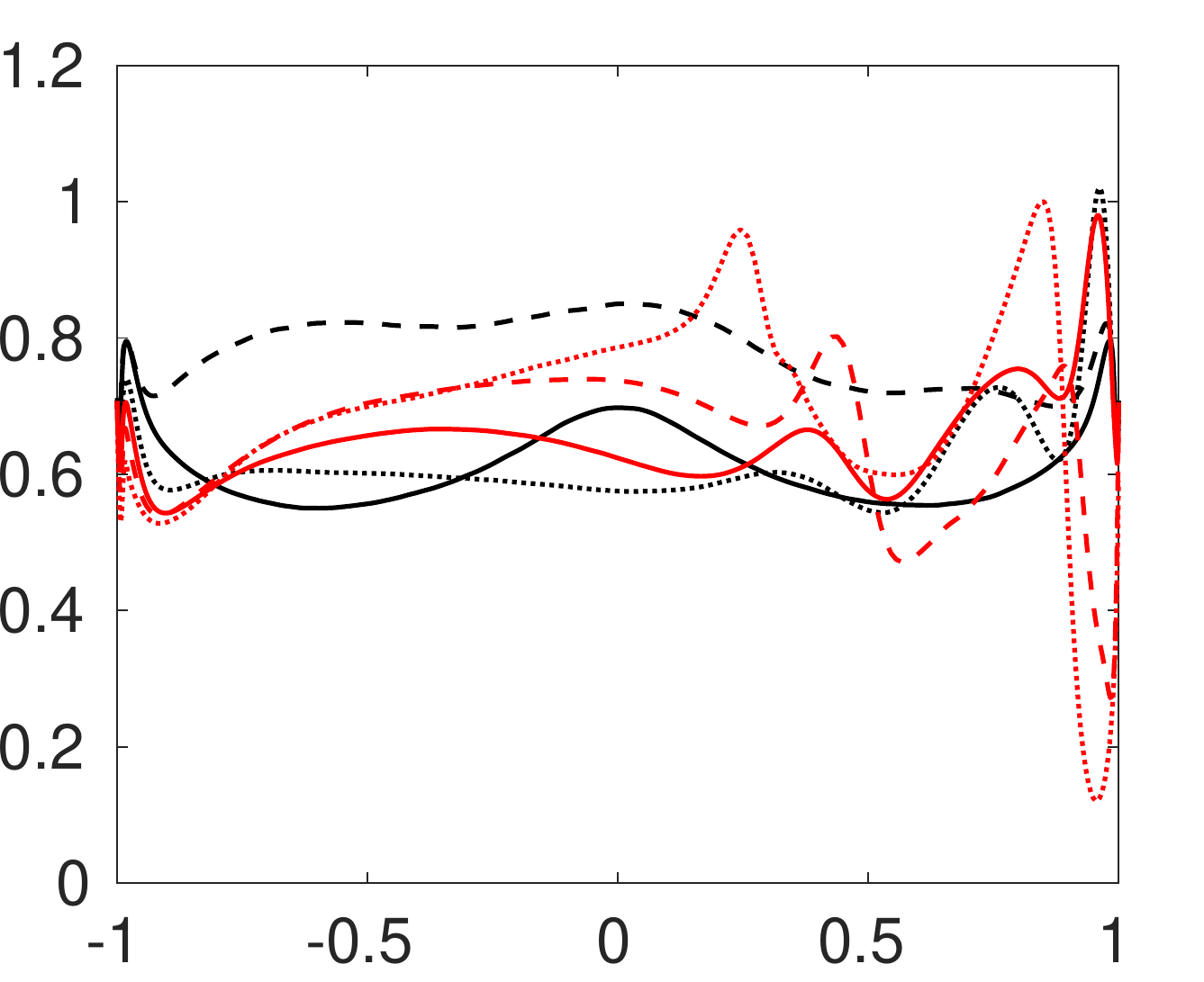}
\put(-3.4,-0.3){$y$}
\put(-7.2,2.6){$r$}
\hskip4mm
(b)\includegraphics[height=5.5cm]{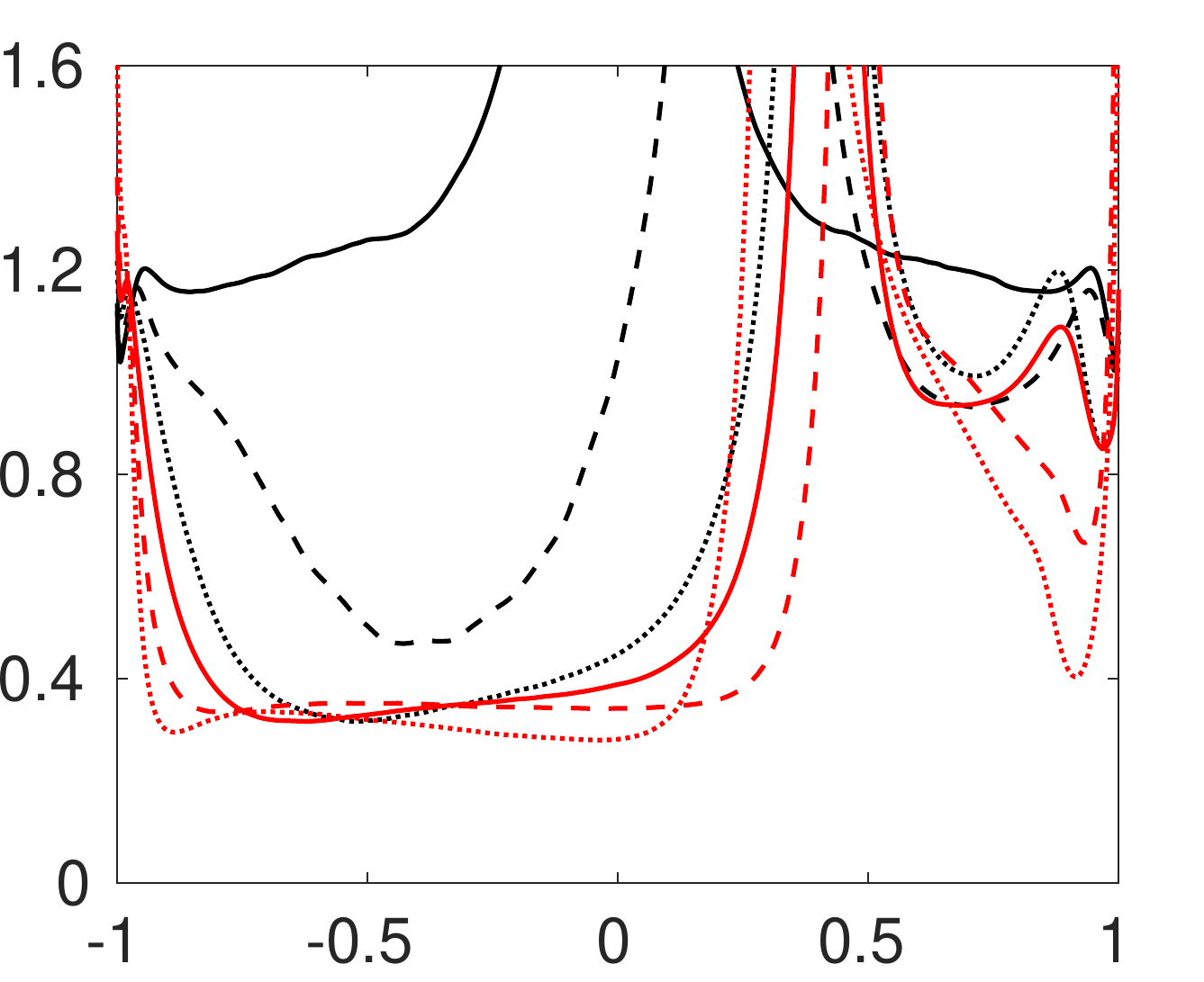}
\put(-3.4,-0.3){$y$}
\put(-7.4,2.6){$B$}

\vskip2mm
(c)\includegraphics[height=5.5cm]{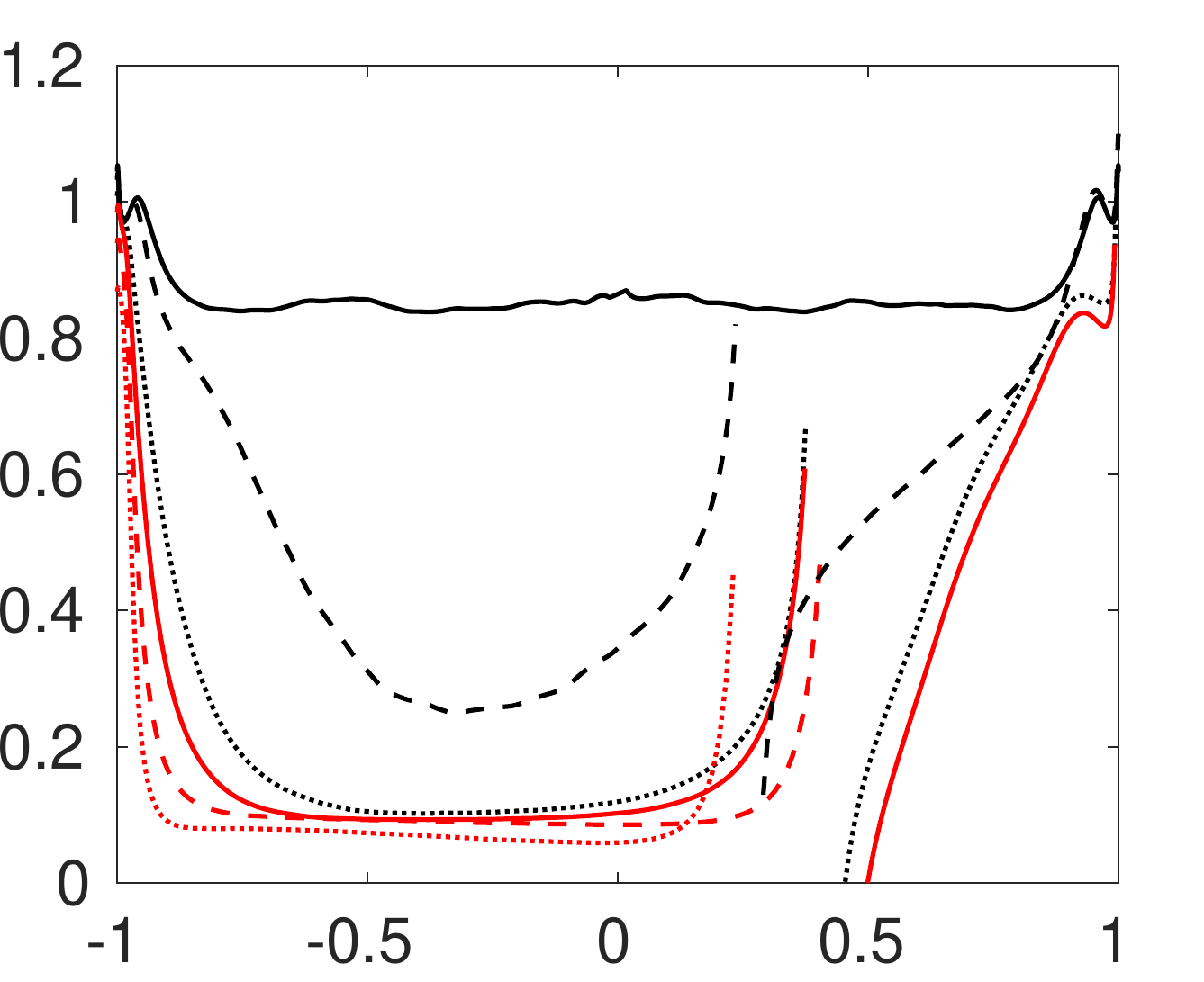}
\put(-3.4,-0.3){$y$}
\put(-7.6,2.6){$Pr_T$}
\hskip4mm
\end{center}
\caption{
Profiles of (a) $r$, (b) $B$ and (c) $Pr_T$.
Lines as in figure \ref{ustat}.
}
\label{turb_pr}
\end{figure}
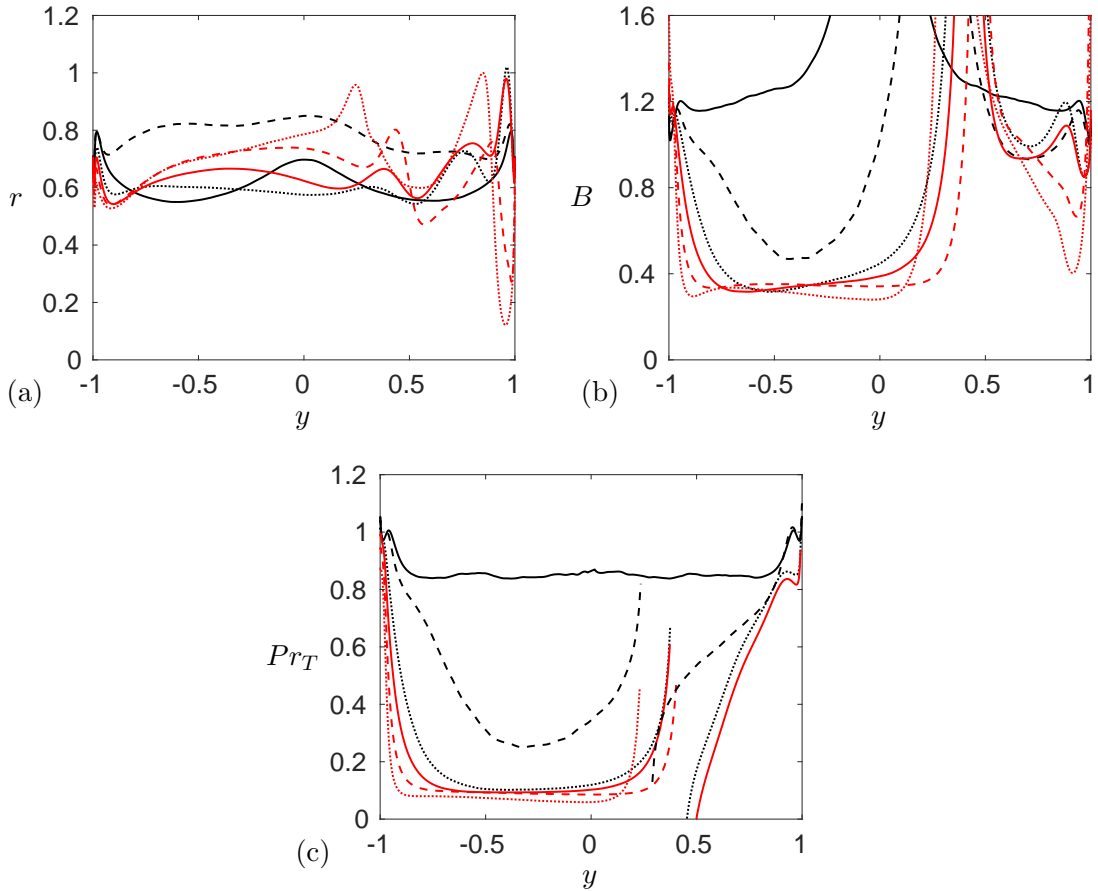

The ratio of scalar to turbulence time-scale
\begin{equation}
r = \frac{K_\theta / \varepsilon_\theta}{K/\varepsilon}
\end{equation}
is in some turbulence models assumed to be a constant. Antonia \etal (2009)
found that in non-rotating turbulent channel flow with a passive scalar with $Pr=0.71$,
$r$ is indeed nearly constant; its value 
is between 0.5 and 0.6 in the outer region. 
In rotating homogeneous shear flow, however, $r$ varies with the imposed 
system rotation rate (Brethouwer 2005).
Figure \ref{turb_pr}(a) shows the profiles of $r$ in rotating channel flow.
The variation of $r$ across the channel in the present DNS at $Ro=0$ is larger than in the DNS
by Antonia \etal (2009), which is the consequence of a different forcing
of the scalar field in those two DNS. Nevertheless, in the outer region
but outside the centre region $r$ is close to that observed by Antonia \etal (2009).
In rotating channel flow, $r$ at first grows with $Ro$ and varies
between 0.7 and 0.85 at $Ro=0.15$, but then declines somewhat on the
unstable channel side and varies there between approximately 0.6 and 0.7.
Thus, the time-scale ratio $r$ displays some variations with $Ro$ but these are not very large.

A measure of turbulence to scalar fluctuation intensity is given by the
dimensionless parameter
\begin{equation}
B = \frac{\sqrt{K} / \diff U /\diff y}{\sqrt{K}_\theta / \diff \Theta / \diff y}
\end{equation}
Figure \ref{turb_pr}(b) shows the profiles of $B$ in the present DNS. 
The values of $B$ of approximately 1.2 to 1.3 at $Ro=0$
in the outer regions $-0.95 \lesssim y \lesssim -0.6$ and $0.4 \lesssim y \lesssim 0.95$
are the same as those observed by Antonia \etal (2009) in their DNS of non-rotating
turbulent channel flow at similar $Re$.
In the centre region, $B$ is much larger in the present DNS
since the mean velocity gradient is small,
unlike the fluctuations and mean scalar gradient.
With increasing $Ro$, however, $B$ declines strongly in the outer region of the unstable
channel side to less than 0.4, meaning that rotation augments scalar fluctuations
relative to velocity fluctuations. The same trend is observed
in rotating homogeneous shear flow with a passive scalar (Brethouwer 2005).
Especially when the mean shear $\diff U/\diff y$ equals $2 \Omega$, as in
the absolute mean vorticity region in the present DNS, scalar fluctuations
are relatively strong, although neither in the governing equation of $K$ nor $K_\theta$
there is a direct influence of rotation. 
The small values of $B$ in rotating channel flow, however, 
suggest that $P_\theta$ is relatively
large compared to the production of $K$.

An important parameter in turbulence modelling, often assumed to be a constant, is
the ratio of turbulent viscosity to scalar diffusivity given by the turbulent Prandtl number
\begin{equation}
Pr_T = \frac{\nu_T}{\alpha_T}=\frac{\overline{uv}}{\overline{v\theta}}
\frac{\diff \Theta /\diff y}{\diff U / \diff y}.
\end{equation}
It can also be considered as a relative measure of the production term of $K$
to $K_\theta$.
Figure \ref{turb_pr}(c) shows the profiles of $Pr_T$ obtained from the present DNS.
At $Ro=0$, $Pr_T \simeq 0.85$ in the outer region, consistent with the values
computed by Antonia \etal (2009) and Pirozzoli \etal (2016) for non-rotating channel flow.
On the other hand, in rotating channel flow $Pr_T$ is considerably smaller in the outer region
of the unstable channel side; it is less than 0.2 for $Ro \geq 0.45$.
Also on the stable channel side at $Ro=0.15$, $Pr_T$ is smaller than in
the non-rotating case away from the wall. 
These results show, like the ratio $2St/C_f$ in figure \ref{nus}.(b), 
that the Reynolds analogy for scalar-momentum transfer does
not hold for rotating channel flow and that turbulent scalar transport
is very efficient compared to momentum transfer on the unstable channel side.
The low value of $Pr_T$ also implies that
the production of $K_\theta$ is relatively strong to that of $K$,
consistent with the small values of $B$ observed in rotating channel flow.
Small values of $Pr_T$ are also observed in rotating homogeneous shear flow
when the imposed shear $\diff U/\diff y = 2 \Omega$, where $\Omega$ is the imposed
system rotation rate (Brethouwer 2005), consistent with the present results.

The small values of $B$ and $Pr_T$ suggest that 
the scalar and velocity field are dissimilar in rotating channel flow,
as already indicated by the correlation coefficients in figure \ref{correl}. 
This dissimilarity is further explored 
by spectra in the next section.

\section{8. Spectra}

\begin{figure}
\begin{center}
\setlength{\unitlength}{1cm}
(a)\includegraphics[height=4.2cm]{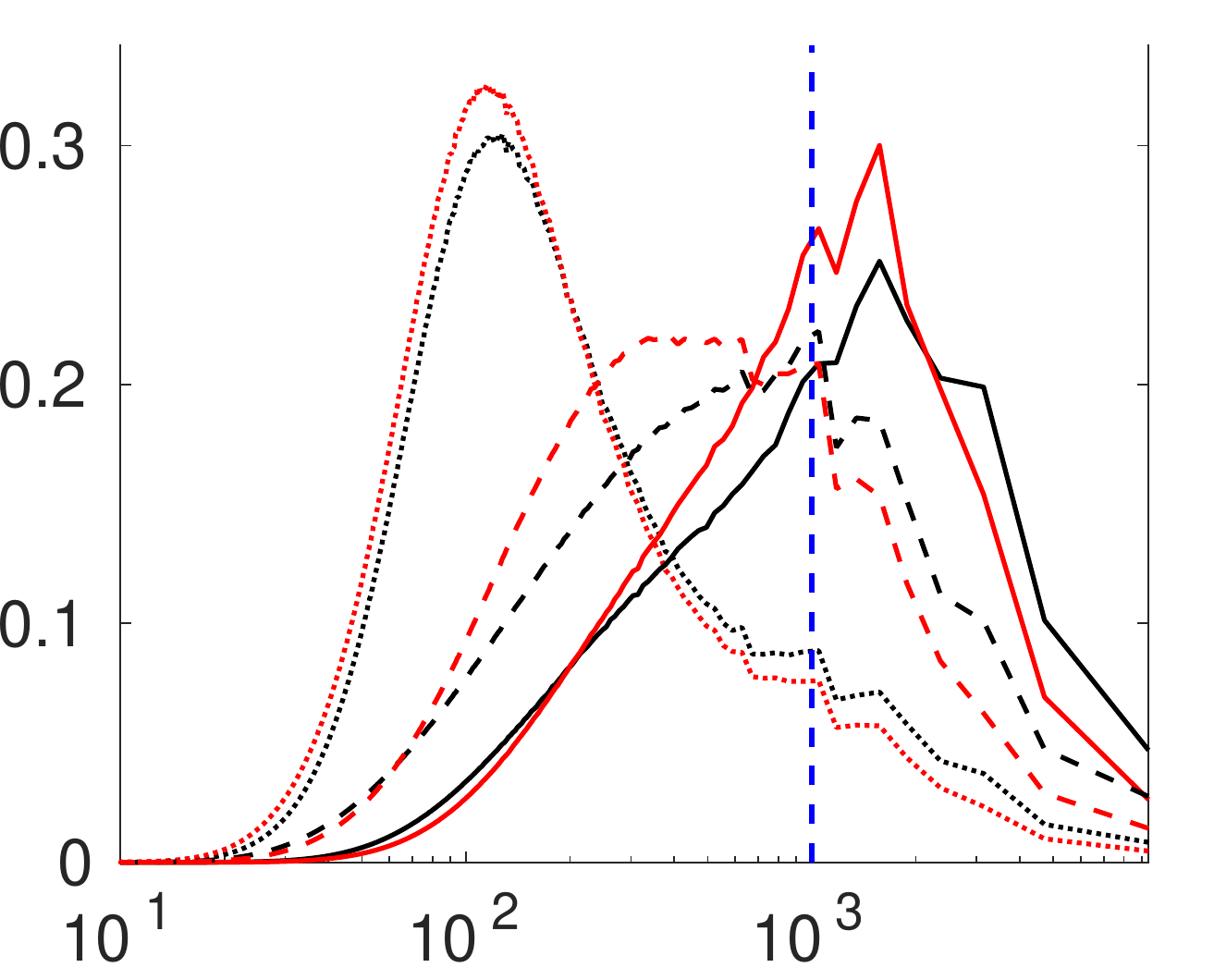}
\put(-2.7,-0.3){$\lambda^*_z$}
\put(-7.0,2.4){$k_z E_K(k_z)$}
\put(-7.0,1.7){$k_z E_{\theta\theta}(k_z)$}
\hskip16mm
(b)\includegraphics[height=4.2cm]{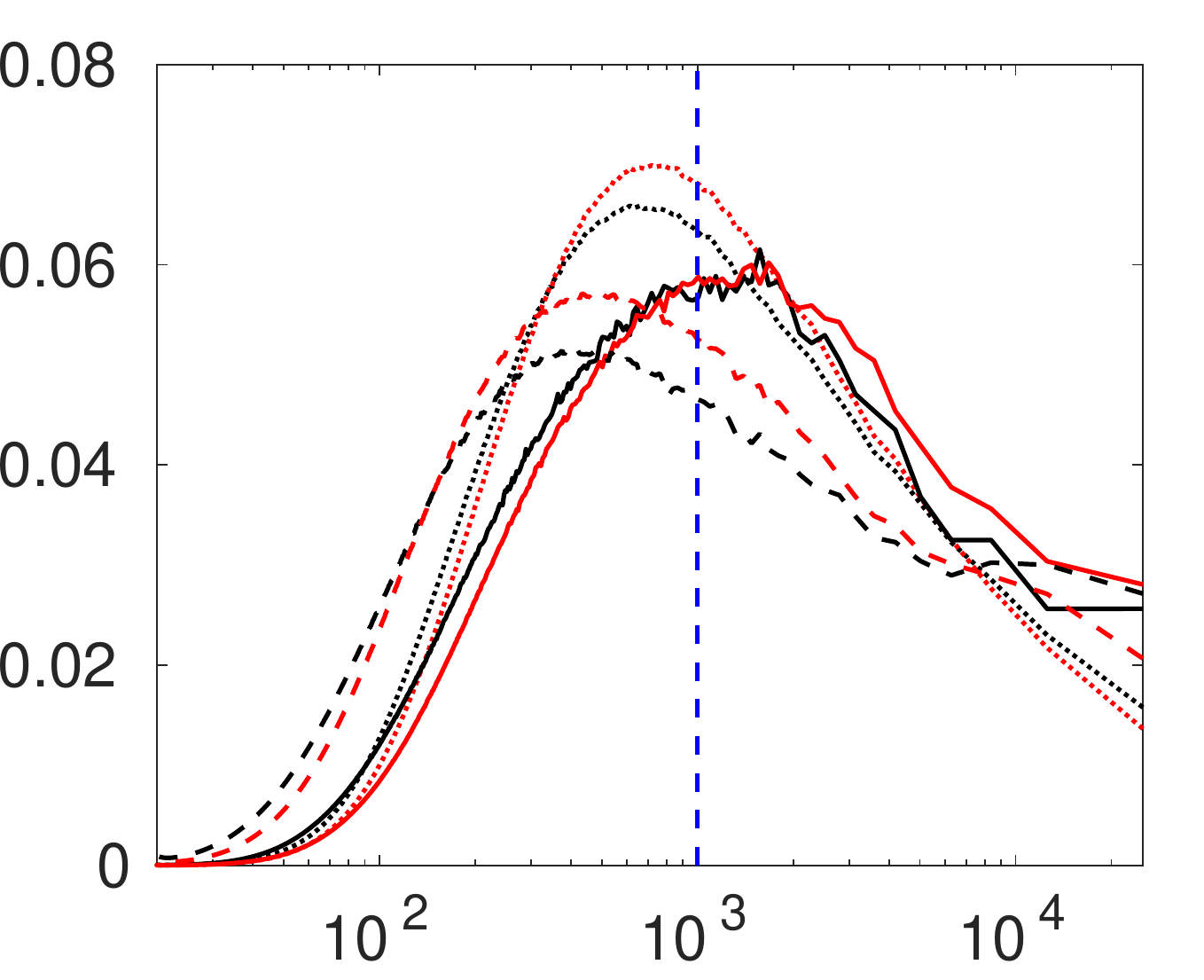}
\put(-2.9,-0.3){$\lambda^*_x$}
\put(-7.0,2.4){$k_x E_K(k_x)$}
\put(-7.0,1.7){$k_x E_{\theta\theta}(k_x)$}

(c)\includegraphics[height=4.2cm]{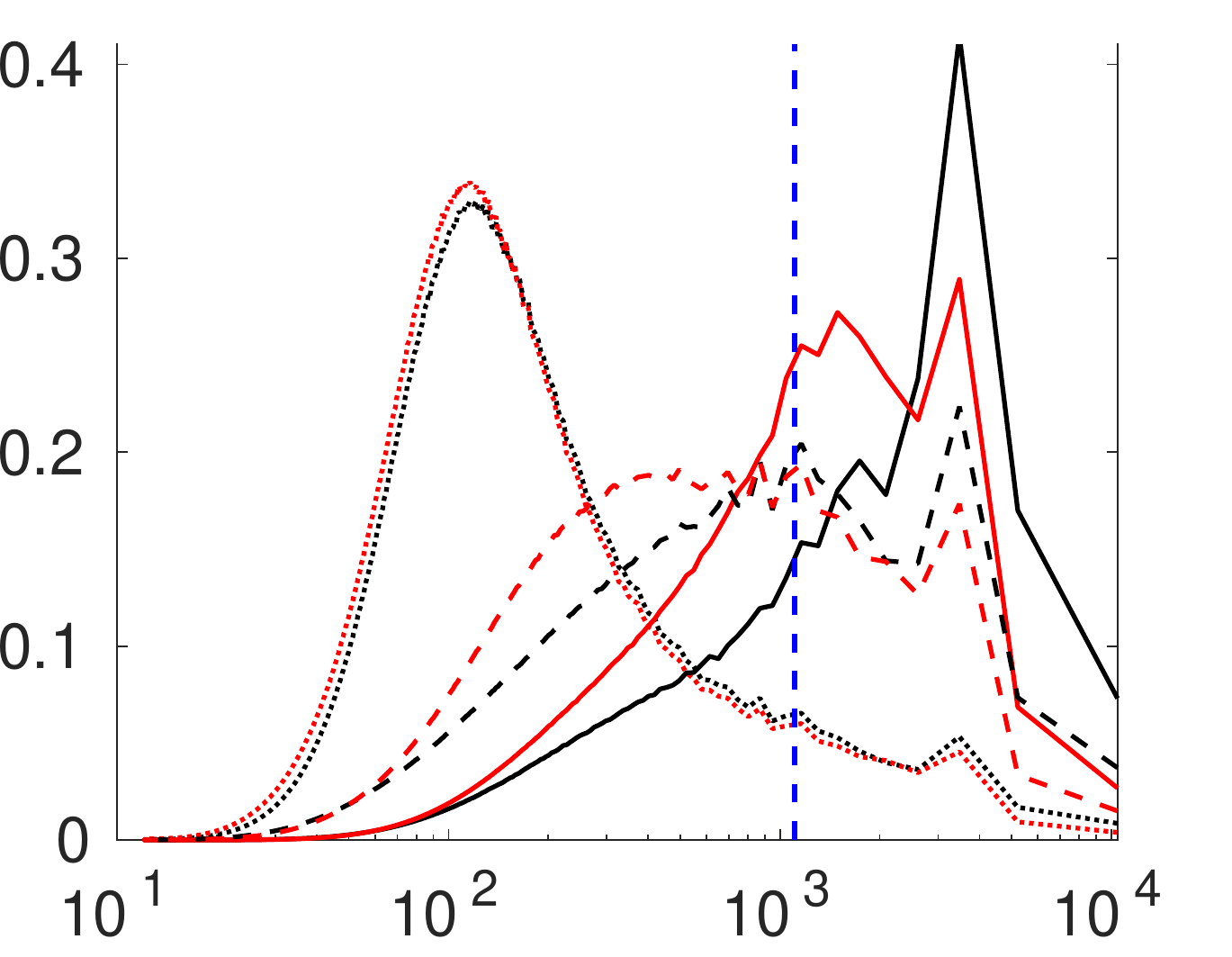}
\put(-2.7,-0.3){$\lambda^*_z$}
\put(-7.0,2.4){$k_z E_K(k_z)$}
\put(-7.0,1.7){$k_z E_{\theta\theta}(k_z)$}
\hskip16mm
(d)\includegraphics[height=4.2cm]{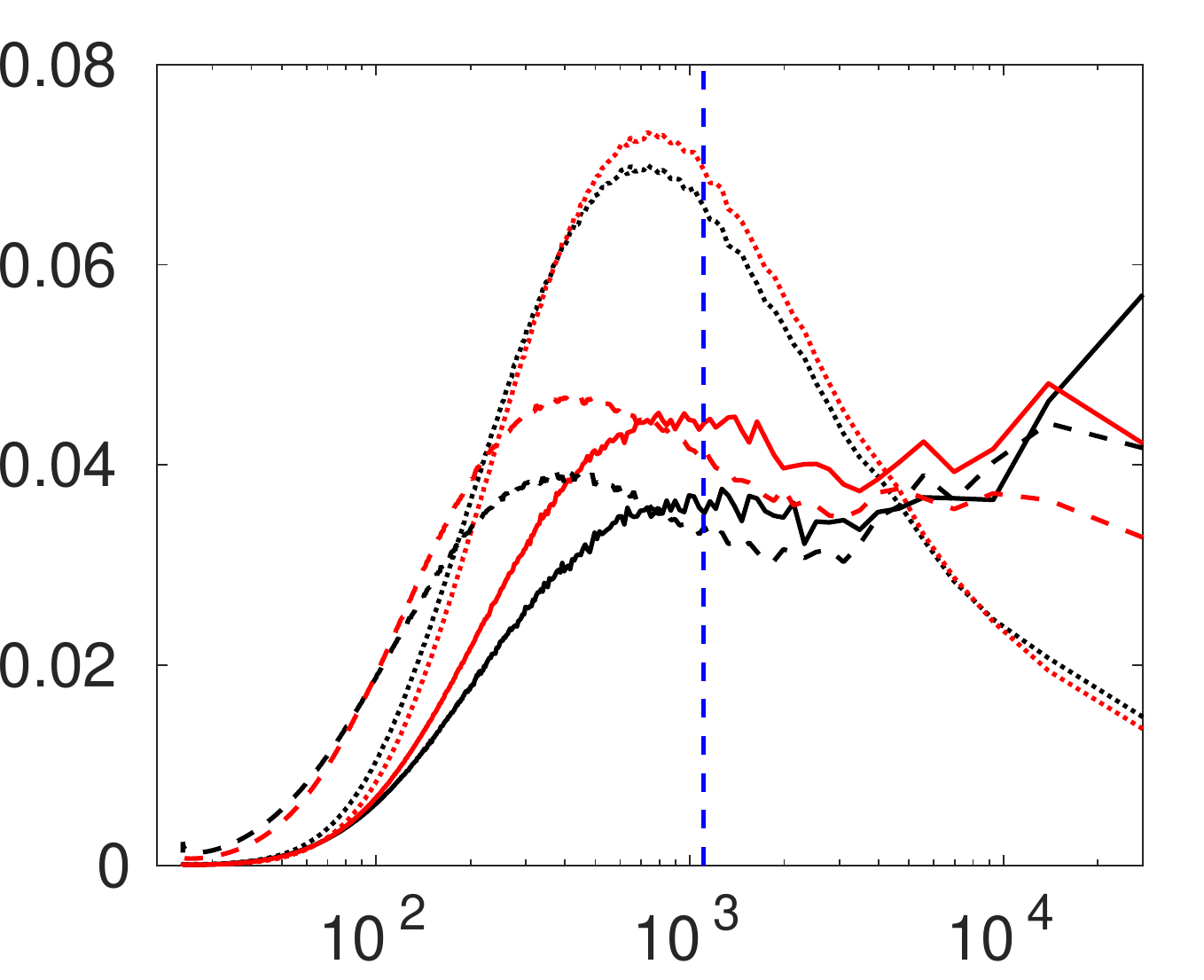}
\put(-2.9,-0.3){$\lambda^*_x$}
\put(-7.0,2.4){$k_x E_K(k_x)$}
\put(-7.0,1.7){$k_x E_{\theta\theta}(k_x)$}

(e)\includegraphics[height=4.2cm]{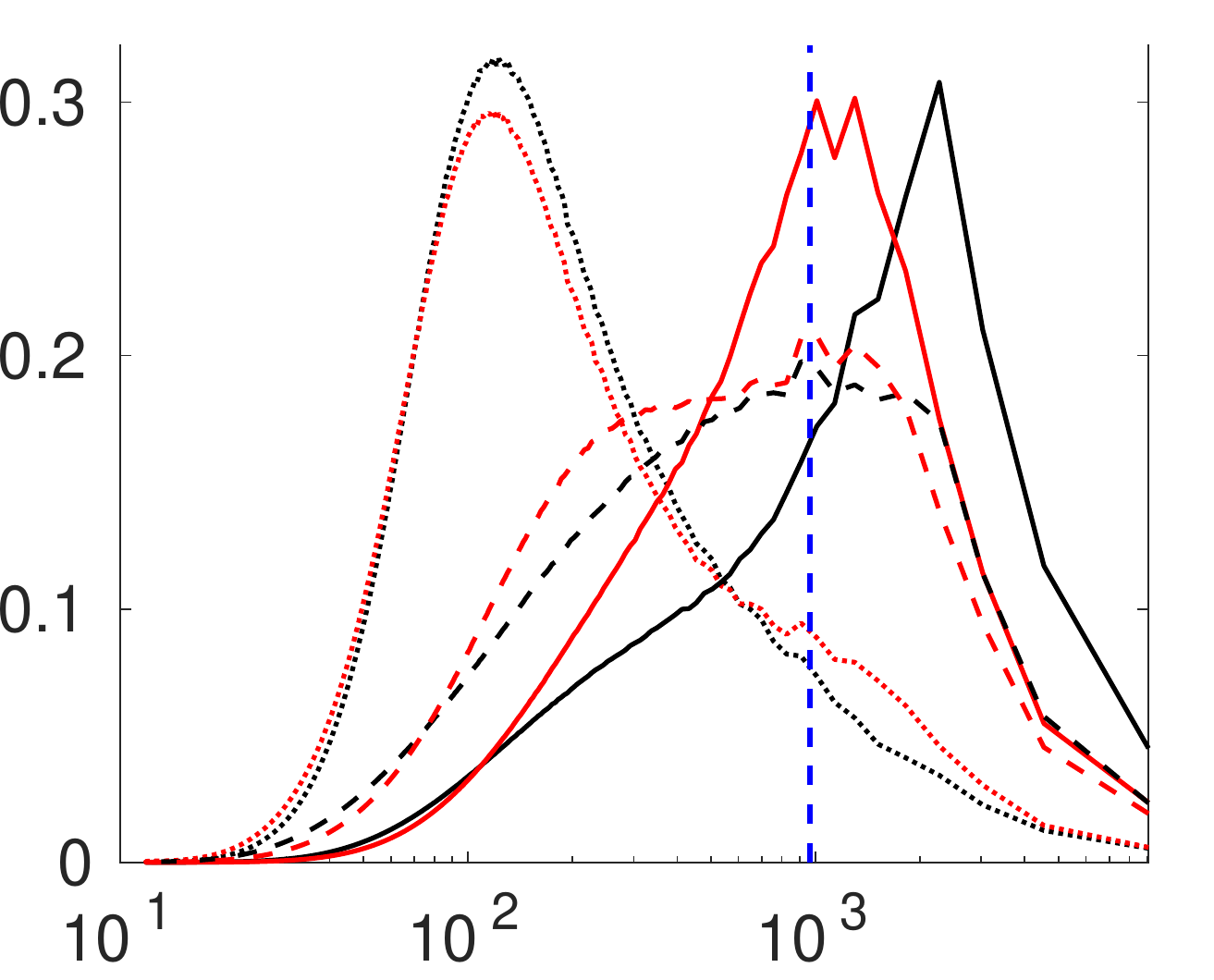}
\put(-2.7,-0.3){$\lambda^*_z$}
\put(-7.0,2.4){$k_z E_K(k_z)$}
\put(-7.0,1.7){$k_z E_{\theta\theta}(k_z)$}
\hskip16mm
(f)\includegraphics[height=4.2cm]{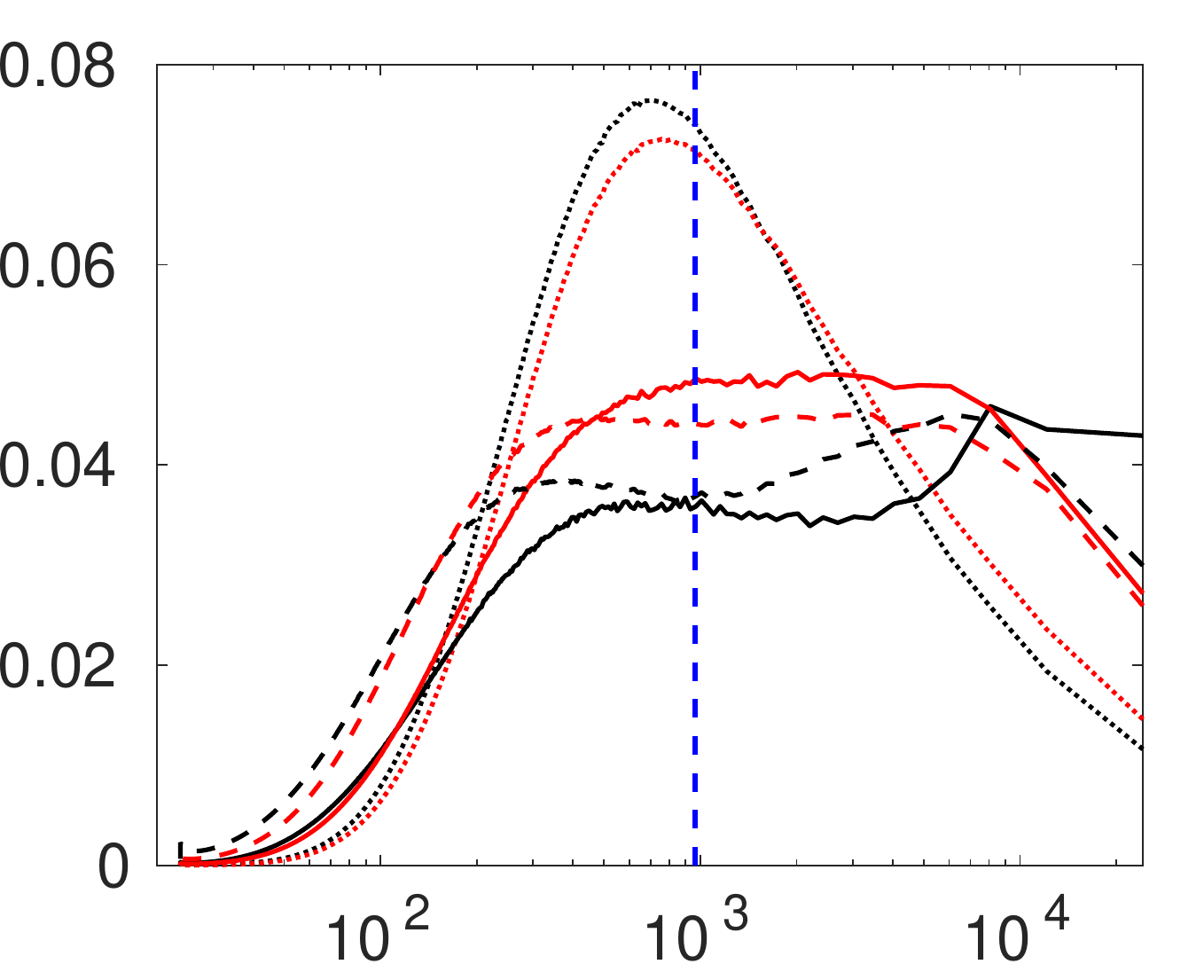}
\put(-2.9,-0.3){$\lambda^*_x$}
\put(-7.0,2.4){$k_x E_K(k_x)$}
\put(-7.0,1.7){$k_x E_{\theta\theta}(k_x)$}

(g)\includegraphics[height=4.2cm]{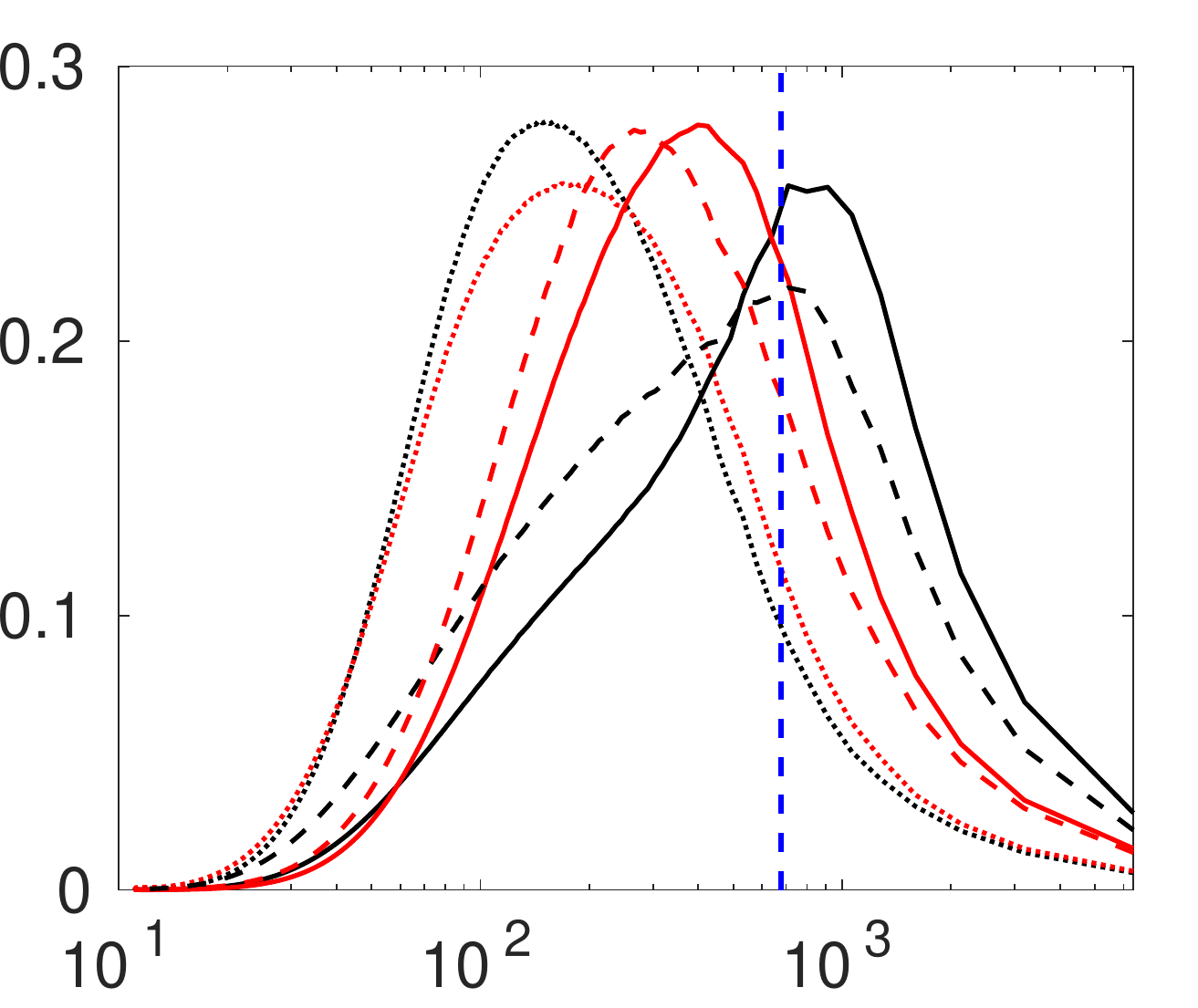}
\put(-2.7,-0.3){$\lambda^*_z$}
\put(-7.0,2.4){$k_z E_K(k_z)$}
\put(-7.0,1.7){$k_z E_{\theta\theta}(k_z)$}
\hskip16mm
(h)\includegraphics[height=4.2cm]{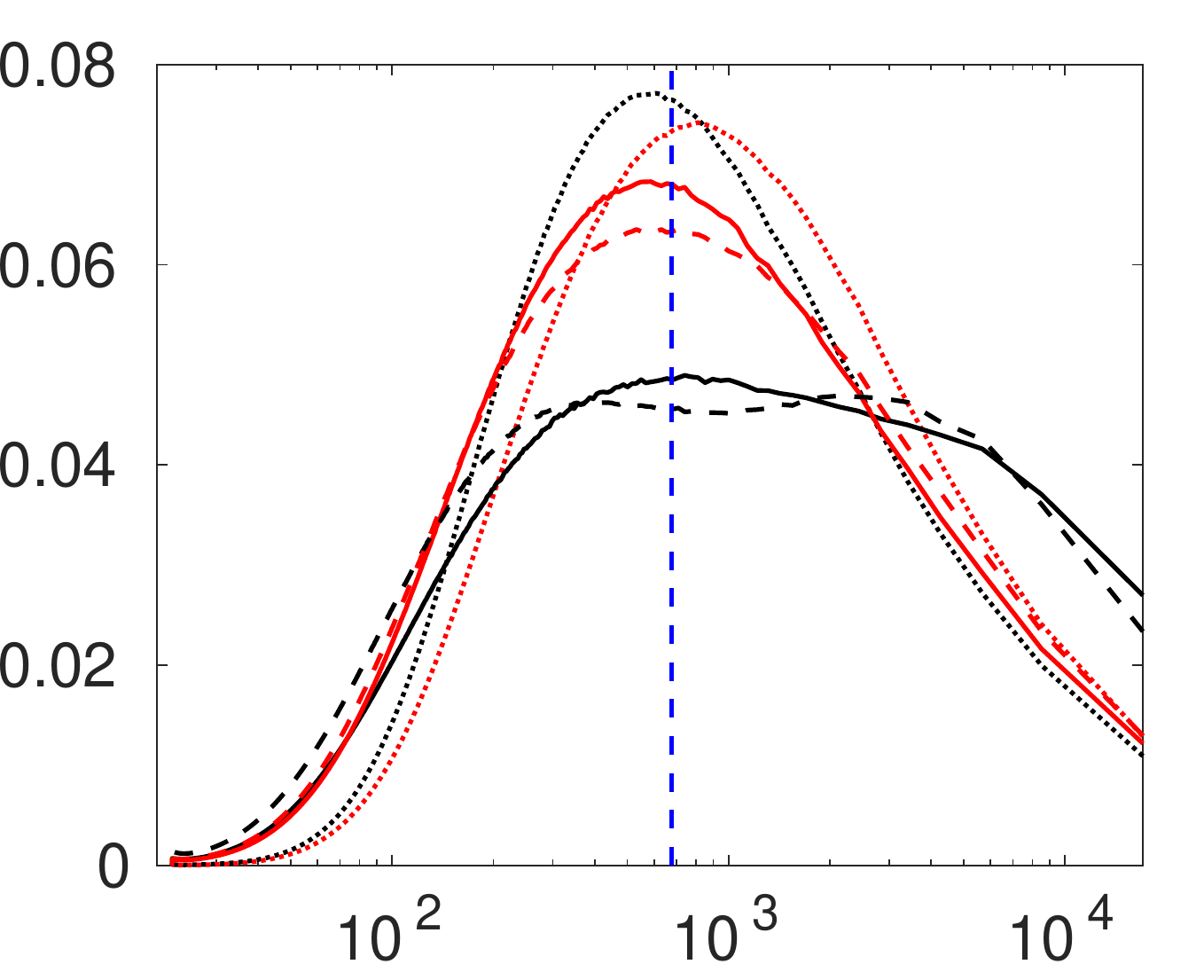}
\put(-2.9,-0.3){$\lambda^*_x$}
\put(-7.0,2.4){$k_x E_K(k_x)$}
\put(-7.0,1.7){$k_x E_{\theta\theta}(k_x)$}
\end{center}
\caption{Premultiplied one-dimensional spanwise and streamwise
spectra as function of $\lambda^*_z$ respective $\lambda^*_x$
at $Ro=0$ ({\it a, b}), $Ro=0.15$ ({\it c, d}), $Ro=0.45$ ({\it e, f})
and $Ro=0.9 $ ({\it g, h}).
Left column shows the spanwise spectra
$k_z E_{K}(k_z)$ (red lines) and $k_z E_{\theta\theta}(k_z)$ (black lines).
Right column shows streamwise spectra
$k_x E_{K}(k_x)$ (red lines) and $k_x E_{\theta\theta}(k_x)$ (black lines).
The spectra are at $y^* \approx 10$ (dotted lines), $y^* \approx 100$ (dashed lines)
and $y \approx -0.55$ (solid lines).
The vertical dashed line indicates scales with $\lambda_x = \lambda_z = 1$.
}
\label{spec}
\end{figure}
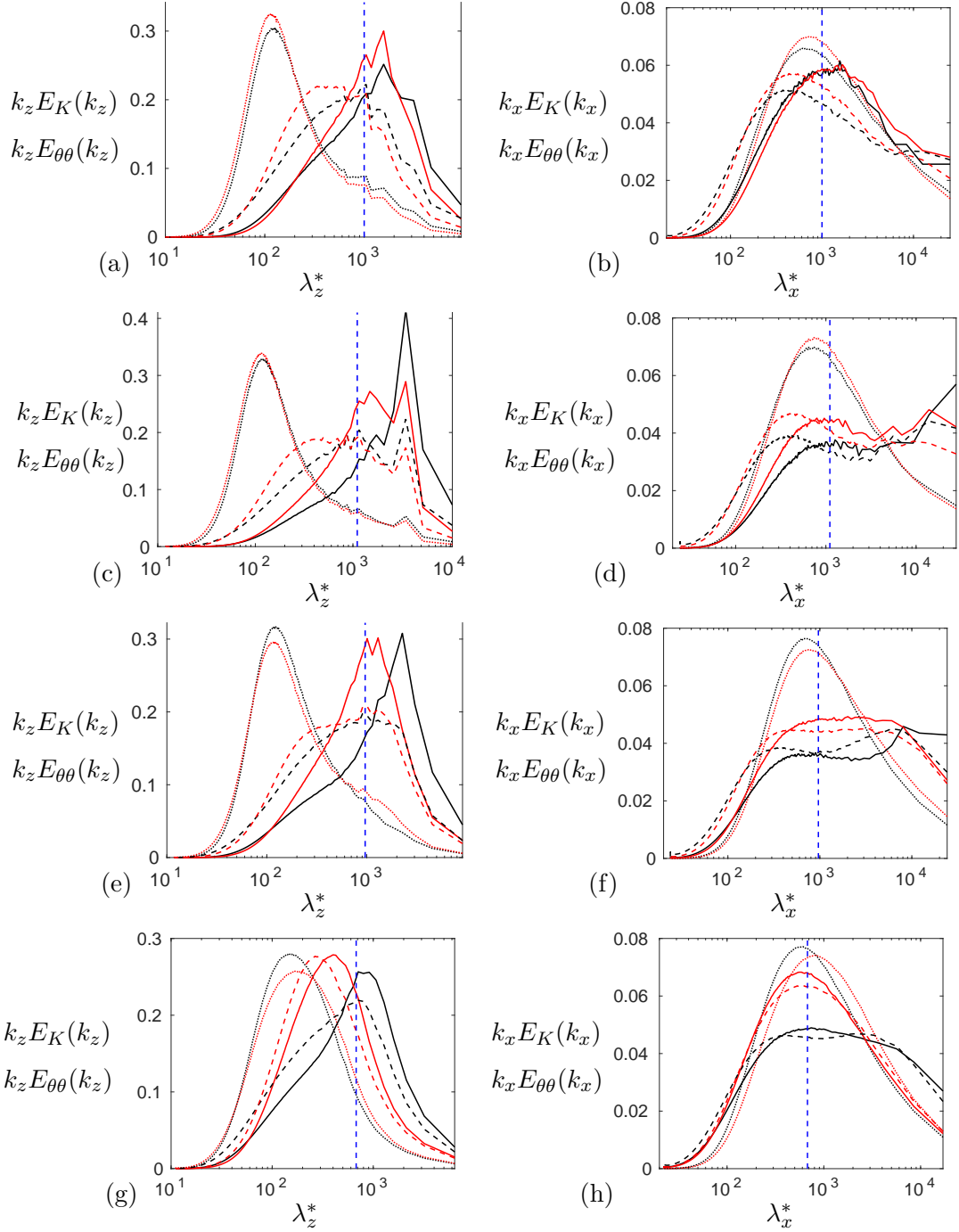

Premultiplied one-dimensional spanwise spectra 
of turbulent kinetic energy $k_z E_K(k_z)$ 
and scalar variance $k_z E_{\theta\theta}(k_z)$ as well as
premultiplied one-dimensional streamwise spectra 
of turbulent kinetic energy $k_x E_K(k_x)$ 
and scalar variance $k_x E_{\theta\theta}(k_x)$ 
at $Ro=0$, 0.15, 0.45 and 0.9 are shown in figure \ref{spec}
as function of the spanwise and streamwise non-dimensional wave length
$\lambda^*_z$ and $\lambda^*_x$, respectively.
Here, $k_x$ and $k_z$ are the streamwise and spanwise wave number, respectively.
The streamwise and spanwise spectra are normalized such that 
\begin{equation}
\int^\infty_0 E_\beta (k_x)\diff k_x =
\int^\infty_0 E_\beta (k_z)\diff k_z = 1,
\end{equation}
respectively ($\beta$ stands for $K$ or $\theta\theta$),
and are computed at three different wall-normal positions on the unstable channel side,
i.e. near the wall at $y^* \approx 10$,
at $y^* \approx 100$ and in the outer region at $y \approx -0.55$.
A superscript $*$ implies scaling by $\nu/u_{\tau u}$,
which is appropriate for the spectra on the unstable channel side.

At $Ro=0$, $k_z E_{\theta\theta}(k_z)$
shows near the wall a peak at $\lambda^*_z \approx 100$, corresponding to the
turbulent near-wall cycle, while in the outer region 
it has a peak at scales larger than $\lambda_z =h$ (figure \ref{spec}.a), 
like the scalar spectra presented by Pirozzoli \etal (2016) for non-rotating channel flow.
Also $k_x E_{\theta\theta}(k_x)$ reveals large energetic scalar scales in the outer region
(figure \ref{spec}.b). 
Antonia \etal (2009) observed that the scalar variance spectra
$k_x E_{\theta\theta}(k_x)$ and $k_z E_{\theta\theta}(k_z)$  
display considerable similarities with the turbulent kinetic energy spectra
$k_x E_K(k_x)$ and $k_z E_K(k_z)$, respectively, especially near the wall.  
This similarity is also reflected by the present spectra, see
figure \ref{spec}.(a) and (b), although the spanwise spectra indicate
that away from the wall the scalar scales are somewhat wider than the turbulent scales. 
Overall, the scalar and turbulent length scales are thus similar in non-rotating
channel flow according to the spectra.
Antonia \etal (2009) found that 
the similarity between the spanwise spectra 
$k_z E_{\theta\theta}(k_z)$ and $k_z E_K(k_z)$ is mostly the consequence of
a similarity between $\theta$ and streamwise velocity fluctuations $u$,
whereas the streamwise scalar variance spectrum 
$k_x E_{\theta\theta}(k_x)$ is more similar to $k_x E_K(k_x)$,
especially at larger scales, 
than the spectrum of the streamwise velocity component only.
In the outer region, the large scales of the scalar field display in fact quite large correlations
with wall-normal velocity fluctuations.

In rotating channel flow on the unstable side, 
$k_x E_{\theta\theta}(k_x)$ and $k_z E_{\theta\theta}(k_z)$
show a high similarity
with $k_x E_K(k_x)$ respective $k_z E_K(k_z)$ near the wall at $y^* \approx 10$ 
(figure \ref{spec}.c-h), although at $Ro=0.9$ the scalar scales appear somewhat shorter
than the turbulence scales 
(figure \ref{spec}.h).
By contrast, away from the wall, at 
$y^* \approx 100$ and in the outer region at $y \approx -0.55$,
$k_x E_{\theta\theta}(k_x)$ and $k_z E_{\theta\theta}(k_z)$ are more skewed towards
longer and wider scales
than $k_x E_K(k_x)$ respective $k_z E_K(k_z)$
(figure \ref{spec}.c-h).
This difference appears to become stronger at higher $Ro$, implying that the scalar field
contains larger scales than the turbulence field in rotating channel flow.
This is another demonstration of the growing difference between 
the scalar and turbulence field on the unstable channel side as a consequence
of rotation.
The sharp peaks at $\lambda_z = \pi h$ and $3\pi h/4$ in
$k_z E_{\theta\theta}(k_z)$ at $Ro=0.15$ and 0.45, respectively
(figure \ref{spec}.c,e) and the energetic long scales observed in
$k_x E_{\theta\theta}(k_x)$ at the same $Ro$
(figure \ref{spec}.d,f) are most likely caused by large streamwise roll cells
present at these $Ro$ (Brethouwer 2017), confirming that roll cells
have a large impact on the scalar field.

The spectra at $Ro=0.15$ on the stable channel side (not shown) indicate that
the scalar field is more affected by the roll cells than the turbulence
field, but otherwise the turbulence and scalar scales appear similar.

\section{9. Conclusions}

This paper reports a DNS study of passive scalar transport in turbulent channel flow
subject to spanwise system rotation at a constant $Re = 20\,000$ and $Pr=0.71$
while $Ro$ is varied from 0 to 1.2. The scalar value is constant but different
at the two walls, leading to a constant mean scalar flux in the wall-normal direction 
in the statistically stationary state. The DNS show
that rotation has a large impact on scalar transport. In rotating
channel flow turbulence and turbulent scalar transport are relatively strong
on the unstable channel side,
whereas on the stable channel side with weak turbulence or laminar-like flow,
turbulent scalar transport is weak.
Although the turbulence is weaker,
the scalar fluctuations are intense on the stable side or near the
border between the stable and unstable channel side since the mean scalar gradient
is here steep. 

The main conclusions of the study are that
({\em i}) rotation weakens the similarity between the scalar and velocity field,
({\em ii}) the Reynolds analogy for scalar-momentum transfer does not hold for
rotating turbulent channel flow.
This is in obvious contrast to scalar transport in non-rotating turbulent channel flow
where a stronger similarity between the scalar and velocity field exists
and the Reynolds analogy is valid, see. e.g. Antonia \etal (2009) and Pirozzoli \etal (2016).
In rotating channel flow the correlations between streamwise respective wall-normal
velocity fluctuations and scalar fluctuations are weaker on the unstable respective
stable channel side and the turbulent Prandtl number in the outer region
of the unstable side is considerably smaller, below 0.2 at higher $Ro$, 
than in non-rotating channel flow.
This implies that turbulent scalar transfer is very efficient on the unstable
side of rotating channel flow compared to momentum transfer.
Owing to rotation, 
the Nusselt number and the ratio of the Stanton number to skin friction both change,
and the streamwise turbulent scalar flux is reduced on the unstable channel,
leading to a stronger alignment between the turbulent scalar flux 
vector and mean scalar gradient.
One-dimensional spectra of the scalar variance and turbulent kinetic energy show
that the scalar scales grow relatively to the turbulence scales in the 
outer region of the unstable side as a result of rotation,
which is another manifestation of the growing dissimilarity between
velocity and scalar field. 
Budgets of the governing equations for the scalar variance and streamwise
and wall-normal turbulent scalar fluxes are presented and discussed. These
show that the Coriolis terms in the scalar flux equations reduce the streamwise
and wall-normal turbulent scalar flux on the unstable and stable channel
side, respectively.

Further studies of scalars in rotating shear
flows are motivated given the significant
effect of rotation on turbulent scalar transport and the prevalence of
mass and heat transfer in rotating turbulent flows in engineering applications.
However, modelling the effect of rotation on turbulent heat and mass transfer
might pose a challenge since simple and convenient assumptions like
the Reynolds analogy are not necessarily valid, as demonstrated by the present study.


\end{document}